\documentclass[MSNbibl,number,citesort,dvips]{arxbj}
\usepackage{upgreek}
\usepackage{graphicx}

\aid{0}
\volume{20}
\issue{4}
\pubyear{2014}
\firstpage{1930}
\lastpage{1978}
\doi{10.3150/13-BEJ546} 

\makeatletter
\newcommand{\rrvert}{\vert}
\newcommand{\llvert}{\vert}
\newcommand{\iint}{\int\!\!\!\int}
\renewcommand{\emptyset}{\varnothing}
\newcommand{\eqref}[1]{(\ref{#1})}

\newcommand{\acc}{\operatorname{acc}}
\newcommand{\sign}{\operatorname{sign}}
\newcommand{\eqsp}{}
\newcommand{\cG}{\mathcal{G}}

\newcommand{\tN}{\mathbb{N}}
\newcommand{\R}{\mathbb{R}}
\newcommand{\Z}{\mathbb{Z}}
\renewcommand{\P}{\mathbb{P}}
\newcommand{\E}{\mathbb{E}}
\newcommand{\argmax}{\arg\max}
\newproclaim{definition}{Definition}
\newtheorem{theorem}{Theorem}
\newtheorem{proposition}{Proposition}
\newtheorem{lemma}{Lemma}
\newremark{remark}{Remark}
\newtheorem{Lemma}{Lemma}[section]
\makeatother

\begin{document}
\begin{frontmatter}

\title{Optimal scaling for the transient phase of Metropolis Hastings
algorithms: The~longtime behavior}
\runtitle{Optimal scaling for the transient phase of MH algorithms}

\begin{aug}
\author{\inits{B.}\fnms{Benjamin} \snm{Jourdain}\thanksref{e1}\ead[label=e1,mark]{jourdain@cermics.enpc.fr}},
\author{\inits{T.}\fnms{Tony} \snm{Leli\`{e}vre}\thanksref{e2}\ead[label=e2,mark]{lelievre@cermics.enpc.fr}} \and
\author{\inits{B.}\fnms{B\l{}a\.{z}ej} \snm{Miasojedow}\corref{}\thanksref{e3}\ead[label=e3,mark]{bmia@mimuw.edu.pl}}
\address{Universit\'{e} Paris-Est, CERMICS, 6 \& 8, avenue Blaise Pascal,
77455 Marne-La-Vall\'ee, France.\\ \printead{e1,e2,e3}}
\end{aug}

\received{\smonth{1} \syear{2013}}

%
\begin{abstract}
We consider the Random Walk Metropolis algorithm on $\mathbb{R}^n$
with Gaussian
proposals, and when the target probability measure is the $n$-fold
product of a one-dimensional law. It is well known (see
Roberts \textit{et al.}
(\textit{Ann. Appl. Probab.} \textbf{7} (1997) 110--120))
that, in the limit
$n \to\infty$, starting at equilibrium and for an appropriate scaling
of the variance and of the timescale as a function of the dimension
$n$, a
diffusive limit is obtained for each component of the Markov chain. In
Jourdain \textit{et al.}
(Optimal scaling for the transient phase of the random walk
Metropolis algorithm: The mean-field limit (2012) Preprint), we
generalize this result when the initial distribution
is not the
target probability measure. The obtained diffusive limit is the
solution to a stochastic differential equation nonlinear in the sense
of McKean. In the present paper, we prove convergence to equilibrium
for this equation. We discuss practical counterparts in order to
optimize the variance of the proposal distribution to accelerate
convergence to equilibrium. Our analysis confirms the interest of the
constant acceptance rate strategy (with acceptance rate between $1/4$
and $1/3$) first suggested in
Roberts \textit{et al.} (\textit{Ann. Appl. Probab.} \textbf{7} (1997) 110--120).

We also address scaling of the Metropolis-Adjusted Langevin
Algorithm. When starting at equilibrium, a diffusive limit for an
optimal scaling of the variance is
obtained in
Roberts and Rosenthal (\textit{J. R. Stat. Soc. Ser. B. Stat. Methodol.} \textbf{60} (1998)
255--268). In the transient case, we
obtain formally that the optimal variance scales very differently in
$n$ depending on the sign of a moment of the distribution, which
vanishes at equilibrium. This suggest that it is difficult to derive
practical recommendations for MALA from such asymptotic results.
\end{abstract}

%

\end{frontmatter}

\section{Introduction}\label{sec:intro}

Many Markov Chain Monte Carlo (MCMC) methods are based on the
Metropolis--Hastings algorithm,
see \cite{metropolis-rosenbluth-rosenbltuh-teller-teller-53,hastings-70}.
To set up the notation, let us recall this well-known
sampling technique. Let us consider a target probability distribution
on $\R^n$ with\vadjust{\goodbreak} density~$p$. Starting from an initial random variable
$X_0$, the Metropolis--Hastings algorithm generates iteratively a Markov
chain $(X_k)_{k \ge0}$ in two steps. At time $k$, given $X_k$, a
candidate $Y_{k+1}$ is sampled using a proposal distribution with
density $q(X_k,y)$. Then, the proposal $Y_{k+1}$ is accepted with
probability $\alpha(X_k,Y_{k+1})$, where
\[
\alpha(x,y)= 1\wedge\frac{p(y)q(y,x)}{p(x)q(x,y)}.
\]
Here and in the following, we use the standard notation $a \wedge b =
\min(a,b)$. If the
proposed value is accepted, then $X_{k+1}=Y_{k+1}$ otherwise
$X_{k+1}=X_{k}$. The Markov Chain $(X_k)_{k \ge0}$ is by construction
reversible with respect to the target density $p$, and thus admits
$p$ as an invariant distribution. The efficiency of this algorithm
highly depends on the choice of the proposal distribution $q$. One
common choice is a Gaussian proposal centered at point $x \in\R^n$
with variance $\sigma^2 \mathrm{Id}_{n \times n}$:
\[
q(x,y)=\frac{1}{(2 \uppi\sigma^2)^{n/2}} \exp \biggl(-\frac
{|x-y|^2}{2\sigma^2} \biggr).
\]
Since the proposal is symmetric ($q(x,y)=q(y,x)$), the acceptance
probability reduces to
%
%
\begin{equation}
\label{def:acceptanceRWM} \alpha(x,y)= 1\wedge\frac{p(y)}{p(x)}.
\end{equation}
Metropolis--Hastings algorithms with symmetric kernels are called
random walk Metropolis (RWM). Another popular choice yields the so
called Metropolis adjusted Langevin algorithm (MALA). In this case,
the proposal distribution is a Gaussian random variable\vspace*{-1pt} with variance
$\sigma^2 \mathrm{Id}_{n \times n}$ and centered at point
$x+\frac{\sigma^2}{2}\nabla\ln(p(x))$ (in particular, it is not
symmetric). It\vspace*{-1pt} corresponds to one step of a time-discretization with
timestep $\sigma^2$ of the (overdamped) Langevin dynamics: $\mathrm{d}X_t
=\mathrm{d}B_t + \frac{1}{2} \nabla\ln p (X_t) \,\mathrm{d}t$ which
is ergodic with
respect to $p(x) \,\mathrm{d}x$ (here, $B_t$ is a standard $n$-dimensional
Brownian motion).

In both cases (RWM and MALA), the variance $\sigma^2$ remains to be
chosen. It should be sufficiently large to ensure a good exploration
of the state space, but not too large otherwise the rejection rate
becomes typically very high since the proposed moves fall in low
probability regions, in particular in high dimension. It is expected
that the higher the dimension, the smaller the variance of the
proposal should be. The first theoretical results to optimize the
choice of $\sigma^2$ in terms of the dimension $n$ can be found in
\cite{roberts-gelman-gilks-97}. The authors study
the RWM algorithm under two fundamental (and somewhat restrictive)
assumptions: (i) the target probability distribution is the $n$-fold tensor
product of a one-dimensional density:
%
%
\begin{equation}
\label{eq:pi} p(x) = \prod_{i=1}^n \exp
\bigl(-V(x_i)\bigr),
\end{equation}
where $x=(x_1, \ldots, x_n)$ and $ \int_{\R} \exp(-V) =
1$, and (ii) the initial distribution is the target probability (what
we refer to as \emph{the stationarity assumption} in the following):
\[
X^n_0 \sim p(x) \,\mathrm{d}x.
\]
The superscript $n$ in the Markov chain $(X^n_k)_{ k \ge0}$
explicitly indicates the dependency on the dimension $n$.
Then, under additional regularity assumption on $V$, the authors prove
that for a proper scaling of the variance as a function of the
dimension, namely
\[
\sigma_n^2=\frac{\ell^2}{n},
\]
where $\ell$ is a fixed constant, the Markov process $
(X^{1,n}_{\lfloor nt \rfloor} )_{t \ge0}$ (where $X^{1,n}_k \in
\R$ denotes the first component of $X^n_k \in\R^n$) converges in law
to a diffusion process:
%
%
\begin{equation}
\label{eq:Cas_Eq} \mathrm{d}X_t = \sqrt{ h(I,\ell)} \,
\mathrm{d}B_t - h(I,\ell) \tfrac{1}{2} V'
(X_t) \,\mathrm{d}t,
\end{equation}
where
%
%
\begin{equation}
\label{eq:h} h(I,\ell)=2\ell^2 \Phi \biggl( - \frac{\ell\sqrt{I}}{2}
\biggr) \quad\mbox{and}\quad I=\int_{\R}
\bigl(V'\bigr)^2 \exp(-V).
\end{equation}
Here and in the following, $\lfloor\cdot\rfloor$ denotes the integer
part (for $y \in\R$, $\lfloor y \rfloor\in\Z$ and $\lfloor y
\rfloor\le y < \lfloor y \rfloor+1$) and $\Phi$ is the cumulative
distribution function of the normal distribution ($\Phi(x) =
\frac{1}{\sqrt{2 \uppi}} \int_{-\infty}^x \exp(-y^2/2) \,\mathrm
{d}y$). The
scalings of the variance and of the
time as a function of the dimension are indications on how to make the
RWM algorithm
efficient in high dimension. Moreover, a
practical counterpart of this result is that $\ell$ should be chosen such
that $h(I,\ell)$ is maximum (the optimal value of $\ell$ is
$\frac{c_0}{\sqrt{I}}$ with $c_0 \simeq2.38$), in order to optimize
the time
scaling in~\eqref{eq:Cas_Eq}. This optimal value of $\ell$ corresponds
equivalently to a constant average
acceptance rate, with approximate value $0.234$: for this choice of
$\ell$, in the limit $n$ large,
\[
\iint\alpha(x,y) p(x) q(x,y) \,\mathrm{d}x \,\mathrm{d}y \simeq0.234.
\]
Notice that the optimal average acceptance rate does not depend on
$I$, and is thus the same whatever the
target probability.
Thus, the practical way to choose $\sigma^2$ is to scale it in such a
way that the average acceptance rate is roughly $1/4$. Similar results
have been obtained\vspace*{-1pt} for the MALA algorithm in \cite
{roberts-rosental-97}. In this
case, the scaling for the variance is
$\sigma_n^2=\frac{\ell^2}{n^{1/3}}$, the time scaling is
$(X^n_{\lfloor
n^{1/3} t \rfloor})_{t \ge0}$ and the optimal average acceptance
rate is $0.574$.

There exists several extensions of such results for various
Metropolis--Hastings algorithms, see
\cite
{roberts-rosental-97,roberts-rosenthal-01,breyer-piccioni-scarlatti-04,neal-roberts-11,neal-roberts-yuen-12,bedard-douc-moulines-12,bedard-douc-moulines-12b,beskos-roberts-sanz-serna-stuart-12},
and some of them relax in particular the first main assumption
mentioned above about
the separated form of the target distribution,
see \cite{breyer-roberts-00,bedard-07,bedard-08,beskos-roberts-stuart-09}.
Extensions to infinite dimensional settings
have also been explored, see
\cite
{mattingly-pillai-stuart-12,pillai-struart-thiery-12,beskos-roberts-stuart-09}.

All these results assume stationarity: the initial
measure is the target probability. To the best of the authors'
knowledge, the only works which deal with a nonstationary case
are \cite{christensen-roberts-rosenthal-05} where the RWM and the MALA
algorithms are analyzed in
the Gaussian case and \cite{PST11}. In the latter paper, the target
measure is assumed
to be absolutely continuous with respect to the law of an infinite
dimensional Gaussian random field and this measure is approximated in
a space of dimension $n$ where the MCMC algorithm is performed. The
authors consider a modified RWM algorithm (called preconditioned
Crank--Nicolson walk) started at a deterministic initial condition and
prove that when $\sigma_n$ tends to $0$ as $n$ tends to $\infty$ (with
no restriction on the rate of convergence of $\sigma_n$ to $0$), the
rescaled algorithm converges to a stochastic partial differential equation,
started at the same initial condition.

The aim of this article is to discuss extensions of the previous
results for the RWM and the MALA algorithms, \emph{without assuming
stationarity}. The main findings are the following.

Concerning the RWM algorithm, in the companion
paper \cite{JLM1}, we prove that, using
the same scaling\vspace*{-1pt} for the variance and the time as in the
stationary case (namely $\sigma_n^2=\frac{\ell^2}{n}$ and considering
$ (X^{1,n}_{\lfloor nt \rfloor} )_{t \ge0}$), one obtains in
the limit $n$ goes to infinity a diffusion process nonlinear in the
sense of McKean
(see Equation \eqref{edsnonlin} below).
This is discussed in Section~\ref{sec:RWM}. Contrary to \eqref
{eq:Cas_Eq}, this diffusion process cannot be obtained from the simple
Langevin dynamics $\mathrm{d}X_t=\mathrm{d}B_t-\frac{V'(X_t)}{2}\,
\mathrm{d}t$ by a deterministic
time-change and its long-time behavior is not obvious. In Section~\ref
{sec:RWM_longtime}, we first prove that its unique stationary
distribution is $\mathrm{e}^{-V(x)}\,\mathrm{d}x$. Assuming that this
measure satisfies a
logarithmic Sobolev inequality, we prove that the Kullback--Leibler
divergence of the marginal distribution at time $t$ with respect to
$\mathrm{e}^{-V(x)}\,\mathrm{d}x$ converges to $0$ at an exponential rate.
In Section~\ref{sec:optim_RWM}, we discuss optimizing
strategies which take into account the transient phase. Roughly\vspace*{-1pt}
speaking, the usual strategy which consists in choosing $\ell$ (recall that
$\sigma_n^2=\frac{\ell^2}{n}$) such that the average acceptance rate is
constant (with value between $1/4$ and $1/3$) seems to be a very good
strategy. This
is numerically
illustrated in Section~\ref{sec:numerics}.

Concerning the MALA algorithm, the situation is much more
complicated. The scaling to be used seems to depend on the sign
of an average quantity (see Section~\ref{sec:MALA_practice}). In
particular, the scaling
$\sigma_n^2=\frac{\ell^2}{n^{1/3}}$ which has been identified
in \cite{roberts-rosental-97} under the stationary assumption is not
adapted to the transient
phase. It seems difficult to draw any practical recommendation from
this analysis. This is explained with details in Section~\ref{sec:MALA}.

\section{Scaling limit for the RWM algorithm}\label{sec:RWM}

In this section, we state the diffusion limit for the RWM algorithm,
and explain formally why this result holds. A rigorous proof can be
found in \cite{JLM1}.

\subsection{The RWM algorithm and the convergence result}

We consider a Random Walk Metropolis algorithm using Gaussian\vspace*{-1pt}
proposal with variance $\sigma_n^2=\frac{\ell^2}{n}$, and with
target $p$ defined by \eqref{eq:pi}. The Markov
chain generated using this algorithm writes:
%
%
\begin{equation}
\label{rwm} X^{i,n}_{k+1}=X^{i,n}_k+
\frac{\ell}{\sqrt{n}}G^i_{k+1}1_{{\mathcal
A}_{k+1}},\qquad1\leq i
\leq n
\end{equation}
with
\[
{\mathcal A}_{k+1}= \bigl\{U_{k+1}\leq\mathrm{e}^{\sum_{i=1}^n
(V(X^{i,n}_k)-V(X^{i,n}_k+ ({\ell}/{\sqrt{n}})G^i_{k+1}))}
\bigr\},
\]
where $(G^i_k)_{i,k\geq1}$ is a sequence of independent and
identically distributed (i.i.d.) normal random
variables independent from a sequence $(U_k)_{k\geq1}$ of i.i.d.
random variables with uniform law on $[0,1]$. We assume that the initial
positions $(X^{1,n}_0,\ldots,X^{n,n}_0)$ are exchangeable
(namely the law of the vector is invariant under permutation of the indices)
and independent from all the previous random
variables. Exchangeability is preserved by the dynamics: for all $k
\ge1$, $(X^{1,n}_k,\ldots,X^{n,n}_k)$ are exchangeable. We denote by
${\mathcal F}^n_k$ the sigma field generated by
$(X^{1,n}_0,\ldots,X^{n,n}_0)$ and $(G^1_l,\ldots,G^n_l,U_l)_{1\leq
l\leq k}$.\vadjust{\goodbreak}

For $t>0$ and $i \in\{1,\ldots,n\}$, let
\begin{eqnarray*}
Y^{i,n}_t&=&\bigl(\lceil nt\rceil-nt\bigr)X^{i,n}_{\lfloor
nt\rfloor}+
\bigl(nt-\lfloor nt\rfloor\bigr)X^{i,n}_{\lceil
nt\rceil}
\\
&=&X^{i,n}_{\lfloor nt\rfloor}+\bigl(nt-\lfloor nt\rfloor\bigr)
\frac{\ell}{\sqrt{n}}G^i_{\lceil nt\rceil}1_{{\mathcal
A}_{\lceil nt\rceil}}
\end{eqnarray*}
be the linear
interpolation of the Markov chain obtained by rescaling time
(the characteristic time scale is $1/n$). This is the classical
diffusive time-scale for a random walk, since the variance of each
move is of order $1/n$.

Let us define the notion of convergence (namely the propagation of
chaos) that will be useful to study
the convergence of the interacting particle system
$((Y^{1,n}_t, \ldots, Y^{n,n}_t )_{t \ge0})_{ n \ge1}$ in the limit
$n$ goes to infinity.
%
%
\begin{definition}
Let $E$ be a separable metric space. A sequence
$(\chi^n_1,\ldots,\chi_n^n)_{n\geq1}$ of exchangeable $E^n$-valued
random variables is said to be $\nu$-chaotic where $\nu$ is a
probability measure on $E$ if for fixed $j\in\tN^*$, the law of
$(\chi^n_1,\ldots,\chi_j^n)$ converges in distribution to $\nu
^{\otimes j}$ as
$n$ goes to $\infty$.
\end{definition}

According to \cite{sznitman-91}, Proposition~2.2, the $\nu$-chaoticity
is equivalent to a law of large numbers result, namely the convergence
in probability of the empirical measures $\mu^n=\frac{1}{n}\sum_{i=1}^n\delta_{\chi^n_i}$ to the constant $\nu$ when the space of
probability measures on $E$ is endowed with the weak convergence topology.

We are now in position to state the convergence result for the RWM
algorithm, taken from \cite{JLM1}. Here and in the following,
the bracket notation refers to the duality bracket for
probability measures on $\R$: for $\mu$ a probability measure and
$\phi$ a bounded measurable function,
\[
\langle\mu, \phi\rangle= \int\phi\,\mathrm{d}\mu.
\]

%
\begin{theorem}\label{chaos}
Let $m$ be a probability measure on $\R$
such that $\langle m,(V')^4\rangle<+\infty$. Let us also assume that
%
%
\begin{equation}
\label{eq:hyp_V} %
V\mbox{ is a }C^3\mbox{ function
on }\R\mbox{ with bounded second and third order derivatives}. 
\end{equation}
If the initial positions $(X^{1,n}_0,\ldots,X^{n,n}_0)_{n\geq1}$ are
$m$-chaotic and such that
\[
\sup_n\mathbb{E}\bigl[\bigl(V'\bigl(X_0^{1,n}\bigr)\bigr)^4\bigr]< +\infty,
\]
then the processes $((Y^{1,n}_t,\ldots,Y^{n,n}_t)_{t\geq0})_{n\geq
1}$ are $P$-chaotic where $P$ denotes the law (on the space
${\mathcal C}(\R_+,\R)$ of continuous function with values in $\R$)
of the unique solution to the stochastic differential equation
nonlinear in the sense of McKean
%
%
\begin{eqnarray}\label{edsnonlin}
 \mathrm{d}X_t&=& \Gamma^{1/2}\bigl(\E\bigl[
\bigl(V'(X_t)\bigr)^2\bigr],\E
\bigl[V''(X_t)\bigr],\ell\bigr) \,
\mathrm{d}B_t
\nonumber
\\[-8pt]
\\[-8pt]
&&{}- {\cG}\bigl(\E\bigl[\bigl(V'(X_t)
\bigr)^2\bigr],\E\bigl[V''(X_t)
\bigr],\ell\bigr)V'(X_t) \,\mathrm{d}t,
\nonumber
\end{eqnarray}
where $(B_t)_{t\geq1}$ is a Brownian motion independent from the
initial position $X_0$ distributed according to $m$. The functions
$\Gamma$ and $\cG$ are, respectively, defined by: for $\ell\in
(0,+\infty)$, $a \in[0,+\infty]$ and $b \in\R$,
%
%
\begin{equation}
\label{eq:Gamma} \Gamma(a,b,\ell)= %
\cases{ \ell^2\Phi
\biggl(-\displaystyle\frac{\ell b}{2\sqrt{a}} \biggr)+\ell^2
\mathrm{e}^{{\ell^2(a-b)}/{2}}\Phi \biggl(\ell \biggl(\displaystyle\frac
{b}{2\sqrt{a}}-
\sqrt{a} \biggr) \biggr),&\quad\mbox{if }$a\in (0,+\infty )$,\vspace*{1pt}
\cr
\displaystyle
\frac{\ell^2}{2},&\quad\mbox{if }$a=+\infty$,\vspace*{1pt}
\cr
\ell^2 \mathrm{e}^{-{\ell^2b^+}/{2}},&\quad\mbox{if }$a=0$,} %
\end{equation}
where $b^+=\max(b,0)$, and
%
%
\begin{equation}
\label{eq:G} {\mathcal G}(a,b,\ell)= %
\cases{ \ell^2
\mathrm{e}^{{\ell^2(a-b)}/{2}}\Phi \biggl(\ell \biggl(\displaystyle\frac{b}{2\sqrt{a}}-
\sqrt{a} \biggr) \biggr),&\quad \mbox{if }$a\in (0,+\infty)$,\vspace*{1pt}
\cr
0,&\quad\mbox{if }$a=+\infty$,
\cr
1_{\{b>0\}}\ell^2 \mathrm{e}^{-{\ell^2b}/{2}},&
\quad\mbox{if }$a=0\eqsp$.} %
\end{equation}
\end{theorem}

Notice\vspace*{-1pt} that the assumption on $(X^{1,n}_0,\ldots,X^{n,n}_0)_{n\geq
1}$ is for example satisfied
when the random variables $X^{1,n}_0,\ldots,X^{n,n}_0$ are
i.i.d. according to some probability measure $m$ on~$\R$.

This convergence result generalizes the previous
result by Roberts \textit{et al.} \cite{roberts-gelman-gilks-97} where the same
diffusive limit
is obtained under
the restrictive assumption that the vector of initial positions
$(X^{1,n}_0,\ldots,X^{n,n}_0)$ is distributed according to the target
distribution $p(x) \,\mathrm{d}x$. In this case, $(X_t)_{t\geq0}$
indeed solves the
stochastic differential equation \eqref{eq:Cas_Eq}--\eqref{eq:h} with
time-homogeneous coefficients (here, we use the fact that
$\Gamma(I,I,\ell)=2\cG(I,I,\ell)=h(I,\ell)$ where\vspace*{-1pt} $I=\int_\R
(V'(x))^2\mathrm{e}^{-V(x)}\,\mathrm{d}x=\int_\R
V''(x)\mathrm{e}^{-V(x)}\,\mathrm{d}x < \infty$, see \cite{JLM1},
Lemma~1). Moreover, by
taking $V(x)=\frac{x^2}{2}+\frac{1} 2 \ln(2 \uppi)$, this theorem also
yields similar results
as \cite{christensen-roberts-rosenthal-05}, where the authors
consider a nonstationary case, but restrict their analysis to the
evolution of $k\mapsto\frac{1}{n}\sum_{i=1}^n (X^{i,n}_k)^2$ for
Gaussian targets.

In addition to the previous convergence result, we are able to
identify the limiting average acceptance rate.
%
%
\begin{proposition}\label{prop:limiting-acceptance} Under the
assumptions of Theorem~\ref{chaos}, the function
\[
t\mapsto \E\biggl\llvert \P\bigl({\mathcal A}_{\lfloor nt\rfloor+1}|{\mathcal
F}^n_{\lfloor
nt\rfloor}\bigr)-\frac{1}{\ell^2}\Gamma\bigl(\E\bigl[
\bigl(V'(X_t)\bigr)^2\bigr],\E
\bigl[V''(X_t)\bigr],\ell\bigr)\biggr
\rrvert
\]
converges locally uniformly to $0$ and in particular, the average
acceptance rate $t\mapsto
\P({\mathcal A}_{\lfloor nt\rfloor+1})$ converges locally uniformly
to $t\mapsto\acc(\E[(V'(X_t))^2],\E[V''(X_t)],\ell)$ where for any
$a\geq0$ and $b\in\R$,
%
%
\begin{equation}
\label{def:limitacc} \acc(a,b,\ell)=\frac{\Gamma(a,b,\ell)}{\ell^2}.
\end{equation}
\end{proposition}

\subsection{A formal derivation of the limiting process (\texorpdfstring{\protect\ref{edsnonlin}}{7})}\label{sec:RWM_formal}

Let us introduce the infinitesimal generator $L_\mu$ associated
to \eqref{edsnonlin}:
%
%
\begin{equation}
\label{deflmu} L_\mu\varphi(x)=\tfrac{1}{2}\Gamma\bigl(\bigl
\langle\mu,\bigl(V'\bigr)^2\bigr\rangle ,\bigl\langle
\mu,V''\bigr\rangle,\ell\bigr)\varphi''(x)-
\cG\bigl(\bigl\langle\mu ,\bigl(V'\bigr)^2\bigr\rangle,
\bigl\langle\mu,V''\bigr\rangle,\ell
\bigr)V'(x)\varphi'(x).
\end{equation}
For a probability measure $\mu$ on $\R$,
$\langle\mu,V''\rangle$ is well defined by boundedness of $V''$ (see
\eqref{eq:hyp_V}), and
$\langle\mu,(V')^2\rangle$ is also well defined in $[0,+\infty]$.

The relationship between \eqref{edsnonlin} and \eqref{deflmu} is the
following: if $X_t$ satisfies \eqref{edsnonlin}, then for any
smooth test function $\varphi$, $\varphi(X_t) - \int_0^t L_{P_s}
\varphi(X_s)
\,\mathrm{d}s$ is a
martingale, where $P_t$ denotes the law of $X_t$: for any $s < t$,
%
%
\begin{equation}
\label{eq:martingale} \E \biggl( \varphi(X_t) - \int_s^t
L_{P_r} \varphi(X_r) \,\mathrm {d}r \big| {\mathcal
F}_s \biggr) = \varphi(X_s).
\end{equation}
Actually, as explained in \cite{JLM1}, Section~3.1, the martingale
representation of the solution is
a weak formulation of \eqref{edsnonlin}: solutions
to \eqref{eq:martingale} are solutions in distribution to \eqref{edsnonlin}.

Let us now present formally how \eqref{edsnonlin} is derived. First,
let us
explain how the scaling of $\sigma_n$ as a function of
$n$ is chosen. The idea (see \cite{roberts-rosenthal-01}) is to
choose $\sigma_n$ in such a way that the limiting acceptance rate
(when $n \to\infty$) is neither zero nor one. In the first case, this
would mean that the variance of the proposal is too large, so that all
proposed moves are rejected. In the second case, the variance of the
proposal is too small, and the rate of convergence to equilibrium is
thus not optimal. In particular, it is easy to check that $\sigma_n$
should go to zero as $n$ goes to infinity. Now, notice that the
limiting acceptance rate is:
%
%
\begin{eqnarray}
\label{eq:accept_RWM} \E \bigl( 1_{{\mathcal
A}_{k+1}} | {\mathcal F}^n_k
\bigr) &=&\E \bigl( \mathrm{e}^{\sum_{i=1}^n
(V(X^{i,n}_k)-V(X^{i,n}_k+\sigma_n G^i_{k+1}))} \wedge1 | {\mathcal
F}^n_k \bigr)
\nonumber
\\
& =& \E \bigl(\mathrm{e}^{- \sum_{i=1}^n
(V'(X^{i,n}_k) \sigma_n G^i_{k+1} +
V''(X^{i,n}_k)( {\sigma_n^2}/{2}) ) } \wedge1 | {\mathcal F}^n_k
\bigr) + {\mathcal O}\bigl(n \sigma_n^{3}\bigr) + {
\mathcal O}\bigl(\sqrt{n} \sigma_n^{2}\bigr)\nonumber
\\
&=& \exp \biggl(\frac{a_n-b_n}{2} \biggr) \Phi \biggl( \frac{b_n}{2
\sqrt{a_n}} -
\sqrt{a_n} \biggr) + \Phi \biggl(-\frac{b_n}{2\sqrt{a_n}} \biggr) \nonumber
\\
&&{}+ {
\mathcal O}\bigl(n \sigma _n^{3}\bigr) + {\mathcal O}\bigl(
\sqrt{n} \sigma_n^{2}\bigr),
\end{eqnarray}
where $a_n=\sigma_n^2 \sum_{i=1}^n
(V'(X^{i,n}_k))^2 $ and $b_n=\sigma_n^2 \sum_{i=1}^n
V''(X^{i,n}_k)$. The formula \eqref{eq:accept_RWM} is obtained by
explicit computations (see \cite{roberts-gelman-gilks-97}, Proposition~2.4).
From this expression, assuming a propagation of chaos (law of large
numbers) on the random\vspace*{-2pt} variables $(X^{i,n}_k)_{1 \le i \le n}$, one can
check that
the correct scaling for the variance is
$\sigma_n^2=\frac{\ell^2}{n}$ in order to obtain a nontrivial
limiting acceptance rate (see \cite{JLM1}, Section~2.3).

Using this scaling, we observe that,
for a test function $\varphi\dvtx \R\to\R$,
%
%
\begin{eqnarray}
\label{eq:gen_inf} &&\E \bigl(\varphi\bigl(X^{1,n}_{k+1}\bigr) | {
\mathcal F}^n_k \bigr)
\nonumber
\\
&&\quad=\E \biggl(\varphi \biggl(X^{1,n}_k+
\frac{\ell}{\sqrt {n}}G^1_{k+1}1_{{\mathcal
A}_{k+1}} \biggr) \big| {
\mathcal F}^n_k \biggr)
\nonumber
\\
&&\quad= \varphi \bigl(X^{1,n}_k \bigr) +
\varphi' \bigl(X^{1,n}_k \bigr)
\frac{\ell}{\sqrt{n}} \E \bigl( G^1_{k+1}1_{{\mathcal
A}_{k+1}} | {
\mathcal F}^n_k \bigr)
\nonumber
\\
&&\qquad{} +\frac{\ell^2}{2n} \varphi''
\bigl(X^{1,n}_k \bigr) \E \bigl( \bigl(G^1_{k+1}
\bigr)^2 1_{{\mathcal
A}_{k+1}} | {\mathcal F}^n_k
\bigr) + {\mathcal O}\bigl(n^{-3/2}\bigr).
\end{eqnarray}
We compute (by conditioning with respect to $G^1_{k+1}$):
%
%
\begin{eqnarray}
\label{eq:drift}
&& \E\bigl( G^1_{k+1} 1_{{\mathcal
A}_{k+1}} | {
\mathcal F}^n_k \bigr) \nonumber
\\
&&\quad=\E \bigl( G^1_{k+1}
\mathrm{e}^{\sum_{i=1}^n
(V(X^{i,n}_k)-V(X^{i,n}_k+({\ell}/{\sqrt{n}})G^i_{k+1}))} \wedge1 | {\mathcal F}^n_k
\bigr)\nonumber
\\
&&\quad = \E \bigl( G^1_{k+1} \mathrm{e}^{- \sum_{i=1}^n
(V'(X^{i,n}_k) ({\ell}/{\sqrt{n}})G^i_{k+1} +
V''(X^{i,n}_k)({\ell^2}/({2n})) ) } \wedge1
| {\mathcal F}^n_k \bigr) + {\mathcal O}
\bigl(n^{-1}\bigr)\nonumber
\\
&&\quad=- V'\bigl(X^{1,n}_k\bigr)
\frac{1}{\ell\sqrt{n}} \cG \bigl( \bigl\langle\nu ^n_{k},
\bigl(V'\bigr)^2 \bigr\rangle, \bigl\langle
\nu^n_{k}, V'' \bigr\rangle,
\ell \bigr) + {\mathcal O}\bigl(n^{-1}\bigr),
\end{eqnarray}
where
\[
\nu^n_{k}=\frac{1}{n} \sum
_{i=1}^n \delta_{X^{i,n}_k}
\]
denotes the empirical distribution associated to the interacting
particle system.
The equation \eqref{eq:drift} is again a consequence of explicit
computations (see \cite{JLM1}, Equation (A.3)), and the fact that the
remainder is of order $n^{-1}$ requires
a detailed analysis (see \cite{JLM1}, Proposition~7). Likewise, for
the diffusion term, we get
%
%
\begin{eqnarray}
\label{eq:diff}
&&\E\bigl( \bigl(G^1_{k+1}\bigr)^2
1_{{\mathcal
A}_{k+1}} | {\mathcal F}^n_k \bigr) \nonumber
\\
&&\quad=\E \bigl(
\bigl(G^1_{k+1}\bigr)^2 \mathrm{e}^{\sum_{i=1}^n
(V(X^{i,n}_k)-V(X^{i,n}_k+({\ell}/{\sqrt{n}})G^i_{k+1}))}
\wedge1 | {\mathcal F}^n_k \bigr)
\nonumber
\\
&&\quad = \E \bigl( \bigl(G^1_{k+1}\bigr)^2
\mathrm{e}^{- \sum_{i=1}^n
(V'(X^{i,n}_k) ({\ell}/{\sqrt{n}})G^i_{k+1} +
V''(X^{i,n}_k)({\ell^2}/({2n})) ) } \wedge1 | {\mathcal F}^n_k
\bigr) + {\mathcal O}\bigl(n^{-1}\bigr)\nonumber
\\
&&\quad = \frac{1}{\ell^2} \Gamma \bigl( \bigl\langle\nu^n_k,
\bigl(V'\bigr)^2 \bigr\rangle, \bigl\langle
\nu^n_k, V'' \bigr\rangle,
\ell \bigr) + {\mathcal O}\bigl(n^{-1}\bigr).
\end{eqnarray}
To obtain \eqref{eq:diff}, we again used an explicit computation (see
\cite{JLM1}, Equation (A.5)).

By plugging \eqref{eq:drift} and \eqref{eq:diff}
into \eqref{eq:gen_inf}, we see that the correct scaling in time
is
to consider $Y^{i,n}_t$ such that $Y^{i,n}_{k/n}=X^{i,n}_k$ (diffusive
timescale), and we get:
\begin{eqnarray*}
\E \bigl( \varphi\bigl(Y^{1,n}_{(k+1)/n}\bigr) | {\mathcal
F}^n_k \bigr) & =& \varphi \bigl(Y^{1,n}_{k/n}
\bigr) - \varphi' \bigl(Y^{1,n}_{k/n} \bigr)
\frac{1}{n} V' \bigl(Y^{1,n}_{k/n} \bigr)
\cG \bigl( \bigl\langle\nu^n_k, \bigl(V'
\bigr)^2 \bigr\rangle, \bigl\langle\nu^n_k,
V'' \bigr\rangle,\ell \bigr)
\\
&&{} +\frac{1}{2n} \varphi'' \bigl(
Y^{1,n}_{k/n} \bigr) \Gamma \bigl( \bigl\langle\nu
^n_k, \bigl(V'\bigr)^2 \bigr
\rangle, \bigl\langle\nu^n_k, V''
\bigr\rangle, \ell \bigr) + {\mathcal O}\bigl(n^{-3/2}\bigr)
\\
&=& \varphi \bigl(Y^{1,n}_{k/n} \bigr) + \frac{1}{n} (
L_{
\nu^n_k} \varphi ) \bigl(Y^{1,n}_{k/n}\bigr) + {
\mathcal O}\bigl(n^{-3/2}\bigr),
\end{eqnarray*}
where $L_\mu$ is defined by \eqref{deflmu}. This can be seen as a
discrete-in-time version (over a timestep of size $1/n$) of the
martingale property \eqref{eq:martingale}. Thus, by sending $n$ to
infinity, assuming\vspace*{-1pt} that $\nu^n_{k}$ converges to the law of
$Y^1_{k/n}$, we expect $Y^{1,n}_t$ to converge to a solution
to \eqref{edsnonlin}. For a rigorous proof, we refer to \cite{JLM1}.

\section{Longtime convergence for the RWM nonlinear dynamics}\label{sec:RWM_longtime}

We would like to study the limiting dynamics \eqref{edsnonlin}
obtained for the RWM algorithm, that we recall for convenience
\[
\mathrm{d}X_t= \Gamma^{1/2}\bigl(\E\bigl[
\bigl(V'(X_t)\bigr)^2\bigr],\E
\bigl[V''(X_t)\bigr],\ell\bigr) \,\mathrm
{d}B_t-{\cG}\bigl(\E \bigl[\bigl(V'(X_t)
\bigr)^2\bigr],\E\bigl[V''(X_t)
\bigr],\ell\bigr)V'(X_t) \,\mathrm{d}t,
\]
where $\Gamma$ and $\cG$ are, respectively, defined by \eqref{eq:Gamma}
and \eqref{eq:G}.
The associated Fokker--Planck equation is ($\psi_t$ denotes the density
of the random variable $X_t$):
%
%
\begin{equation}
\label{eq:FP} %
\cases{ \partial_t \psi_t =
\partial_x \bigl(\cG\bigl(a[\psi_t], b[\psi
_t],\ell\bigr) V' \psi_t + \Gamma\bigl(a[
\psi_t] ,b[\psi_t],\ell\bigr)\, \partial_x
\psi_t / 2 \bigr),\vspace*{2pt}
\cr
\mbox{where}\quad a[\psi_t]=
\displaystyle\int_{\R} \bigl(V'
\bigr)^2 \psi _t \quad\mbox{and}\quad b[
\psi_t]= \int_{\R} V''
\psi_t. } %
\end{equation}
Let us denote $\psi_\infty= \exp(-V)$. Notice that $a[\psi_\infty
]=b[\psi_\infty]$ and
$\cG(a,a,\ell)=\Gamma(a,a,\ell)/2$. We thus expect $\psi_\infty$
to be the
longtime limit of $\psi_t$.

\subsection{Stationary solution}

We start the analysis of the limiting process by checking that the
solution of \eqref{edsnonlin} has the expected stationary distribution.
%
%
\begin{proposition}\label{prop:unique_stationary} There exists a
unique stationary distribution $\mu$ for the process $X_t$ defined
by \eqref{edsnonlin}. In addition, this distribution is absolutely continuous
with respect to the Lebesgue measure, with density $\psi_\infty
(x)=\exp(-V(x))$.
\end{proposition}

Before proving Proposition~\ref{prop:unique_stationary}, we need some
preliminary facts the proof of which is postponed to Appendix~\ref{sec:fF}.
%
%
\begin{lemma}\label{lem:fF}
Defining the function $\sign$ by:
%
%
\begin{equation}
\label{eq:sign} \sign(x)= \cases{ %
1, & \quad\mbox{if } $x
> 0$,
\cr
0, & \quad\mbox{if } $x = 0$,
\cr
-1, &\quad\mbox{if } $x < 0$, }
\end{equation}
one has
%
%
\begin{equation}
\label{lem:f} \sign{ \bigl({\Gamma(a,b,\ell)-2\cG(a,b,\ell)} \bigr)}=\sign {
(a-b )}.
\end{equation}

Moreover, the function $F$ defined for $a \ge0$, $b \in\R$ and
$\ell> 0$ by
%
%
\begin{equation}
\label{eq:bGamma-2aG} F(a,b,\ell) 
=\cases{ %
{ \displaystyle\frac{b\Gamma(a,b,\ell)-2a\cG(a,b,\ell
)}{b-a}},&\quad\mbox{if }$a\neq b$,\vspace*{2pt}
\cr
{ 2
\ell^2 \biggl( \biggl(1+\displaystyle\frac{\ell^2a}{4} \biggr)\Phi
\biggl(-\displaystyle\frac
{\ell\sqrt{a}}{2} \biggr)-\displaystyle\frac{\ell\sqrt{a}}{2\sqrt {2\uppi}}
\exp { \biggl(-\frac{\ell^2a}{8} \biggr)} \biggr)},&\quad\mbox{if }$a=b$, }
\end{equation}
is a continuous function satisfying
%
%
\begin{equation}
\label{lem:F} \forall\ell>0, \forall M\in(0,+\infty), \qquad\inf
_{(a,b)\in
[0,M]\times
[-M,M]}F(a,b,\ell)>0.
\end{equation}
\end{lemma}
\begin{pf*}{Proof of Proposition~\ref{prop:unique_stationary}}
Let $c=\int_{\R} (V'(x))^2\psi_\infty(x)\,\mathrm{d}x $.
Since $V''$ is bounded then one can check that $c=\int_{\R}
V''(x)\psi_\infty(x)\,\mathrm{d}x<\infty$ (see \cite{JLM1}, Lemma~1). By
\eqref{lem:f}, we get that $\Gamma(c,c,\ell)=2{\cG}(c,c,\ell)$.
Let us define the Langevin diffusion
\[
\mathrm{d}\tilde{X}_t=\sqrt{2{\cG}(c,c,\ell)}\,\mathrm{d}B_t-{
\cG }(c,c,\ell)V'(\tilde {X}_t)\,\mathrm{d}t
\]
with $X_0$ distributed according to the density $\psi_\infty$. It is
well known that for any $t\geq0$ the density of $\tilde{X}_t$ is
$\psi_\infty$ and therefore
$c=\E[(V'(\tilde{X}_t))^2]=\E[V''(\tilde{X}_t)]$. Then it is clear that
the process $\tilde{X}_t$ satisfies \eqref{edsnonlin}. Hence,
$\psi_\infty(x) \,\mathrm{d}x$ is a stationary probability distribution
for the stochastic differential equation \eqref{edsnonlin}.

Let us now prove the uniqueness of the invariant measure. Assume that
there exists another
stationary probability measure with density $p_\infty$ (the fact
that the stationary measure admits a density is standard, since the
diffusion term is bounded from below). Assume\vspace*{-1pt} $\int_{\R} V'^2
p_\infty=+\infty$.
Since ${\cG}(+\infty,b,\ell)=0$ and $\Gamma(+\infty,b,\ell)=\frac
{\ell^2}{2}$,
the stochastic differential equation \eqref{edsnonlin} with\vspace*{-1pt}
$X_0$ distributed according to the density $p_\infty$ reduces in this
case to
$\mathrm{d}X_t=\frac{\ell}{\sqrt{2}}\,\mathrm{d}B_t$
which does not admit a stationary distribution. Thus, necessarily, we have
\[
\int_{\R} V'^2 p_\infty<
\infty.
\]
Let us denote $a=\int_{\R}
V'^2 p_\infty$ and $b=\int_{\R} V'' p_\infty$. Then, Equation
\eqref{edsnonlin}
with $X_0$ distributed according to the density $p_\infty$ reduces to
\[
\mathrm{d}X_t=\Gamma^{1/2}(a,b,\ell)\,\mathrm{d}B_t-{
\cG}(a,b,\ell )V'(X_t)\,\mathrm{d}t.
\]
The stationary distribution thus writes
\[
p_\infty\propto\exp{ \biggl(-\frac{2{\cG}(a,b,\ell)}{\Gamma
(a,b,\ell)}V \biggr)}.
\]
By integration by parts, we obtain that
\[
b\Gamma(a,b,\ell)=2a{\cG}(a,b,\ell).
\]
Hence,
by definition of $F$
and (\ref{lem:F}),
we
obtain $a=b$ and by \eqref{lem:f} we get
that $\frac{2{\cG}(a,b,\ell)}{\Gamma(a,b,\ell)}=1$. In conclusion,
$p_\infty=\exp(-V)=\psi_\infty$.
\end{pf*}
%

\subsection{Longtime convergence}

It is actually possible to prove that, for fixed $\ell>0$, the law of
$X_t$ solution
to \eqref{edsnonlin} converges exponentially fast to the equilibrium
density $\psi_\infty$. The proof is based on entropy estimates, using
the Fokker--Planck equation \eqref{eq:FP}, and requires the notion of
logarithmic Sobolev inequality.
%
%
\begin{definition}
The probability measure $\nu$ satisfies a logarithmic Sobolev
inequality with constant $\rho>0$ (in short $\operatorname{LSI}(\rho
)$) if and only
if, for any probability measure $\mu$ absolutely continuous with
respect to $\nu$,\vspace*{-1pt}
%
%
\begin{equation}
\label{logsobol} H(\mu| \nu) \le\frac{1}{2 \rho} I( \mu| \nu),
\end{equation}
where $H(\mu| \nu) = {\int\ln ( \frac{\mathrm{d} \mu}{
\mathrm{d} \nu}
) \,\mathrm{d}\mu}$ is the Kullback--Leibler divergence\vspace*{-1pt} (or relative
entropy) of $\mu$ with respect to
$\nu$ and $I(\mu| \nu) = {\int\llvert  \nabla\ln ( \frac
{\mathrm{d}
\mu}{ \mathrm{d} \nu}  )\rrvert ^2 \,\mathrm{d}\mu}$ is the
Fisher information
of $\mu$ with respect to $\nu$.
\end{definition}

With a slight abuse of notation, we will denote in the following
$H(\psi| \phi)$ and $I(\psi|\phi)$ the Kullback--Leibler divergence
and the
Fisher information associated with the continuous probability
distributions $\psi(x)
\,\mathrm{d}x$ and $\phi(x) \,\mathrm{d}x$. We recall that, by the
Csiszar--Kullback
inequality (see, for instance, \cite{ABC-00}, Th\'eor\`eme 8.2.7, page
139), for any probability densities $\psi$\vspace*{-1pt} and~$\phi$,
%
%
\begin{equation}
\label{eq:CK} \int_{\R}|\psi- \phi| \le\sqrt{ 2 H(\psi|
\phi)},
\end{equation}
so that $H(\psi| \phi)$ may be seen as a measure of the ``distance''
between $\psi$
and $\phi$.
%
%
\begin{theorem}\label{convexpo}
Let us assume \eqref{eq:hyp_V}, and that $X_0$ admits a density $\psi
_0$ such that
$\E[(V'(X_0))^2]<+\infty$ and
$H(\psi_0|\psi_\infty) < \infty$. Then, for all $t \ge0$,\vspace*{-1pt}
%
%
\begin{equation}
\label{optirat} \frac{\mathrm{d}}{\mathrm{d}t} H(\psi_t| \psi_\infty)
\leq- \frac{F(a[\psi
_t],b[\psi_t],\ell)}{2} I (\psi_t | \psi_\infty),
\end{equation}
and the function $t \mapsto H(\psi_t| \psi_\infty)$ is
decreasing.\vadjust{\goodbreak}

Let us assume moreover that $\psi_\infty=\mathrm{e}^{-V}$ satisfies a
$\operatorname{LSI}(\rho)$. Then there exists a positive and
nonincreasing function
$\lambda\dvtx[0,+\infty) \to(0,+\infty)$ such that $\forall t\geq0$\vspace*{-1pt}
%
%
\begin{equation}
\label{eq:convexpo} H(\psi_t|\psi_\infty) \leq
\mathrm{e}^{-t\lambda (H(\psi_0|\psi_\infty) )}H(\psi _0|\psi _\infty).
\end{equation}
\end{theorem}

Equation \eqref{eq:convexpo} shows that $\psi_t$ converges
exponentially fast to $\psi_\infty$.
%
%
\begin{remark}
Roughly speaking, $\psi_\infty$ satisfies a LSI if $V$ grows sufficiently
fast at infinity. For example, according to \cite{ABC-00}, Th\'eor\`eme
6.4.3, a sufficient condition for
$\psi_\infty$ to satisfy a LSI, is that $|V'|$ does not vanish
outside of some compact subset of $\R$ and\vspace*{-1pt}
\[
\lim_{|x|\to\infty}\frac{V''(x)}{(V'(x))^2}=0\quad\mbox {and}\quad\limsup
_{|x|\to\infty}\frac{|V(x)+\ln|V'(x)||}{(V'(x))^2}<+\infty.
\]
In the Gaussian case $V(x)=\frac{x^2}{2}+\frac{1}{2}\ln(2\uppi)$,
$\psi_\infty(x)=\frac{1}{\sqrt{2\uppi}}\exp (-\frac
{x^2}{2} )$ satisfies $\operatorname{LSI}(1)$.\vadjust{\goodbreak}
\end{remark}
\begin{pf*}{Proof of Theorem~\ref{convexpo}}
By simple computation, we have (for notational
convenience, we write $a,b$ for $a[\psi_t],b[\psi_t]$):
%
%
\begin{eqnarray}
\label{eq:ent1} \frac{\mathrm{d}}{\mathrm{d}t} \int_\R
\psi_t \ln(\psi_t / \psi _\infty) &=& \int
_\R\partial_t \psi_t \ln
\psi_t + \int_\R V \,\partial_t
\psi_t
\nonumber
\\
&=& \int_\R\partial_x \bigl(\cG(a, b,\ell)
V' \psi_t + \Gamma (a,b,\ell) \,\partial_x
\psi_t / 2 \bigr) \ln\psi_t
\nonumber
\\
&&{} + \int_\R V \,\partial_x \bigl(\cG(a, b,
\ell) V' \psi_t + \Gamma (a,b,\ell) \,
\partial_x \psi_t / 2 \bigr)
\nonumber
\\
&=& - \cG(a, b,\ell) \int_\R V' \,
\partial_x \psi_t - \bigl(\Gamma (a,b,\ell) /2\bigr) \int
_\R(\partial_x \ln\psi_t)^2
\psi_t\nonumber
\\
&&{} - \cG(a, b,\ell) \int_\R\bigl(V'
\bigr)^2 \psi_t - \bigl(\Gamma(a,b,\ell)/2\bigr) \int
_\R V' \,\partial_x
\psi_t
\nonumber
\\
&=& \cG(a, b,\ell) b - \bigl(\Gamma(a,b,\ell) /2\bigr) \int_\R(
\partial_x \ln\psi_t)^2 \psi_t
\nonumber
\\
&&{} - \cG(a, b,\ell) a + \Gamma(a,b,\ell) b /2.
\end{eqnarray}
On the other hand, we have
\begin{eqnarray*}
\int_\R\bigl(\partial_x \ln(
\psi_t / \psi_\infty)\bigr)^2 \psi_t
&=&\int_\R\bigl(\partial_x \ln
\psi_t +V'\bigr)^2 \psi_t
\\
&=&\int_\R(\partial_x \ln
\psi_t)^2 \psi_t + 2 \int_\R(
\partial _x \ln\psi_t) V' \psi_t
+ \int_\R\bigl(V'\bigr)^2
\psi_t
\\
&=&\int_\R(\partial_x \ln
\psi_t)^2 \psi_t - 2 b + a.
\end{eqnarray*}
We thus obtain
\begin{eqnarray*}
&&\frac{\mathrm{d}}{\mathrm{d}t} \int_\R\psi_t \ln(
\psi_t / \psi _\infty)
\\
&&\quad= \cG(a, b,\ell) b - \bigl(
\Gamma(a,b,\ell) /2\bigr) \biggl[\int_\R\bigl(
\partial_x \ln(\psi_t / \psi_\infty)
\bigr)^2 \psi_t + 2b -a \biggr]
\\
&&\qquad{} - \cG(a, b,\ell) a + \Gamma(a,b,\ell)b /2
\\
&&\quad= - \bigl(\Gamma(a,b,\ell) /2\bigr) \int_\R\bigl(
\partial_x \ln(\psi_t / \psi_\infty)
\bigr)^2 \psi_t + (b-a) \bigl( \cG(a, b,\ell)-\Gamma(a,b,
\ell)/2\bigr)
\\
&&\quad= \bigl(\Gamma(a,b,\ell) /2\bigr) \biggl[- \int_\R
\bigl(\partial_x \ln(\psi_t / \psi_\infty)
\bigr)^2 \psi_t + 1_{\{b\neq a\}}(b-a)^2
\frac{\cG(a, b,\ell) -
\Gamma(a,b,\ell) /2}{(b-a)\Gamma(a,b,\ell)/2} \biggr],
\end{eqnarray*}
where the ratio $\frac{\cG(a, b,\ell) -
\Gamma(a,b,\ell) /2}{(b-a)\Gamma(a,b,\ell)/2}$ is nonnegative by
\eqref{lem:f}. We remark that
\begin{eqnarray*}
(a-b)^2&=& \biggl(\int_\R
\bigl(V'\bigr)^2\psi_t-\int
_\R V''\psi_t
\biggr)^2= \biggl(\int_\R V'
\bigl(V'\psi_t+\partial_x\psi_t
\bigr) \biggr)^2
\\
&=& \biggl(\int_\R V' \,
\partial_x\ln\bigl(\psi_t / \mathrm{e}^{-V}
\bigr)\psi_t \biggr)^2\leq a\int_\R
\bigl(\partial_x \ln(\psi_t / \psi_\infty)
\bigr)^2 \psi_t.
\end{eqnarray*}
Using the function $F$ defined in \eqref{eq:bGamma-2aG}, we deduce that
\begin{eqnarray*}
\frac{\mathrm{d}}{\mathrm{d}t} \int_\R\psi_t \ln(
\psi_t / \psi _\infty) &\leq& \bigl(\Gamma(a,b,\ell) /2\bigr)
\biggl[- 1 + a\frac{\cG(a, b,\ell) -
\Gamma(a,b,\ell) /2}{(b-a)\Gamma(a,b,\ell)/2} \biggr]\int_\R \bigl(
\partial_x \ln(\psi_t / \psi_\infty)
\bigr)^2 \psi_t
\\
&\leq& - \frac{F(a,b,\ell)}{2} \int_\R\bigl(
\partial_x \ln(\psi_t / \psi_\infty)
\bigr)^2 \psi_t,
\end{eqnarray*}
which is \eqref{optirat}.
Since by Lemma~\ref{lem:fF}, $F$ is positive, we deduce that
\[
\frac{\mathrm{d}}{\mathrm{d}t} \int_\R\psi_t \ln(
\psi_t / \psi_\infty)\leq0.
\]

Let us now assume that $\psi_\infty$ satisfies a logarithmic-Sobolev
inequality \eqref{logsobol} with parameter $\rho$. We thus have, from
\eqref{optirat},
%
%
\begin{equation}
\label{optirat2} \frac{\mathrm{d}}{\mathrm{d}t} H(\psi_t | \psi_\infty)
\le- \rho F\bigl(a[\psi _t],b[\psi_t],\ell\bigr) H(
\psi_t | \psi_\infty).
\end{equation}
%
Thus, to obtain exponential convergence, in view of
Lemma~\ref{lem:fF} and since $b[\psi_t]\in[-\|V''\|_\infty,\allowbreak  \|V''\|
_\infty]$, we need a (uniform-in-time) upper bound on
$\int_\R(V')^2 \psi_t$, to get a (uniform-in-time) positive lower
bound on $F(a[\psi_t],b[\psi_t],\ell)$. This is the aim of the
next paragraph.

First, notice that by \cite{JLM1}, Lemma~1 and Lemma~3,
$\int_\R(V')^2(\psi_t+\psi_\infty)<+\infty$. Now, according
to \cite{otto-villani-00}, Theorem~1, since $\psi_\infty$ satisfies a
$\operatorname{LSI}(\rho)$, $\psi_\infty$ also satisfies the transport
inequality: for any probability density $\varphi$ on $\R$,
\[
W^2_2(\varphi,\psi_\infty)=\inf
_{\gamma(\mathrm{d}x,\mathrm
{d}y)<^{\varphi
(x)\,\mathrm{d}x}_{\psi_\infty(y)\,\mathrm{d}y}}\int_{\R
^2}(x-y)^2\gamma(
\mathrm{d}x,\mathrm{d}y)\leq \frac{2}{\rho}\int_\R
\varphi\ln(\varphi/ \psi_\infty)=\frac
{2}{\rho}H(\varphi|
\psi_\infty),
\]
where, in the definition of the quadratic Wasserstein distance $W_2$,
the infimum is taken over all coupling measures $\gamma$ on $\R^2$
with marginals $\varphi(x)\,\mathrm{d}x$ and $\psi_\infty(y)\,
\mathrm{d}y$. Moreover, for a
coupling measure $\gamma$ between the probability measures $\psi
_t(x)\,\mathrm{d}x$ and $\psi_\infty(y)\,\mathrm{d}y$, we have,
using Cauchy--Schwarz inequality,
\begin{eqnarray*}
\biggl\llvert \int_\R\bigl(V'
\bigr)^2(\psi_t-\psi_\infty)\biggr\rrvert &=&
\biggl\llvert \int_{\R
^2}\bigl(V'(x)+V'(y)
\bigr) \bigl(V'(x)-V'(y)\bigr)\gamma(\mathrm{d}x,
\mathrm{d}y)\biggr\rrvert
\\
&\leq& \biggl(2\int_\R\bigl(V'
\bigr)^2(\psi_t+\psi_\infty)\bigl \|V''
\bigr \|^2_\infty\int_{\R
^2}(x-y)^2
\gamma(\mathrm{d}x,\mathrm{d}y) \biggr)^{1/2}.
\end{eqnarray*}
By taking the infimum over all coupling measures between $\psi_t(x)\,
\mathrm{d}x$
and $\psi_\infty(y)\,\mathrm{d}y$, using the above transport
inequality and the
monotonicity of the relative entropy with respect to $t$, we deduce that
\begin{eqnarray*}
\biggl|\int_\R\bigl(V'\bigr)^2(
\psi_t -\psi_\infty) \biggr|&\leq& \biggl(\frac
{4}{\rho}
\bigl \|V''\bigr \|^2_\infty H(
\psi_t|\psi_\infty)\int_\R
\bigl(V'\bigr)^2(\psi _t+
\psi_\infty) \biggr)^{1/2}
\\
&\leq& \biggl(\frac{4}{\rho}\bigl \|V''
\bigr \|^2_\infty H(\psi_0|\psi_\infty )
\biggl(\biggl\llvert \int_\R\bigl(V'
\bigr)^2(\psi_t-\psi_\infty)\biggr\rrvert +2\int
_\R \bigl(V'\bigr)^2
\psi_\infty \biggr) \biggr)^{1/2}.
\end{eqnarray*}
Setting\vspace*{-1pt} $c=\frac{4}{\rho}\|V''\|^2_\infty H(\psi_0|\psi_\infty)$
and $d=\frac{8}{\rho}\|V''\|^2_\infty
\int_\R(V')^2\psi_\infty$,
one concludes that\break  $|\int_\R(V')^2\*(\psi_t-\psi_\infty)|\leq\frac
{c+\sqrt{c^2+4d}}{2}$ so that
\[
\forall t\geq0,\qquad\int_\R\bigl(V'
\bigr)^2\psi_t\leq\int_\R
\bigl(V'\bigr)^2\psi _\infty +
\frac{c+\sqrt{c^2+4d}}{2}.
\]
%
By definition of $a[\psi_t]$, this yields an upper bound on
$a[\psi_t]$ which depends on $H(\psi_0|\psi_\infty)$. Now, since
$b[\psi_t]\in[-\|V''\|_\infty,\|V''\|_\infty]$, \eqref{lem:F}
implies that $t\mapsto F(a[\psi_t],b[\psi_t],\ell)$ is bounded from
below by a positive and nonincreasing function of
$H(\psi_0|\psi_\infty)=\int_\R\psi_0\ln(\psi_0/\psi_\infty)$. We
conclude that there exists a positive and nonincreasing function
$\lambda\dvtx \R_+\to\R_+^*$ such that
\[
\frac{\mathrm{d}}{\mathrm{d}t} H(\psi_t|\psi_\infty) \leq -\lambda
\bigl(H(\psi_0|\psi_\infty) \bigr) H(\psi_t|
\psi_\infty),
\]
which yields \eqref{eq:convexpo}.
\end{pf*}
%

\section{Optimization strategies for the RWM algorithm}\label{sec:optim_RWM}

In this section, we discuss how to choose the constant $\ell$ in the
scaling $\sigma_n^2=\frac{\ell^2}{n}$ in order to optimize the
convergence to equilibrium, using the nonlinear diffusion
limit \eqref{edsnonlin}.

As a preliminary remark, notice that we will restrict the discussion to
cases when
%
%
\begin{equation}
\label{eq:V''_positif} b[\psi_t]=\E\bigl(V''(X_t)
\bigr) > 0.
\end{equation}
Indeed, points where $V''$ is negative correspond to neighborhood of
local maxima of
the potential $V$, which are visited with very low probability over
large time intervals
by the dynamics \eqref{edsnonlin}. Moreover, we observe from \eqref
{eq:ent1} that if
$b[\psi_t] \le0$, then, since $\Gamma$ and $\cG$ are nonnegative
functions and $a[\psi_t]\geq0$, $\frac{\mathrm{d}}{\mathrm
{d}t}H(\psi_t | \psi
_\infty) \le-
\frac{\Gamma(a[\psi_t],b[\psi_t],\ell)}{2} \int_{\R} (\partial
_x \ln
\psi_t)^2 \psi_t$ so that, since $\lim_{\ell\to\infty}
\Gamma(a,b,\ell)=+\infty$ (when $b \le0$), $\ell$ should be chosen
as large as
possible in order to leave the concave region.

In the following, we thus assume \eqref{eq:V''_positif}.

\subsection{Maximization of the exponential rate of convergence}\label{sec:optim_rate}

In view of the inequalities \eqref{optirat} and \eqref{optirat2}, it
seems natural to try to choose $\ell$
maximizing (for given values $(a,b)=(a[\psi_t],b[\psi_t])$)
\[
\ell\mapsto F(a,b,\ell),
\]
in order to maximize the exponential rate of convergence to zero of
$H(\psi_t|\psi_\infty)$. In view
of \eqref{eq:bGamma-2aG}, for $a\neq b$, this is equivalent to
maximizing $\ell\mapsto\llvert b\Gamma(a,b,\ell)-2a\cG(a,b,\ell
)\rrvert $.

%
\begin{remark}
We notice that, for $X_t$ solution to \eqref{edsnonlin},
%
%
\begin{equation}
\label{evolev} 2\frac{\mathrm{d}}{\mathrm{d}t} \E\bigl(V(X_t)\bigr) = b
\Gamma(a,b,\ell )-2a\cG(a,b,\ell )
\end{equation}
with $(a,b)=(\E(V'(X_t)),\E(V''(X_t)))$, so that this optimization
procedure has a simple interpretation in terms of the evolution of the
energy: it amounts to maximizing $\llvert  \frac{\mathrm{d}}{\mathrm
{d}t} \E
(V(X_t))\rrvert $, namely making the largest possible moves in terms
of energy. This seems quite a reasonable objective.
\end{remark}

%
\begin{remark}\label{rem:gaussian}
In the Gaussian case (namely when
$V(x)=\frac{x^2}{2}+\frac{1}{2}\ln(2\uppi)$), and assuming that the
initial condition is also Gaussian, the density remains Gaussian for
all time. Let us denote $m(t)=\E(X_t)$ its mean and $s(t)=\E(X_t^2)$
its second order moment, which completely characterize the Gaussian
law at time $t$. Simple computations, still valid for non-Gaussian
initial conditions, yield
%
%
\begin{equation}
\label{eq:ode_gauss} \cases{ %
\displaystyle\frac{\mathrm{d}s}{\mathrm{d}t}=
\Gamma(s,1,\ell )-2s{\cG}(s,1,\ell)=F(s,1,\ell ) (1-s),\vspace*{1pt}
\cr
\displaystyle
\frac{\mathrm{d}m}{\mathrm{d}t}=-{\cG}(s,1,\ell)m, 
}
\end{equation}
where the first equation corresponds to \eqref{evolev}, since
$V'(x)=x$ and $V''(x)=1$. We observe that the optimization procedure in
this case amounts to
maximizing $\llvert \frac{\mathrm{d}s}{\mathrm{d}t}\rrvert $. This
accelerates the
convergence to the equilibrium value $1$ of $s$.
\end{remark}

Let us denote
\[
F_1(s,\ell)=F(s,1,\ell)=\cases{ %
\ell^2\exp{ \biggl(-\displaystyle\frac{\ell^2}{2} \biggr)}, \qquad
\mbox{if }s=0,\vspace*{2pt}
\cr
2\ell^2 \biggl( \biggl(1+\displaystyle
\frac{\ell^2}{4} \biggr)\Phi \biggl(-\displaystyle\frac
{\ell}{2} \biggr)-
\displaystyle\frac{\ell}{2\sqrt{2\uppi}}\exp { \biggl(-\frac{\ell ^2}{8} \biggr)} \biggr),\vspace*{1pt}
\cr
\hphantom{\ell^2\exp{ \biggl(-\displaystyle\frac{\ell^2}{2} \biggr)}, \qquad\,}\mbox{if }s=1,\vspace*{2pt}
\cr
\displaystyle\frac{\ell^2}{1-s} \biggl(\Phi \biggl(-\displaystyle
\frac{\ell
}{2\sqrt{s}} \biggr)+(1-2s)\exp{ \biggl(\displaystyle\frac{\ell^2(s-1)}{2}
\biggr)}\Phi \biggl(\displaystyle\frac{\ell
}{2\sqrt{s}}-\ell\sqrt{s} \biggr) \biggr),\vspace*{2pt}
\cr
\hphantom{\ell^2\exp{ \biggl(-\displaystyle\frac{\ell^2}{2} \biggr)}, \qquad\,}\mbox{if }s\in(0,1)\cup (1,+\infty), 
}
\]
the function to be maximized in the Gaussian case, see Remark~\ref
{rem:gaussian}. We observe that (using the fact that $b>0$),
%
%
\begin{equation}
\label{eq:GG} F(a,b,\ell)= \frac{1}{b}F_1 \biggl(
\frac{a}{b},\ell\sqrt{b} \biggr)
\end{equation}
so that the general maximization problem on $F$ can be reduced to the
maximization problem on~$F_1$. Notice that the function $F_1$ is
${\mathcal C}^\infty$ on $\R_+ \times\R_+$.
%
%
\begin{lemma}\label{lem:lstar}
For any $s\geq0$, the function $\ell\mapsto F_1(s,\ell)$ admits a unique
global maximum at a point
%
%
\begin{equation}
\label{eq:lstar} \ell^\star(s)=\argmax_{\ell\ge0} F_1(s,
\ell).
\end{equation}
\end{lemma}
The proof of this lemma is quite tedious and is given in Appendix~\ref
{seclstar}.
From Lemma~\ref{lem:lstar} and Equation \eqref{eq:GG}, we deduce
that, for
$(a,b)\in\R_+\times\R_+^*$, there
exists a unique $\tilde\ell^\star(a,b)$ such that
%
%
\begin{equation}
\label{eq:ltildestar} \tilde\ell^\star(a,b)=\argmax_{\ell\ge0}F(a,b,\ell),
\end{equation}
and that
%
%
\begin{equation}
\label{eq:ltilde} \tilde\ell^\star(a,b)=\frac{1}{\sqrt{b}}
\ell^\star \biggl(\frac{a} b \biggr).
\end{equation}
In particular, $\ell^\star(s)=\tilde\ell^\star(s,1)$. Notice that these
scaling results show that a constant $\ell$ strategy is far from optimal
in the transient case, since when $a$ and $b$ vary, the optimal value
$\tilde\ell^\star(a,b)$ also varies.

We now consider three regimes: the near equilibrium case $s \to1$
(recall that at equilibrium, $a=b$ and thus $s=a/b=1$), and the
two situations far from equilibrium $s \to0$ and $s \to\infty$ (see
Figure~\ref{fig:lstar} for an illustration). In the Gaussian case (see
Remark~\ref{rem:gaussian}), $s(t)=\E(X_t^2)$ so that these three
regimes are easy to understand in terms of second moment.\looseness=-1
%
%
\begin{lemma}\label{lem:asymptotics}
We have the following asymptotic behaviors for the function $\ell
^\star$:
\begin{itemize}[$\bullet$]
\item[$\bullet$] {$s \to1$}: The function $\ell\mapsto F_1(1,\ell)$
admits a unique maximum at point $\ell^\star(1)\simeq1.85$.
Moreover,\vadjust{\goodbreak}
%
%
\begin{equation}
\label{eq:lstar_eq} \lim_{s \to1} \ell^\star(s)=
\ell^\star(1),
\end{equation}
and thus
$\tilde\ell^\star(a,b) \sim_{a/b \to1} \frac{\ell^\star
(1)}{\sqrt{b}}$.
\item[$\bullet$] {$s \to0$}: The function $\ell\mapsto F_1(0,\ell)$
admits a unique maximum at point $\ell^\star(0)=\sqrt{2}$. Moreover,
%
%
\begin{equation}
\label{eq:lstar_0} \lim_{s \to0} \ell^\star(s)=
\ell^\star(0)=\sqrt{2},
\end{equation}
and thus
$\tilde\ell^\star(a,b) \sim_{a/b \to0} \frac{\sqrt{2}}{\sqrt{b}}$.
\item[$\bullet$] {$s \to\infty$}:
Let us introduce
$\psi(x)=x\sqrt{\frac{2}{\uppi}}\mathrm{e}^{-{x^2}/{8}}-x^2\Phi
(-\frac{x}{2} )$. The
function $\psi$ admits a unique maximum at point $x^\star\simeq1.22$.
Moreover,
%
%
\begin{equation}
\label{eq:lstar_infty} \lim_{s\to+\infty}\frac{\ell^\star(s)}{\sqrt{s}}=x^\star
\end{equation}
so that
$\tilde\ell^\star(a,b) \sim_{a/b \to\infty}\frac{x^\star\sqrt{a}}{b}$.
\end{itemize}
\end{lemma}
\begin{pf}
The first two statements for $s=1$ and $s=0$ are
simple consequences
of Lemma~\ref{lem:lstar} and the implicit function theorem applied to
$F_1(s,\ell)$, respectively, at
point $(1,\ell^\star(1))$ and $(0,\ell^\star(0))$, using the fact that
$\frac{\partial^2 F_1}{\partial\ell^2}(1,\ell^\star(1)) \neq
0$ and $\frac{\partial^2 F_1}{\partial\ell^2}(0,\ell^\star(0))
\neq
0$ (see Equations \eqref{eq:d2F1_s1} and \eqref{eq:d2F1_s0} below).

%
\begin{figure}

\includegraphics{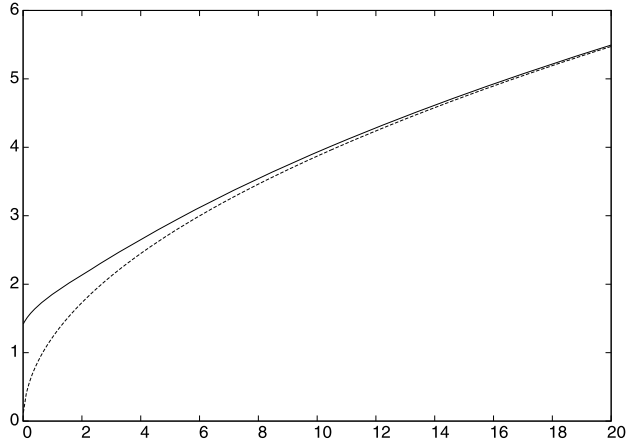}

\caption{Solid line: the function $s \mapsto\ell^\star(s)$. Dashed
line: the function: $s \mapsto
x^\star\sqrt{s}$.}\label{fig:lstar}
\end{figure}

Let us now consider the case $s \to\infty$ and recall the well-known
Mill's inequalities:
%
%
\begin{equation}
\label{contPhi} \forall x<0, \qquad\frac{-x}{\sqrt{2\uppi}(1+x^2)}\exp{ \biggl(-
\frac {x^2}{2} \biggr)}<\Phi(x)<\frac{1}{-x\sqrt{2\uppi}}\exp { \biggl(-
\frac{x^2}{2} \biggr)}.
\end{equation}
One\vspace*{-2pt} has
$\psi'(x)=\sqrt{\frac{2}{\uppi}}\mathrm{e}^{-{x^2}/{8}}-2x\Phi
(-\frac{x}{2} )$
so that $\psi'(0)=\sqrt{\frac{2}{\uppi}}>0$ and, by \eqref{contPhi},
$\psi'(x)\sim-\sqrt{\frac{2}{\uppi}}\mathrm{e}^{-{x^2}/{8}}<0$ as
$x\to+\infty$. Moreover, by the lower-bound in \eqref{contPhi},
\[
\bigl(\mathrm{e}^{{x^2}/{8}}\psi'(x)\bigr)'=-2
\biggl(1+\frac{x^2}{4} \biggr)\mathrm{e}^{{x^2}/{8}}\Phi \biggl(-
\frac{x}{2} \biggr)+\frac
{x}{\sqrt {2\uppi}}<0 \qquad\mbox{for } x>0.
\]
Whence the existence and uniqueness of $x^\star$.

For $s>1/2$, by the upper-bound in \eqref{contPhi}, $(2s-1)\mathrm{e}^{
{\ell^2(s-1)}/{2}}
\Phi (\frac{\ell(1-2s)}{2\sqrt{s}} )\leq\frac{\sqrt
{2s}}{\ell\sqrt{\uppi}}\mathrm{e}^{-{\ell^2}/({8s})}$\vspace*{-2pt}\break  so that, for $s>1$,
$F_1(s,\ell)\leq\frac{s}{s-1}\psi(\ell/\sqrt{s})$. For
$\varepsilon>0$, one deduces that\break
$\lim_{s\to\infty}\sup_{\ell\notin\sqrt{s}[x^\star-\varepsilon
,x^\star+\varepsilon]}F_1(s,\allowbreak  \ell)\leq
\sup_{x\notin[x^\star-\varepsilon,x^\star+\varepsilon]}\psi(x)$.
On the
other hand,
\[
\limsup_{s\to+\infty}F_1\bigl(s,\ell^\star(s)
\bigr)\geq \lim_{s\to+\infty}F_1\bigl(s,x^\star
\sqrt{s}\bigr)=\psi\bigl(x^\star\bigr)>\sup_{x\notin[x^\star-\varepsilon,x^\star+\varepsilon]}
\psi(x).
\]
Hence, for $s$ large enough,
$F_1(s,\ell^\star(s))>\sup_{x\notin[x^\star-\varepsilon,x^\star
+\varepsilon]}
\psi(x)$ and
$\frac{\ell^\star(s)}{\sqrt{s}}\in[x^\star-\varepsilon,x^\star
+\varepsilon]$. Since
$\varepsilon>0$ is arbitrary, this yields \eqref{eq:lstar_infty}.
\end{pf}
%

\subsection{Comparison with the constant average acceptance rate strategy}\label{sec:constant_ratio}

Under the stationarity assumption, it is standard (see \cite
{roberts-gelman-gilks-97}) to associate to the
optimal value of $\ell\simeq\frac{2.38}{\sqrt{I}}$ an average acceptance
rate (see the \hyperref[sec:intro]{Introduction}). Indeed, in this
case, there is a
one-to-one correspondence
between $\ell$ and the limiting acceptance
rate
\[
\acc(I,I,\ell)=\frac{\Gamma(I,I,\ell)}{\ell^2}=\frac{h(I,\ell
)}{\ell^2}=2 \Phi \biggl( -
\frac{\ell\sqrt{I}}{2} \biggr).
\]
More precisely, $\ell\simeq\frac{2.38}{\sqrt{I}}$ is equivalent to
\[
\acc(I,I,\ell)\simeq2 \Phi \biggl( -\frac{2.38}{2} \biggr)\simeq0.23
\]
which does not depend on $I$. A natural strategy is thus to adjust the
variance in such a way that the average acceptance rate is 23\%. In
this section, we discuss how to use an equivalent approach in the
transient phase. Of course, the interest of the constant
average acceptance rate strategy is that it can be implemented using
the so-called adaptive scaling Metropolis algorithm (see
\cite{AndrieuRobert,Atchade}): at iteration $k$, the standard deviation
$\sigma$ is chosen equal to $\exp(\theta_k)$ where $\theta_k$ is
updated using the Robbins--Monro procedure
$\theta_{k+1}=\theta_k+\gamma_{k+1}(\alpha_k-\alpha)$ where
$\alpha_{k+1}$ is the observed acceptance rate
\eqref{def:acceptanceRWM} at iteration $k$, $\alpha\in(0,1)$ is the target
acceptance rate and $(\gamma_{k})$ is a deterministic fixed sequence
of step sizes.

The first question is: for given values of $a$ and $b$, does an
acceptance rate $\alpha\in(0,1)$ corresponds in a one-to-one way to a
value $\ell>0$?
The average acceptance rate is (see Proposition~\ref{prop:limiting-acceptance})
\[
\acc(a,b,\ell)=\frac{\Gamma(a,b,\ell)}{\ell^2}=\Phi \biggl(-\frac
{\ell
b}{2\sqrt{a}} \biggr)+
\mathrm{e}^{{\ell^2(a-b)}/{2}}\Phi \biggl(\ell \biggl(\frac{b}{2\sqrt{a}}-\sqrt{a}
\biggr) \biggr) .
\]
We recall that we only consider the case $b>0$, see the discussion at
the beginning of
Section~\ref{sec:optim_RWM}. (Actually, if $b \le0$, $\acc(a,b,\ell
)\ge\Phi (-\frac{\ell
b}{2\sqrt{a}} ) \ge1/2$ for all $a \ge0$ and $\ell>0$, so
that it is not possible to solve
$\acc(a,b,\ell)=\alpha$ for any values of $\alpha$, which
is again an indication of the ill-posedness of the optimization
procedure when $b \le0$.)

Now, for $b >0$, observe that
\[
\acc(a,b,\ell)=J \biggl(\frac{a}{b},\ell\sqrt{b} \biggr),
\]
where
%
%
\begin{equation}
\label{eq:H} J(s,\ell)=\Phi \biggl(-\frac{\ell}{2\sqrt{s}} \biggr)+
\mathrm{e}^{
{\ell^2(s-1)}/{2}}\Phi \biggl(\ell \biggl(\frac{1}{2\sqrt{s}}-\sqrt {s}
\biggr) \biggr).
\end{equation}
Solving $\acc(a,b,\ell)=\alpha$ amounts to solving
$J (\frac{a}{b},\ell\sqrt{b} )=\alpha$.
%
%
\begin{lemma}\label{lem:existence-acc} Let $s> 0$ be fixed. The
function $\ell\mapsto J(s,\ell)$ is decreasing. Moreover, for all
$\alpha\in(0,1)$ there exists a unique solution to
the equation $J(s,\ell)=\alpha$. This solution is denoted
$\ell^\alpha(s)$ in the following.
\end{lemma}
\begin{pf}
Let us first prove that, for a given $s>0$, $\ell
\mapsto J(s,\ell)$ is strictly
decreasing. We compute
\[
\frac{\partial J}{\partial\ell}(s,\ell) 
= - \sqrt{\frac{s}{2\uppi}} \exp \biggl(-
\frac{\ell^2}{8s} \biggr) + \ell(s-1) \exp \biggl(\frac{\ell^2(s-1)}{2} \biggr)\Phi
\biggl(\frac{\ell
}{2\sqrt{s}} (1-2s ) \biggr).
\]
The right-hand side is negative for $s \in(0,1]$. For $s>1$, we
have, using the upper-bound in \eqref{contPhi},
\[
\frac{\partial J}{\partial\ell}(s,\ell) 
\le-
\sqrt{\frac{s}{2\uppi}} \frac{1 }{
(2s-1)} \exp \biggl(-\frac{\ell^2}{8s}
\biggr) <0.
\]
This shows that $\ell\mapsto J(s,\ell)$ is strictly
decreasing.

It is easy to see that $J(s,0)=1$. Now, using again the upper-bound in
\eqref{contPhi} for $s>1/2$, one has
%
%
\begin{equation}
\label{majoh} J(s,\ell)\leq\Phi{ \biggl(\frac{-\ell}{2\sqrt{s}}
\biggr)}+1_{\{
s\leq1/2\}
}\exp{ \biggl(\frac{\ell^2}{2}(s-1)
\biggr)}+1_{\{s>1/2\}}\frac
{\sqrt {2s}}{\ell(2s-1)\sqrt{\uppi}}\exp{ \biggl(-\frac{\ell^2}{8s}
\biggr)}
\end{equation}
so that $\lim_{\ell\to\infty} J(s,\ell)=0$. By continuity and strict
monotonicity of $J$, we
then get that for any
$\alpha\in(0,1)$ there exists a unique $\ell^\alpha(s)$ such that
$J(s,\ell^\alpha(s))=\alpha$.
\end{pf}

As a corollary of this lemma, we get that for any $a \ge0$, $b>0$,
$\alpha\in(0,1)$,
there exists a unique $\tilde\ell^\alpha(a,b)>0$ such that\vadjust{\goodbreak}
\[
\acc\bigl(a,b,\tilde\ell^\alpha(a,b)\bigr)=\alpha
\]
and that
%
%
\begin{equation}
\label{eq:alphatilde} \tilde\ell^\alpha(a,b)=\frac{1}{\sqrt{b}}
\ell^\alpha \biggl( \frac{a}{b} \biggr).
\end{equation}
In particular, $\ell^\alpha(s)= \tilde\ell^\alpha(s,1)$.

Let us now compare the strategy based on the maximization of the
exponential rate of convergence, presented in
Section~\ref{sec:optim_rate}, with a strategy based on a constant
average acceptance rate. By
comparing \eqref{eq:ltilde} and \eqref{eq:alphatilde}, we observe that
the scalings of $\tilde\ell^\star$ and $\tilde\ell^\alpha$ in
terms of $a$ and
$b$ are the same, which is already an indication of the fact that a
constant acceptance rate strategy is very natural.

Near equilibrium, namely in the limit $a/b \to1$, the
two strategies are the same if $\alpha$ is chosen such that
$\ell^\alpha(1)=\ell^\star(1)$ which corresponds to
%
%
\begin{equation}
\label{eq:alpha_eq} \alpha\simeq0.35.
\end{equation}
Notice that this value is not far (but different, since we take into
account the transient phase around equilibrium) from the acceptance
probability 0.23
obtained under the stationarity assumption.

To study the two limits $s\to0$ and $s \to\infty$, we need the
following lemma.\vspace*{-3pt}
%
%
\begin{lemma} We have the following asymptotic behaviors for the
function $\ell^\alpha$:
\begin{itemize}[$\bullet$]
\item[$\bullet$]$s \to0$:
For any $\alpha\in(0,1)$,
%
%
\begin{equation}
\label{eq:lalpha_0} \lim_{s\to0} \ell^\alpha(s)=\sqrt{-2\ln(
\alpha)},
\end{equation}
and thus $\tilde\ell^\alpha(a,b) \sim_{a/b \to0} \frac{\sqrt {-2\ln(\alpha)}}{\sqrt{b}}$.
\item[$\bullet$]$s \to\infty$: For any $\alpha\in (0,\frac
{1}{2} )$,
%
%
\begin{equation}
\label{eq:lalpha_infty} \lim_{s\to\infty}\frac{\ell^\alpha(s)}{\sqrt{s}}=-2\Phi
^{-1}(\alpha),
\end{equation}
and thus $\tilde\ell^\alpha(a,b) \sim_{a/b \to\infty}
-2\Phi^{-1}(\alpha) \frac{\sqrt{a}}{b}$.\vspace*{-3pt}
\end{itemize}
\end{lemma}
\begin{pf} Let us first consider the case $s\to0$.
Observe that for any given $\ell\geq0$, it holds
%
%
\begin{equation}
\label{eq:Limit-s-0} \lim_{s\to0}J(s,\ell)=\exp{ \biggl(
\frac{-\ell^2}{2} \biggr)}.
\end{equation}
Let $\varepsilon\in(0,-2\ln(\alpha))$. By the monotonicity property
of $\ell\mapsto J(s,\ell)$ stated in Lemma~\ref{lem:existence-acc},
\[
\sup_{\ell\geq\sqrt{-2\ln(\alpha)+\varepsilon}}J(s,\ell)\leq J\bigl(s,\sqrt{-2\ln(\alpha)+
\varepsilon}\bigr)\longrightarrow_{s\to0} \alpha\exp{ \biggl(
\frac{-\varepsilon}{2} \biggr)}.\vadjust{\goodbreak}
\]
In the same way, $\liminf_{s\to0}\inf_{\ell\leq\sqrt{-2\ln
(\alpha)-\varepsilon}}\geq\alpha\exp{ (\frac{\varepsilon
}{2} )}$. Therefore, for $s$ close enough to $0$, $\ell^\alpha
(s)\in
[\sqrt{-2\ln(\alpha)-\varepsilon},\sqrt{-2\ln(\alpha
)+\varepsilon}]$ and \eqref{eq:lalpha_0} holds.

Let us now consider the case $s \to\infty$. Observe that
$J(s,\ell)>\Phi{ (-\frac{\ell}{2\sqrt{s}} )}$. Hence, if
\[
\liminf_{s\to\infty}\frac{\ell^\alpha(s)}{\sqrt{s}}=0
\]
 then
$\liminf_{s\to\infty}J(s,\ell^\alpha(s))\geq\frac{1}{2}$. Therefore,
there exists a constant $C_1>0$ such that, for large enough $s$, $\ell
^\alpha(s)\geq C_1\sqrt{s}$. Using \eqref{majoh} for the
upper-bound, we get that, for $s$ large enough,
\[
\Phi{ \biggl(-\frac{\ell^\alpha(s)}{2\sqrt{s}} \biggr)}\leq\alpha \leq\Phi { \biggl(-
\frac{\ell^\alpha(s)}{2\sqrt{s}} \biggr)}+\frac{\sqrt {2}}{\sqrt {\uppi}C_1(2s-1)}\exp{ \biggl(-\frac{C_1^2}{8}
\biggr)}.
\]
Therefore,
\[
\lim_{s\to\infty}\Phi{ \biggl(-\frac{\ell^\alpha(s)}{2\sqrt {s}} \biggr)}=\alpha.
\]
The continuity of $\Phi^{-1}$ concludes the proof of \eqref{eq:lalpha_infty}.
\end{pf}

By comparing \eqref{eq:lstar_0} with \eqref{eq:lalpha_0}, we observe
that in the regime $a/b \to0$, the two strategies are the same if
$\alpha$ is chosen such that $\sqrt{-2 \ln(\alpha)}=\sqrt{2}$ namely
%
%
\begin{equation}
\label{eq:alpha_0} \alpha=\mathrm{e}^{-1}\sim0.37.
\end{equation}

Finally, by comparing \eqref{eq:lstar_infty} with \eqref
{eq:lalpha_infty}, we observe
that in the regime $a/b \to\infty$, the two strategies are the same if
$\alpha$ is chosen such that $-2 \Phi^{-1}(\alpha)=x^\star$ namely
%
%
\begin{equation}
\label{eq:alpha_infty} \alpha\sim0.27.
\end{equation}

In view of \eqref{eq:alpha_eq}, \eqref{eq:alpha_0} and \eqref
{eq:alpha_infty}, the constant average acceptance rate strategy with
target value between $1/4$ and $1/3$ seems to be a
very good strategy, since it is almost equivalent to the optimal
exponential rate strategy.

\section{Numerical experiments}\label{sec:numerics}

In this section, we present numerical experiments to illustrate results
from Section~\ref{sec:optim_RWM}.

\subsection{On the choice of the target average acceptance rate}

In this section, we would like to discuss the choice of the average
acceptance rate $\alpha$ in the constant average acceptance rate
strategy. As mentioned above, we identified three different values of
$\alpha$ for the constant average acceptance rate to be equivalent to
the optimization of the exponential rate of convergence, depending on
the regimes: $\frac{a}{b} \to1$ ($\alpha\simeq0.35$); $\frac{a}{b}
\to0$ ($\alpha\simeq0.37$); $\frac{a}{b} \to\infty$ ($\alpha
\simeq0.27$).

\begin{figure}

\includegraphics{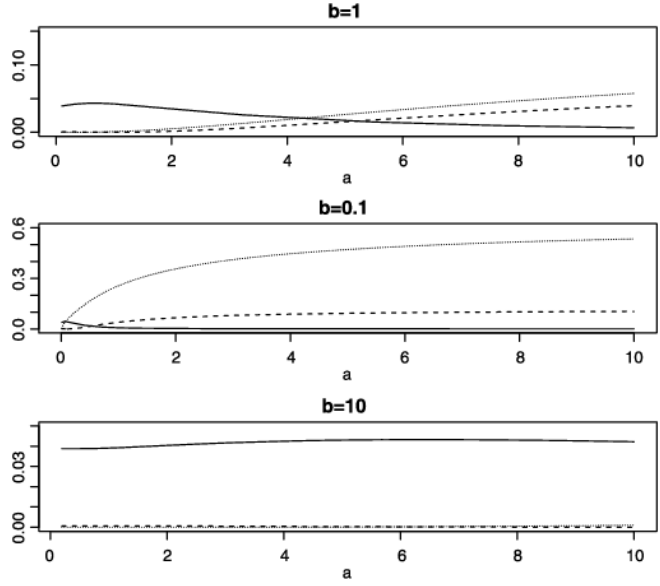}

\caption{$\frac{F(a,b,\tilde\ell^\star(a,b))-F(a,b,\tilde\ell
^\alpha(a,b))}{F(a,b,\tilde\ell^\star(a,b))}$ as function of $a$
for $b=1,0.1,10$ and $\alpha\simeq0.27$ (solid line), $\alpha\simeq
0.35$ (dashed line), $\alpha=\mathrm{e}^{-1}\simeq0.37$ (dotted
line).}\label{fig:diff}
\end{figure}

In practice, a value has to be chosen for $\alpha$. On
Figure~\ref{fig:diff}, we plot as a function of $a$ and $b$ the
relative loss in terms of exponential rate of convergence, for the
constant average acceptance rate strategy compared to the
optimization of the exponential rate of convergence:
$\frac{F(a,b,\tilde\ell^\star(a,b))-F(a,b,\tilde
\ell^\alpha(a,b))}{F(a,b,\tilde\ell^\star(a,b))}$, for the three values
of $\alpha$ mentioned above.

The main output of these numerical experiments is that the choice
$\alpha\simeq0.27$ seems to be the most robust, namely the one
which leads to an exponential rate of convergence the closest to the
optimal one, over the largest range of variation of $a$ and $b$. This
confirms the interest of the constant acceptance rate strategy.

%

\subsection{Gaussian case}

Let us first consider the Gaussian target
$V(x)=\frac{1}{2} (x^2+\ln(2\uppi) )$ (see Remark~\ref
{rem:gaussian}), with
a Gaussian initial condition $X_0$ such that $m(0)=\E(X_0)$ and
$s(0)=\E(X_0^2)$. At time $t$,
the law of $X_t$ solution to the
limiting stochastic differential equation \eqref{edsnonlin} is
Gaussian with mean $m(t)$ and second moment $s(t)$, where $m$ and $s$
satisfies \eqref{eq:ode_gauss}. The Kullback--Leibler divergence admits
an analytical expression in terms of $m$ and $s$:
\[
H(\psi_t|\psi_\infty)=\tfrac{1}{2} \bigl(s(t)- \ln
\bigl(s(t)-m(t)^2\bigr)-1 \bigr),
\]
and its derivative writes
\begin{eqnarray*}
\frac{\mathrm{d}}{\mathrm{d}t} H(\psi_t| \psi_\infty) &=&
\frac{1}{2} \biggl(\frac{\mathrm{d}s}{\mathrm{d}t}-\frac{
{\mathrm{d}s}/{\mathrm{d}t}-2m
{\mathrm{d}m}/{\mathrm{d}t}}{s-m^2} \biggr)
\\
&=&\frac{1}{2} \biggl(F_1(s,\ell) (1-s)-\frac{F_1(s,\ell)(1-s)+2m
{\mathcal G}(s,1,\ell)}{s-m^2}
\biggr).
\end{eqnarray*}
In the Gaussian case, it is thus possible, for each time
$t$ (and thus for fixed values of $m(t)$ and $s(t)$), to
minimize
$  \frac{\mathrm{d}}{\mathrm{d}t} H(\psi_t| \psi_\infty)
  $ in $\ell$. This yields the best strategy that we could think
of and
implement numerically, in terms of the speed of convergence of the
Kullback--Leibler divergence to 0.
In the following, let us denote
\[
\ell^\mathrm{ent}(m,s)=\arg\min_{\ell}
\biggl(F_1(s,\ell ) (1-s)-\frac{F_1(s,\ell)(1-s)+2m {\mathcal G}(s,1,\ell
)}{s-m^2} \biggr)   .
\]

In the numerical experiments, we thus compare four strategies: (i) the
constant $\ell$ strategy, with $\ell=2.38$ (which is the optimal
value under
stationarity assumption since $I=1$ in the Gaussian case); (ii) the
constant average acceptance rate strategy,
using $\ell^\alpha(a,b)$ (for
$\alpha\simeq0.27$ and $\alpha=\mathrm{e}^{-1} \simeq0.37$); (iii) the
optimal exponential rate of convergence using $\ell^\star(a,b)$;
(iv) the
optimal strategy for the convergence of the entropy using $\ell
^\mathrm{ent}(m,s)$.
Notice that in the Gaussian case, $a=\E(X_t^2)=s(t)$ and $b=1$, so
that $\ell^\alpha$ and $\ell^\star$ are actually functions of $s$
only. Let
us also mention that there are actually two ways to implement
(ii): either using a numerical approximation for $\ell^\alpha(a,1)$
(and an estimator $\hat{a}$ of $a$), or using the adaptive scaling Metropolis
algorithm mentioned at the beginning of
Section~\ref{sec:constant_ratio} (see \cite{AndrieuRobert,Atchade}).

The dimension is fixed to $n=100$. To assess the convergence, we
observe, as a function of the so-called
burn-in time $t_0$, the convergence to zero of the square biases:
%
%
\begin{equation}
\label{eq:MSB} \bigl(\E \bigl(\hat I^s_{t_0,T+t_0} \bigr) - 1
\bigr)^2 \quad \mbox{and}\quad \bigl(\E \bigl(\hat
I^m_{t_0,T+t_0} \bigr) \bigr)^2,
\end{equation}
where
%
%
\begin{equation}
\label{eq:Is} \hat I^s_{t_0,T+t_0}=\frac{1}{T}\sum
_{k=t_0+1}^{T+t_0}\frac
{(X_k^1)^2+\cdots+ (X_k^{100})^2}{100}
\end{equation}
and
%
%
\begin{equation}
\label{eq:Im} \hat I^m_{t_0,T+t_0}=\frac{1}{T}\sum
_{k=t_0+1}^{T+t_0}\frac
{X_k^1+\cdots+X_k^{100}}{100}.
\end{equation}
The expectations in \eqref{eq:MSB} are approximated by empirical
averages over 200 independent realizations of
$(X_k^1,\ldots,X_k^{100})_{0 \le k \le t_0+T}$. The size of the time
window is
$T=1500$. When needed, we estimate the values of $s=a$ and $m$ using
empirical averages over the $n=100$ components of the process.

\begin{figure}

\includegraphics{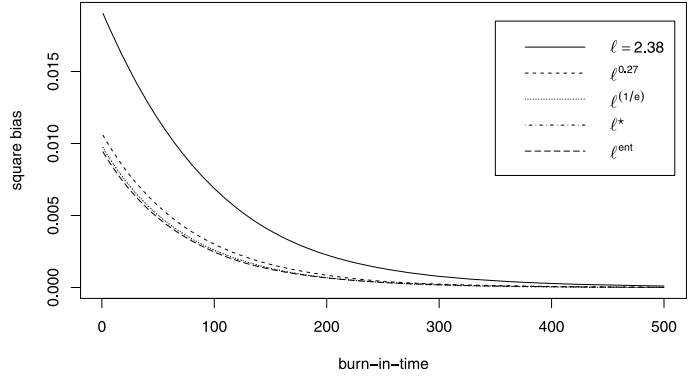}

\caption{Square bias of $\hat I^s_{t_0,t_0+T}$ as a function of the
burn-in-time $t_0$ for various strategies. The initial condition is
$(0,\ldots,0)$, and the constant acceptance rate strategies are
implemented using an adaptive scaling Metropolis algorithm.}\label{fig:gauss_0}
\end{figure}

On Figure~\ref{fig:gauss_0}, we first consider the initial condition
$X_0=(0,\ldots,0)$. The first moment is thus already at equilibrium,
and we only observe the convergence of the second moment. Clearly,
the\vadjust{\goodbreak}
constant $\ell$ strategy is the worst. Using $\ell^\star$ yields a
convergence which is almost the optimal one, obtained for $\ell=\ell
^\mathrm{ent}$. And
the constant average rate strategies also lead to excellent results
in terms of convergence compared to the optimal scenario, even though
it is here implemented using an adaptive scaling Metropolis algorithm.

%

%
\begin{figure}

\includegraphics{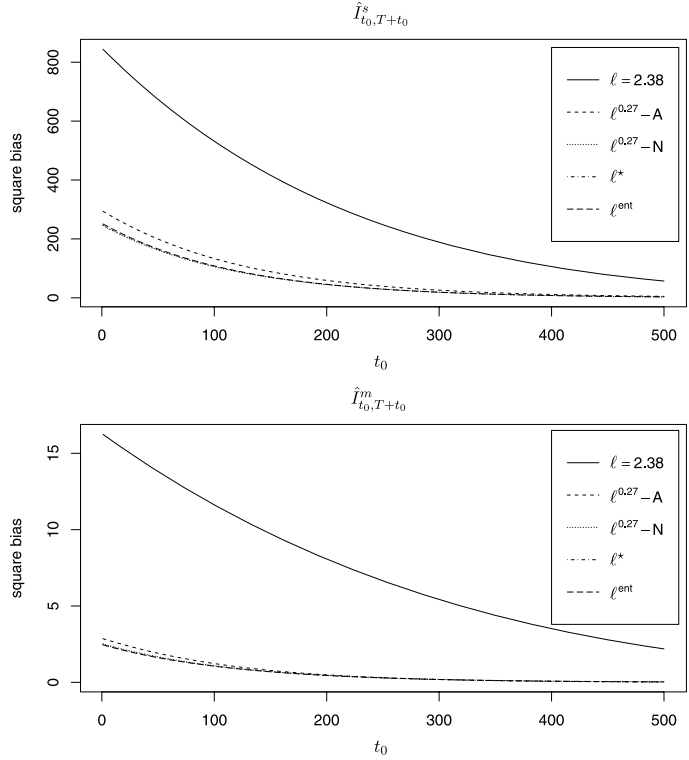}

\caption{Square bias of $\hat I^s_{t_0,T+t_0}$ (top) and $\hat
I^m_{t_0,T+t_0}$ (bottom) as a function of the burn-in-time\vspace*{-1pt} $t_0$ for
various strategies. The initial condition is
$(10,\ldots,10)$. The notations $\ell^{0.27} - A $ and $\ell^{0.27}
- N$
refer to the two implementations of the constant average acceptance
rate: $A$ for the adaptive scaling Metropolis algorithm and $N$ for the
numerical approximation of $\ell^\alpha(a,1)$.}\label{fig:gauss_10}
\end{figure}

On Figure~\ref{fig:gauss_10}, we perform similar experiments with the
initial condition $X_0=(10, \ldots, 10)$. We observe the convergence
of the first and second moment. It is clear that the constant $\ell$
strategy is
outperformed by all the other strategies. We notice also that the adaptive
scaling Metropolis implementation leads to slightly slower
convergences compared to an implementation using
$\ell^\alpha(\hat{a},1)$. This difference could certainly be reduced
by optimizing
the parameters in the adaptive scaling Metropolis algorithm.

In conclusion, we observed that: (i) The constant $\ell$ strategy is bad;
(ii) The constant average acceptance rate strategy (using $\ell
^\alpha$) leads to convergence curves which are very close to the ones
obtained with the optimal exponential rate of
convergence strategy (using $\ell^\star$); (iii) the optimal
exponential rate of
convergence strategy is as good as the most optimal strategy one could
design in terms of entropy decay (using $\ell^\mathrm{ent}$).

\subsection{Non-Gaussian case}

Let us now consider a non-Gaussian target, and more precisely a
double-well potential. In order to satisfy the
assumptions of Theorem~\ref{chaos}, we consider the function $V$ given
up to a normalizing additive constant by:
\[
V(x)= %
\cases{ (x-1)^2(x+1)^2,&\quad\mbox{if } $|x|\leq1$,
\cr
4x^2-8|x|+4,&\quad\mbox{otherwise}. } %
\]
Simple calculations yield
\[
V'(x)= %
\cases{ 2(x-1) (x+1)^2+2(x-1)^2(x+1),
&\quad\mbox{if } $|x|\leq1$,
\cr
8x-8 \sign(x),&\quad\mbox{otherwise}, }
\]
and
\[
V''(x)= %
\cases{ 2(x+1)^2+8(x-1)
(x+1)+2(x-1)^2, &\quad\mbox{if } $|x|\leq1$,
\cr
8,&\quad
\mbox{otherwise}. } %
\]
Of course, no analytical expression for the entropy is available in
this context, and we thus concentrate on the three following
strategies: (i) the
constant $\ell$ strategy; (ii) $\ell=\ell^\alpha(a,b)$ and (iii)
$\ell=\ell^\star(a,b)$. For the constant $\ell$ strategy, we use
$\ell=\frac{2.38}{\sqrt{I}}=1.18$ (where we recall, $I$ is defined
by \eqref{eq:h}). When
needed, $a$ and $b$ are approximated by the estimators over the $n$
components $\hat a=
\frac{1}{n}\sum_{i=1}^{n} V'(X_t^i)^2$ and $\hat b=\frac{1}{n}\sum_{i=1}^{n}V''(X_t^i)$.
The parameters $n=100$ and $T=1500$ are the same as in the Gaussian case.

\begin{figure}

\includegraphics{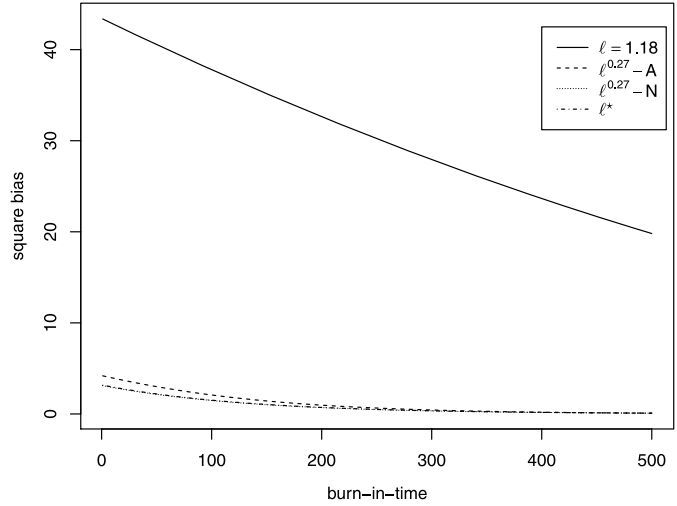}

\caption{Square\vspace*{-1pt} bias of $\hat I^m_{t_0,T+t_0}$ for non-Gaussian target
as a function of the burn-in-time $t_0$ for various strategies. The
initial condition is $(10,\ldots,10)$. The notations $\ell^{0.27} - A
$ and $\ell^{0.27} - N$
refer to the two implementations of the constant average acceptance
rate: $A$ for the adaptive scaling Metropolis algorithm and $N$ for the
numerical approximation of $\ell^\alpha(a,b)$.}\label{fig:ngauss_10}
\end{figure}

Let us first consider as an initial condition $X_0=(10, \ldots,
10)$. On Figure~\ref{fig:ngauss_10}, we observe the convergence of the
first moment to its equilibrium value (namely 0). Again, the constant
$\ell$ strategy appears to be very bad, and the other strategies perform
approximately equally well.

%

\begin{figure}

\includegraphics{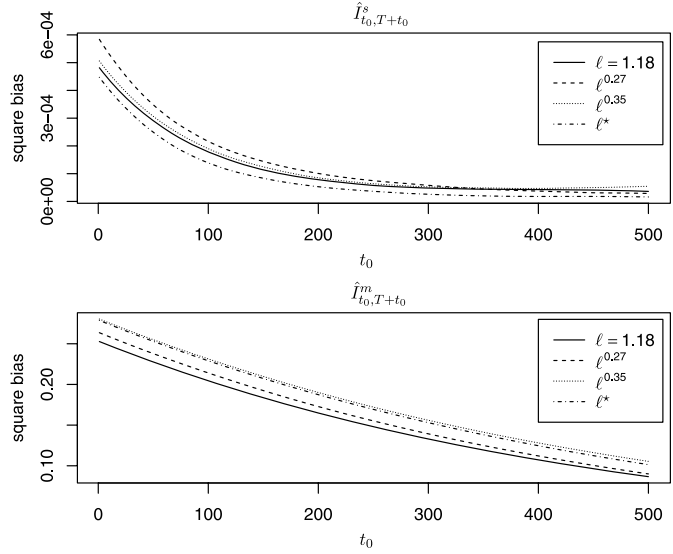}

\caption{Square\vspace*{1pt} bias of $\hat I^s_{t_0,T+t_0}$ (top) and $\hat
I^m_{t_0,T+t_0}$ (bottom) for non-Gaussian target as a function of the
burn-in-time $t_0$ for various strategies and Gaussian initial
condition. The constant acceptance rate strategies are implemented
using an adaptive scaling Metropolis algorithm.}\label{fig:ngauss_eq}
\end{figure}

Finally, let us consider $X_0$ distributed according to a Gaussian
distribution with mean $1$ and variance $0.143 \mathrm{Id}$. The mean and
the variance are chosen in such a way that $a=\E(V'(X_0)^2)=5.24$ and
$b=\E(V''(X_0))=5.24$. On Figure~\ref{fig:ngauss_eq}, we observe the
convergence of the
first and second moments to their equilibrium values (namely 0 and
0.96). For the constant acceptance rate strategy, we compare the
results obtained with $\alpha=0.35$ and $\alpha=0.27$. Here, it is
much more complicated to draw general conclusions from these
plots. Basically, all strategies yield comparable results. One could
wonder why $\ell^\star$ performs poorly for the first moment. The
reason is probably that its bias cannot be encoded into $a$ and $b$
which are integrals of even functions with respect to the current
marginal distribution.\looseness=-1

%

In conclusion, we observed that the results obtained with the constant
acceptance rate strategy (even when it is implemented using an adaptive
scaling Metropolis algorithm) are very similar to those obtained with
the optimal exponential rate of convergence strategy.\looseness=-1

\section{Scaling limits for the MALA algorithm}\label{sec:MALA}

The aim of this section is to derive a diffusive limit for the MALA
algorithm, following the same reasoning as for the RWM algorithm in
Section~\ref{sec:RWM}.

The Markov chain generated by the MALA algorithm writes:
%
%
\begin{eqnarray}
\label{eq:MALA} &&X^{i,n}_{k+1}=X^{i,n}_k
+Z^{i,n}_{k+1}1_{{\mathcal A}_{k+1}}
\\[-1pt]
&&\quad\mbox{where } Z^{i,n}_ {k+1}=\sigma_n
G^i_{k+1}-\frac{\sigma_n^2}{2}V'
\bigl(X^{i,n}_k\bigr),\qquad1\leq i\leq n,\nonumber
\\[-1pt]
&&\quad\mbox{and } {\mathcal A}_{k+1}= \Biggl
\{U_{k+1}\leq\exp\Biggl\{ \sum_{i=1}^n
\biggl( V\bigl(X^{i,n}_k\bigr)-V\bigl(X^{i,n}_k
+Z^{i,n}_{k+1}\bigr)\nonumber
\\[-1pt]
&&\hphantom{\quad\mbox{and } {\mathcal A}_{k+1}= \Biggl
\{}{}+\frac{1}{2} \biggl[
\bigl(G^i_{k+1}\bigr)^2-\biggl(G^i_{k+1}-
\frac{\sigma
_n}{2}\bigl(V'\bigl(X^{i,n}_k
\bigr)+V'\bigl(X^{i,n}_k +Z^{i,n}_{k+1}
\bigr)\bigr) \biggr)^2 \biggr] \biggr) \Biggr\} \Biggr\}
\nonumber
\end{eqnarray}
is the accepting event.
Here again, $(G^i_k)_{i,k\geq1}$ is a sequence of i.i.d. normal random
variables independent from a sequence $(U_k)_{k\geq1}$ of i.i.d.
random variables uniform on $[0,1]$. In
Section~\ref{sec:MALA_formal}, we formally derive a limiting diffusion
process. It appears that the scaling to be used depend on the sign of
$\E ( (
(V')^2V''+V^{(4)}-2V^{(3)}V'-(V'')^2
)(X^{1,n}_k) )$. This is more rigorously discussed in
Section~\ref{sec:MALA_Gauss} for a Gaussian target probability measure.

\subsection{A formal derivation of the limiting process}\label{sec:MALA_formal}

\subsubsection{Asymptotic analysis and limiting process}

We adapt the same strategy as for the RWM algorithm, in Section~\ref
{sec:RWM_formal}.
Let us first discuss how to choose the proper scaling for $\sigma_n$.
Using a Taylor expansion, one obtains:
\begin{eqnarray*}
&&V\bigl(X^{i,n}_k\bigr)-V\bigl(X^{i,n}_k
+Z^{i,n}_{k+1}\bigr)+\frac{1}{2} \biggl[
\bigl(G^i_{k+1}\bigr)^2- \biggl(G^i_{k+1}-
\frac{\sigma_n}{2}\bigl(V'\bigl(X^{i,n}_k
\bigr)+V'\bigl(X^{i,n}_k +Z^{i,n}_{k+1}
\bigr)\bigr) \biggr)^2 \biggr]
\\[-1pt]
&&\quad=\sigma_n^3 \biggl(\frac
{V^{(3)}(X^{i,n}_k)}{12}
\bigl(G^i_{k+1}\bigr)^3-\frac
{V'V''(X^{i,n}_k)}{4}G^i_{k+1}
\biggr)
\\
&&\qquad{} +\sigma_n^4 \biggl(\frac{(V')^2V''(X^{i,n}_k)}{8}+
\frac
{V^{(4)}(X^{i,n}_k)}{24}\bigl(G^i_{k+1}\bigr)^4
 -
\frac
{2V^{(3)}V'(X^{i,n}_k)+(V''(X^{i,n}_k))^2}{8}\bigl(G^i_{k+1}\bigr)^2
\biggr)
\\[-1pt]
&&\qquad{} +{\mathcal{\mathcal O}}\bigl(\sigma_n^5
\bigr).
\end{eqnarray*}
Setting, as above, $\nu^n_{k}=\frac{1}{n} \sum_{i=1}^n
\delta_{X^{i,n}_k}$, one has by Gaussian computations
\begin{eqnarray*}
&&\E \Biggl( \Biggl(\sum_{i=1}^n \biggl(
\frac
{V^{(3)}(X^{i,n}_k)}{12}\bigl(G^i_{k+1}\bigr)^3-
\frac
{V'V''(X^{i,n}_k)}{4}G^i_{k+1} \biggr) \Biggr)^2
\Biggr)
\\
&&\quad=\frac{n}{48}\E \bigl(\bigl\langle\nu ^n_k,5
\bigl(V^{(3)}\bigr)^2-6V'V''V^{(3)}+3
\bigl(V'V''\bigr)^2\bigr
\rangle \bigr)
\end{eqnarray*}
so that one expects that $\sum_{i=1}^n (\frac
{V^{(3)}(X^{i,n}_k)}{12}(G^i_{k+1})^3-\frac
{V'V''(X^{i,n}_k)}{4}G^i_{k+1} )={\mathcal O}(\sqrt{n})$ and
similarly that
$\sum_{i=1}^n (\frac
{V^{(4)}(X^{i,n}_k)}{24}((G^i_{k+1})^4-3)-\frac
{2V^{(3)}V'(X^{i,n}_k)+(V''(X^{i,n}_k))^2}{8}((G^i_{k+1})^2-1)
)={\mathcal O}(\sqrt{n})$. If\vspace*{1pt} this holds and $\lim_{n\to\infty
}\sigma_n=0$, then
\begin{eqnarray*}
&&\sum_{i=1}^n \biggl(V
\bigl(X^{i,n}_k\bigr)-V\bigl(X^{i,n}_k
+Z^{i,n}_{k+1}\bigr)
\\[4pt]
&&\hphantom{\sum_{i=1}^n \biggl(}{}+\frac{1}{2} \biggl[
\bigl(G^i_{k+1}\bigr)^2- \biggl(G^i_{k+1}-
\frac{\sigma_n}{2}\bigl(V'\bigl(X^{i,n}_k
\bigr)+V'\bigl(X^{i,n}_k +Z^{i,n}_{k+1}
\bigr)\bigr) \biggr)^2 \biggr] \biggr)
\\
&&\quad=\frac{n\sigma_n^4}{8}\bigl\langle\nu ^n_k,
\bigl(V'\bigr)^2V''+V^{(4)}-2V^{(3)}V'-
\bigl(V''\bigr)^2\bigr\rangle+ {\mathcal
O}\bigl(\sqrt{n}\sigma_n^3\bigr)+{\mathcal O}\bigl(n
\sigma_n^5\bigr).
\end{eqnarray*}
From this,
%
%
\begin{equation}
\label{eq:accept_MALA} \E \bigl( 1_{{\mathcal
A}_{k+1}} | {\mathcal F}^n_k
\bigr) 
=\mathrm{e}^{({n \sigma_n^4}/{8}) \langle\nu^n_k,  (
(V')^2V''+V^{(4)}-2V^{(3)}V'-(V'')^2  ) \rangle} \wedge 1+{\mathcal O}\bigl(\sqrt{n}
\sigma_n^3\bigr)+{\mathcal O}\bigl(n \sigma
_n^{5}\bigr).
\end{equation}
Here, we have assumed that $\langle\nu^n_k,
(V')^2V''+V^{(4)}-2V^{(3)}V'-(V'')^2 \rangle\neq0$. From this
formula, we get the correct scaling for the variance, in order to
obtain a nontrivial limiting acceptance rate (in accordance with \cite
{christensen-roberts-rosenthal-05}, Section~5):
\[
\sigma_n^4 =\frac{\ell^4}{n}.
\]

Now, following the same reasoning as in Section~\ref{sec:RWM_formal},
we have: for a test
function $\varphi\dvtx \R\to\R$,
\begin{eqnarray*}
&&\E \bigl(\varphi\bigl(X^{1,n}_{k+1}\bigr) | {\mathcal
F}^n_k \bigr)
\\
&&\quad=\E \bigl(\varphi \bigl(X^{1,n}_k
+Z^{1,n}_{k+1}1_{{\mathcal
A}_{k+1}} \bigr) | {\mathcal
F}^n_k \bigr)
\\
&& \quad= \E \biggl(\varphi \bigl(X^{1,n}_k \bigr) +
\varphi' \bigl(X^{1,n}_k \bigr)
Z^{1,n}_{k+1} 1_{{\mathcal
A}_{k+1}} + \tfrac{1}{2}
\varphi'' \bigl(X^{1,n}_k \bigr)
\bigl(Z^{1,n}_{k+1}\bigr)^2 1_{{\mathcal
A}_{k+1}} | {
\mathcal F}^n_k \biggr) + {\mathcal O}
\bigl(n^{-3/4}\bigr)
\\
&& \quad= \varphi \bigl(X^{1,n}_k \bigr) +
\varphi' \bigl(X^{1,n}_k \bigr) \E \bigl(
Z^{1,n}_{k+1} 1_{{\mathcal
A}_{k+1}} | {\mathcal
F}^n_k \bigr) + \tfrac{1}{2} \varphi''
\bigl(X^{1,n}_k \bigr) \E \bigl( \bigl(Z^{1,n}_{k+1}
\bigr)^2 1_{{\mathcal
A}_{k+1}} | {\mathcal F}^n_k
\bigr) + {\mathcal O}\bigl(n^{-3/4}\bigr).
\end{eqnarray*}
Using the Lipschitz continuity of $y\mapsto\mathrm{e}^{y}\wedge1$,
one may
remove the contribution of the $i$th coordinate in the acceptance ratio
and then introduce it again after using conditional independence to
check that
\begin{eqnarray*}
\E\bigl(G^{i}_{k+1}1_{{\mathcal A}_{k+1}}|{\mathcal
F}^n_k\bigr)&=&\E\bigl(G^{i}_{k+1}
\bigr) \E\bigl(1_{{\mathcal A}_{k+1}}|{\mathcal F}^n_k\bigr)+{
\mathcal O}\bigl(\sigma_n^3\bigr)={\mathcal O}
\bigl(n^{-3/4}\bigr),
\\
\E\bigl(\bigl(G^{i}_{k+1}\bigr)^21_{{\mathcal A}_{k+1}}|{
\mathcal F}^n_k\bigr)&=&\E\bigl[\bigl(G^{i}_{k+1}
\bigr)^2\bigr] \E\bigl(1_{{\mathcal A}_{k+1}}|{\mathcal
F}^n_k\bigr)+{\mathcal O}\bigl(\sigma_n^3
\bigr)=\P\bigl({\mathcal A}_{k+1}|{\mathcal F}^n_k
\bigr)+{\mathcal O}\bigl(n^{-3/4}\bigr).
\end{eqnarray*}
From this, one obtains
\begin{eqnarray*}
&&\E \bigl(\varphi\bigl(X^{1,n}_{k+1}\bigr) | {\mathcal
F}^n_k \bigr)
\\
&&\quad= \varphi \bigl(X^{1,n}_k \bigr) +
\varphi' \bigl(X^{1,n}_k \bigr) \E \biggl(
\biggl( \sigma_nG^i_{k+1}-\frac{\sigma_n^2}{2}V'
\bigl(X^{1,n}_k\bigr) \biggr) 1_{{\mathcal
A}_{k+1}} \big| {\mathcal
F}^n_k \biggr)
\\
&& \qquad{} + \frac{1} 2 \varphi''
\bigl(X^{1,n}_k \bigr) \E \biggl( \biggl(
\sigma_nG^i_{k+1}-\frac{\sigma
_n^2}{2}V'
\bigl(X^{1,n}_k\bigr) \biggr)^2
1_{{\mathcal
A}_{k+1}}\big | {\mathcal F}^n_k \biggr) + {\mathcal
O}\bigl(n^{-3/4}\bigr)
\\
&&\quad= \varphi \bigl(X^{1,n}_k \bigr) +
\frac{\ell^2}{2\sqrt {n}} \bigl( \mathrm{e}^{({\ell^4}/{8}) \langle\nu^n_k,  (
(V')^2V''+V^{(4)}-2V^{(3)}V'-(V'')^2  ) \rangle}\wedge1 \bigr)
\\
&&\qquad{}\times \bigl( -
V'\bigl(X^{1,n}_k\bigr) \varphi'
\bigl(X^{1,n}_k \bigr)+\varphi''
\bigl(X^{1,n}_k \bigr) \bigr)
  + {\mathcal O}\bigl(n^{-3/4}\bigr).
\end{eqnarray*}
The correct scaling\vspace*{-2pt} in time is thus to consider a piecewise linear
process $Y^{1,n}_t$ such that $Y^{1,n}_{k/\sqrt{n}}=X^{1,n}_{k}$ (this
is again the standard diffusive timescale), and
the expected propagation of chaos limit is solution to the nonlinear
stochastic differential equation:
%
%
\begin{eqnarray}
\label{eq:edsnonlin_MALA} &&\mathrm{d}X_t= \sqrt{w(t,\ell)} \,
\mathrm{d}B_t - w(t,\ell) \tfrac{1}{2} V'(X_t)
\,\mathrm{d}t,
\nonumber
\\[-8pt]
\\[-8pt]
&&\quad\mbox{where }w(t,\ell)=\ell^2 \bigl( \mathrm{e}^{({\ell^4}/{8}) \E ( (
(V')^2V''+V^{(4)}-2V^{(3)}V'-(V'')^2  )(X_t) )}
\wedge1 \bigr).
\nonumber
\end{eqnarray}
This equation is obtained by a deterministic (and nonlinear in the
sense of McKean) change of time applied to the standard overdamped
Langevin stochastic differential equation with reversible density
$\mathrm{e}^{-V}$.
Under appropriate assumptions on the potential $V$, we believe that a
rigorous proof of this result could be done using similar
techniques as for the RWM algorithm in \cite{JLM1}.

\subsubsection{Relation to previous results in the literature}

These results are related to previous ones in the literature. First,
in the Gaussian case $V(x)=\frac{x^2}{2} + \frac{1} 2 \ln(2 \uppi)$, one
obtains from \eqref{eq:edsnonlin_MALA} that $(\E(X_t^2))_{t \ge0}$
solves the ordinary differential equation
\[
\frac{\mathrm{d}}{\mathrm{d}t} \E\bigl(X_t^2\bigr)=
\ell^2 \bigl(\mathrm{e}^{({\ell^4}/{8})(\E(X_t^2)-1)}\wedge 1 \bigr) \bigl(1-\E
\bigl(X_t^2\bigr)\bigr).
\]
%
We recover here a result from \cite{christensen-roberts-rosenthal-05}, Theorem~2, where it is shown
that the process
$
 (\frac{1}{n} \sum_{i=1}^n(X^{i,n}_{\lfloor
\sqrt{n}t\rfloor})^2 )_{t\geq0}$,
 in the limit $n \to
\infty$ satisfies this ordinary differential equation.

Second, in the stationary case, namely when
$(X^{1,n}_0,\ldots,X^{n,n}_0)$ are distributed according to the target
density $p$ defined by \eqref{eq:pi}, the equalities
\begin{eqnarray*}
\int_{\R} V^{(3)}V'(x)
\mathrm{e}^{-V(x)}\,\mathrm{d}x&=&\int_\R
V^{(4)}(x)\mathrm{e}^{-V(x)}\,\mathrm{d}x,
\\
\int_{\R} V''
\bigl(V'\bigr)^2(x)\mathrm{e}^{-V(x)}\,
\mathrm{d}x&=&\int_{\R
}\bigl[V^{(3)}V'+
\bigl(V''\bigr)^2\bigr](x)
\mathrm{e}^{-V(x)}\,\mathrm{d}x
\end{eqnarray*}
imply that $\langle\nu^n_k,  (
(V')^2V''+V^{(4)}-2V^{(3)}V'-(V'')^2  ) \rangle=0$ and this
changes the scaling of the limiting acceptance rate
in \eqref{eq:accept_MALA}. In \cite{roberts-rosental-97}, it is shown
that in this
case, the correct scaling\vspace*{-1pt} is $\sigma_n^2=\frac{\ell^2}{n^{1/3}}$ and
then $(X^{1,n}_{\lfloor n^{1/3}t\rfloor})_{t\geq0}$ converges in
distribution to the solution $(X_t)_{t\geq0}$ of the stochastic
differential equation
%
%
\begin{eqnarray}
\label{eq:edsnonlin_MALA_eq} && \mathrm{d} X_t=\sqrt{z(\ell)} \,
\mathrm{d}B_t - z(\ell) \frac
{1}{2} V'(X_t)
\,\mathrm{d}t,
\nonumber
\\[-8pt]
\\[-8pt]
&&\quad\mbox{where } z(\ell)= 2 \ell^2 \Phi \biggl(-
\frac{\ell
^3\sqrt {\langle
m,5(V^{(3)})^2-3(V'')^3\rangle/3}}{8} \biggr),
\nonumber
\end{eqnarray}
where $\mathrm{d}m=\mathrm{e}^{-V(x)} \,\mathrm{d}x$.

\subsubsection{Practical counterparts}\label{sec:MALA_practice}

The practical counterparts of the convergence results discussed above
are the following. We can actually distinguish between three regimes:
\begin{itemize}[$\bullet$]
\item[$\bullet$] On time intervals such that $\E ( (
(V')^2V''+V^{(4)}-2V^{(3)}V'-(V'')^2  )(X^{1,n}_k) )<0$, then\vspace*{-2pt}
the correct scaling to obtain a diffusive limit is
$\sigma_n^2=\frac{\ell^2}{n^{1/2}}$ and
there exists an optimal value of $\ell$ to speed up the time scale of the
dynamics of $X_t$, by maximizing $w(t,\ell)$ (see Equation~\eqref
{eq:edsnonlin_MALA}).
\item[$\bullet$] On time intervals such that $\E ( (
(V')^2V''+V^{(4)}-2V^{(3)}V'-(V'')^2  )(X^{1,n}_k) ) = 0$,
then\vspace*{-2pt} the correct scaling to obtain a diffusive limit is
$\sigma_n^2=\frac{\ell^2}{n^{1/3}}$, and again, there exists on optimal
value of $\ell$ to speed up the convergence to equilibrium, by maximizing
$z(\ell)$ (see
Equation~\eqref{eq:edsnonlin_MALA_eq} and \cite{roberts-rosental-97}).
\item[$\bullet$] On time intervals such that $\E ( (
(V')^2V''+V^{(4)}-2V^{(3)}V'-(V'')^2
)(X^{1,n}_k) ) > 0$, with\vspace*{-1pt} the scaling
$\sigma_n^2=\frac{\ell^2}{n^{1/2}}$, we observe that $w(t,\ell
)=\ell^2$ in \eqref{eq:edsnonlin_MALA} so that
one should take $\ell$ as\vspace*{-1pt} large as possible. This is an indication of the
fact that the correct scaling for $\sigma_n^2$ in this case should be
such that $\sigma_n^2 \gg\frac{\ell^2}{n^{1/2}}$. Indeed, in the
Gaussian case, Proposition~\ref{propmalagausfav} below shows that one
should take $\sigma_n$ going to $0$ as slowly as possible.
\end{itemize}

In conclusion, in the MALA case (and contrary to the RWM case), the
correct scaling as a function of the dimension is not the same at equilibrium
and in the transient phase. Moreover, in the transient phase, the
scaling depends on the sign of
\[
\E \bigl( \bigl( \bigl(V'\bigr)^2V''+V^{(4)}-2V^{(3)}V'-
\bigl(V''\bigr)^2 \bigr)
\bigl(X^{1,n}_k\bigr) \bigr).
\]
It seems
thus difficult to draw any general simple recommendation for
practitioners from this analysis. It is likely that the assumption
that the target probability is the product of $n$ one-dimensional laws
is too restrictive to understand correctly the scaling $n \to\infty$
in this case.

\subsection{Rigorous results in the Gaussian case and when $\E((X^{1,n}_k)^2)>1$}\label{sec:MALA_Gauss}

In this section, we consider the case of a Gaussian target, namely
%
%
\begin{equation}
\label{eq:gauss_target} V(x)=\frac{x^2}{2} + \frac{1} 2 \ln(2 \uppi).
\end{equation}
We thus have
\[
\bigl(V'\bigr)^2V''+V^{(4)}-2V^{(3)}V'-
\bigl(V''\bigr)^2 = x^2 - 1.
\]
The aim of this section is to study in details the situation when
\[
\E \bigl( \bigl( \bigl(V'\bigr)^2V''+V^{(4)}-2V^{(3)}V'-
\bigl(V''\bigr)^2 \bigr)
\bigl(X^{1,n}_k\bigr) \bigr)>0\quad\mbox{namely}\quad\E
\bigl(\bigl(X^{1,n}_k\bigr)^2\bigr) > 1.
\]

%
\begin{proposition}\label{propvarconst}
Let us consider $(X_k^{i,n})$ solution to \eqref{eq:MALA} for the
Gaussian target \eqref{eq:gauss_target}, with a
variance independent of $n$:
\[
\sigma_n=\ell\in(0,2).
\]
Let $m$ be a probability measure on $\R$ such that $\langle
m,x^2\rangle>\frac{1}{1-\ell^2/4}$. We\vspace*{-1pt} endow the space $\R^{\mathbb
N}$ with the product topology.
If the initial random variables $(X^{1,n}_0,\ldots,X^{n,n}_0)$ are
exchangeable and $m$-chaotic, then the processes
$(X^{1,n},\ldots,X^{n,n})$ are $P$-chaotic where $P$ denotes the law
of the Markov chain
%
%
\begin{equation}
\label{eq:Yk} Y_{k+1}= \biggl(1-\frac{\ell^2}{2}
\biggr)Y_k+\ell G_{k+1}
\end{equation}
with the sequence $(G_k)$ i.i.d. according to the normal law and
independent from the initial position $Y_0$ distributed according to $m$.
\end{proposition}

A simple case for which the assumption on the initial condition is
satisfied is i.i.d. initial conditions
$(X^{i,n}_0)_{i}$ with law $m$.

Notice\vspace*{-2pt} that $Y_k$ converges in law to ${\mathcal
N} (0,\frac{1}{1-\ell^2/4} )$ as $k\to+\infty$. The asymptotic
distribution converges to the target density when $\ell\to0$. Of
course, for fixed $n$ and $i\in\{1,\ldots,n\}$, $X^{i,n}_k$ converges
in law to ${\mathcal
N} (0,1 )$ as $k\to+\infty$. So the limits $k\to\infty$
and $n\to\infty$ do not commute, meaning that, for large $n$, the
rate of convergence in distribution of $(X^{i,n}_k)_{k\geq1}$ to
${\mathcal
N} (0,1 )$ should deteriorate.

\begin{pf*}{Proof of Proposition~\ref{propvarconst}}
Let $(Y^{1,n},\ldots,Y^{n,n})$ with $Y^{i,n}_0=X^{i,n}_0$ and
$Y^{i,n}_{k+1}= (1-\frac{\ell^2}{2} )Y^{i,n}_k+\ell G^i_{k+1}$
denote the processes obtained when all moves are accepted in the MALA
algorithm \eqref{eq:MALA}. The proof is divided into two steps. We are
first going to
prove that the processes $(Y^{1,n},\ldots,Y^{n,n})$ are $P$-chaotic
(this would be trivial if the initial conditions $(X^{i,n}_0)_{i}$ were
supposed to be i.i.d.). Then, setting
\begin{eqnarray*}
&&\tilde{\mathcal A}^n_{k+1}= \Biggl\{U_{k+1}\leq
\exp\Biggl\{\sum_{i=1}^n \biggl(V
\bigl(Y^{i,n}_k\bigr)-V\bigl(Y^{i,n}_{k+1}
\bigr)
\\
&&\hphantom{\tilde{\mathcal A}^n_{k+1}= \Biggl\{U_{k+1}\leq
\exp\Biggl\{\sum_{i=1}^n \biggl(}{}+\frac{1}{2} \biggl[\bigl(G^i_{k+1}
\bigr)^2-\biggl(G^i_{k+1}-\frac{\ell
}{2}
\bigl(V'\bigl(Y^{i,n}_k\bigr)+V'
\bigl(Y^{i,n}_{k+1}\bigr)\bigr)\biggr)^2 \biggr]\biggr)
\Biggr\} \Biggr\},
\end{eqnarray*}
we will check that $\forall K\in{\mathbb
N}^*, \lim_{n\to\infty}\P (\bigcap_{k=1}^K \tilde{\mathcal
A}^n_{k} )=1$. Since, on the event $\bigcap_{k=1}^K
\tilde{\mathcal A}^n_{k}$, $(X^{1,n}_k,\ldots,\allowbreak  X^{n,n}_k)_{0\leq k\leq
K}=(Y^{1,n}_k,\ldots,Y^{n,n}_k)_{0\leq k\leq K}$ one obtains the
$P$-chaoticity of the processes $(X^{1,n},\ldots,\allowbreak  X^{n,n})$ by
combining the two steps.

For the first step, notice that for fixed $j,K\in{\mathbb N}^*$, the
law of $((Y^{1,n}_k,\ldots,Y^{j,n}_k))_{0\leq k\leq K}$ is
\[
m_j^n\bigl(\mathrm{d}y_0^1,
\ldots,\mathrm{d}y_0^j\bigr)\prod
_{k=0}^{K-1} \bigl(Q\bigl(y^1_k,
\mathrm{d}y^1_{k+1}\bigr)\times\cdots\times Q
\bigl(y^j_k,\mathrm{d}y^j_{k+1}
\bigr) \bigr),
\]
where
$Q(y,\mathrm{d}y')=\frac{1}{\ell\sqrt{2\uppi}}\mathrm{e}^{-{
(y'-y(1-\ell^2/2)
)^2}/({2\ell^2})}\,\mathrm{d}y'$
and the law $m^n_j$ of\vspace*{-1pt} $(X^{1,n}_0,\ldots,X^{j,n}_0)$ converges weakly
to $m^{\otimes j}$ as $n\to\infty$ (since
the initial conditions $(X^{1,n}_0,\ldots,X^{n,n}_0)$ are
$m$-chaotic). Since $y\mapsto Q(y,\mathrm{d}y')\in{\mathcal P}(\R)$
is weakly
continuous, this law converges weakly to $\prod_{i=1}^j
(m(\mathrm{d}y^i_0)\prod_{k=0}^{K-1}Q(y^i_k,\mathrm
{d}y^i_{k+1}) )$ which is the
$j$-fold product of the image of $P$ by the canonical restriction to
the $K+1$ first coordinates. Hence, the processes $(Y^{1,n},\ldots
,Y^{n,n})$ are $P$-chaotic.

For the second step, let us introduce
\[
S^n_k=\frac{8}{n\ell^4}\sum
_{i=1}^n \biggl\{ V\bigl(Y^{i,n}_k
\bigr)-V\bigl(Y^{i,n}_{k+1}\bigr)+\frac{1}{2} \biggl[
\bigl(G^i_{k+1}\bigr)^2-\biggl(G^i_{k+1}-
\frac{\ell
}{2}\bigl(V'\bigl(Y^{i,n}_k
\bigr)+V'\bigl(Y^{i,n}_{k+1}\bigr)\bigr)
\biggr)^2 \biggr] \biggr\}.
\]
One has
%
%
\begin{equation}
\label{contprobrej} \P\bigl(\bigl\{\tilde{\mathcal A}^{n}_{k+1}
\bigr\}^{c}\bigr)\leq\P\bigl(S^n_k<0\bigr).
\end{equation}
Some tedious but simple computations yields (using $V(x)=\frac{x^2}{2}
+ \frac{1} 2 \ln(2 \uppi)$)
%
%
\begin{eqnarray}
\label{eq:accept_gauss} &&V\bigl(Y^{i,n}_k\bigr)-V
\bigl(Y^{i,n}_{k+1}\bigr)+\frac{1}{2} \biggl[
\bigl(G^i_{k+1}\bigr)^2-\biggl(G^i_{k+1}-
\frac{\ell
}{2}\bigl(V'\bigl(Y^{i,n}_k
\bigr)+V'\bigl(Y^{i,n}_{k+1}\bigr)\bigr)
\biggr)^2 \biggr]
\nonumber
\\[-8pt]
\\[-8pt]
&&\quad=\frac{\ell^4}{8}\bigl(\bigl(Y^{i,n}_k
\bigr)^2-\bigl(G^i_{k+1}\bigr)^2
\bigr)+ \biggl(\frac
{\ell
^5}{8}-\frac{\ell^3}{4} \biggr)Y^{i,n}_kG^i_{k+1}-
\frac{\ell
^6}{32}\bigl(Y^{i,n}_k\bigr)^2
\nonumber
\end{eqnarray}
so that (in law)
\[
S_k^n= \biggl(1-\frac{\ell^2}{4} \biggr)\bigl\langle
\mu^n_k,y^2\bigr\rangle + \biggl(\ell-
\frac{2}{\ell} \biggr)\sqrt{\bigl\langle\mu ^n_k,y^2
\bigr\rangle}\frac{\tilde{G}^n_{k+1}}{\sqrt{n}}-\frac{1}{n}\sum
_{i=1}^n\bigl(G^i_{k+1}
\bigr)^2
\]
with $\tilde{G}^n_{k+1}=1_{\{\langle\mu^n_k,y^2\rangle>0\}}\frac
{\sum_{i=1}^nY^i_kG^i_{k+1}}{\sqrt{n\langle\mu^n_k,y^2\rangle
}}+1_{\{\langle\mu^n_k,y^2\rangle=0\}}G^1_{k+1}$ a\vspace*{-3pt} normal random
variable independent from
$\mu^n_k=\frac{1}{n}\sum_{i=1}^n\delta_{Y^{i,n}_k}$. As the
exchangeability of the initial condition
$(Y^{1,n}_0,\ldots,Y^{n,n}_0)$ is preserved by the evolution, the
propagation of chaos result obtained in the first step implies (and is
actually equivalent to) the convergence in probability of
the empirical measures
$\mu^n=\frac{1}{n}\sum_{i=1}^n\delta_{Y^{i,n}}\in{\mathcal P}(\R
^{\mathbb
N})$ to $P$ (see \cite{sznitman-91}, Proposition~2.2). In
particular, $\mu^n_k$ converges in probability to the law $P_k$ of
$Y_k$, solution to \eqref{eq:Yk}.

With\vspace*{-1pt} this law of large numbers, we see that in order to estimate $\P
(S^n_k<0)$ we need to understand the evolution of $\langle
P_k,y^2\rangle=\E((Y_k)^2)$ with $k$. One has
$\langle P_{k+1},y^2\rangle=(1-\frac{\ell^2}{2})^2\langle
P_k,y^2\rangle+\ell^2$, and since $\langle P_{0},y^2\rangle=\langle
m,y^2\rangle>\frac{1}{1-\ell^2/4}$, one easily checks by induction\vspace*{-1pt} that
for all $k\in{\mathbb N}$, $\langle
P_k,y^2\rangle>\frac{1}{1-\ell^2/4}$. Hence for fixed $k\in{\mathbb N}$,
there exists $M<+\infty$ and $\varepsilon>0$ such that $\langle
P_k,y^2\wedge M\rangle\geq\frac{1+3\varepsilon}{1-\ell^2/4}$.
%
%
One has
\begin{eqnarray*}
\P\bigl(S^n_k<0\bigr)& \leq&\P \Biggl(
\frac{1}{n}\sum_{i=1}^n
\bigl(G^i_{k+1}\bigr)^2>1+\varepsilon \Biggr)
\\
&&{}+\P
\biggl(\bigl\langle\mu ^n_k,y^2\bigr\rangle
\biggl(1+\frac{\ell-2/\ell}{1-\ell^2/4}\frac
{\tilde{G}^n_{k+1}}{\sqrt{n\langle\mu^n_k,y^2\rangle}} \biggr)<\frac{1+\varepsilon}{1-\ell^2/4}
\biggr)
\\
&\leq& \P \Biggl(\frac{1}{n}\sum_{i=1}^n
\bigl(G^i_{k+1}\bigr)^2>1+\varepsilon \Biggr)+\P
\biggl(\bigl\langle \mu^n_k,y^2\wedge M\bigr
\rangle<\frac{1+2\varepsilon}{1-\ell^2/4} \biggr)
\\
&&{} +\P \biggl(\bigl\langle\mu^n_k,y^2
\bigr\rangle\geq\frac{1+2\varepsilon
}{1-\ell^2/4}, \frac{\ell-2/\ell}{1-\ell^2/4}\tilde
{G}^n_{k+1}<-\frac{\varepsilon\sqrt{n}}{\sqrt{(1+2\varepsilon
)(1-\ell^2/4)}} \biggr).
\end{eqnarray*}
The first term of the right-hand side converges to $0$ as $n\to
+\infty$, since, by the strong law of large numbers,
$\frac{1}{n}\sum_{i=1}^n(G^i_{k+1})^2$ converges a.s. to $1$. The
second term converges to $0$ since $\langle\mu^n_k,y^2\wedge
M\rangle$ converges in probability to $\langle P_k,y^2\wedge
M\rangle\geq\frac{1+3\varepsilon}{1-\ell^2/4}$. The third term is\vspace*{-3pt}
bounded from above by $\Phi (-\frac{\varepsilon}{|\ell-2/\ell
|}\sqrt{\frac{n(1-\ell^2/4)}{(1+2\varepsilon)}} )$ and also
converges to $0$. Hence, $\P(S^n_k<0)$ tends to $0$ as\vspace*{-2pt} $n\to\infty$
and with \eqref{contprobrej}, one deduces that for fixed $K\in
{\mathbb N}^*$, $\P(\bigcap_{k=1}^K \tilde{\mathcal A}^n_{k})\geq
1-\sum_{k=1}^K\P(\{\tilde{A}^{n}_k\}^{c})$ tends to $1$.
\end{pf*}

As\vspace*{-1pt} is clear from the previous proposition, for a fixed variance
$\sigma_n=\ell$ and if $\E((X^{1,n}_0)^2) > 1$,
then, for sufficiently small $\ell$ (namely $\ell< 2
\sqrt{1-1/\E((X^{1,n}_0)^2)}$) and in the limit\vspace*{-2pt} $n \to\infty$, (i)~the components $(X^{i,n}_k)_{i}$ do
not interact and evolve independently according to the explicit Euler
discretization \eqref{eq:Yk} (with a timestep
$\ell^2$) of the Langevin
dynamics $\mathrm{d}Y_t =\mathrm{d}B_t - Y_t/2 \,\mathrm{d}t $ and
(ii)~the system remains in the
region $\E(X_k^2) > 1$ for
all $k \ge0$.

Based on the previous result, it is natural to look for a diffusive
limit for a $\sigma_n$ which goes to zero at an
arbitrary rate with respect to $n$.
%
%
\begin{proposition}\label{propmalagausfav}
Let us consider $(X_k^{i,n})$ solution to \eqref{eq:MALA} for the
Gaussian target \eqref{eq:gauss_target}, with a
variance $\sigma_n$ satisfying:
\[
\lim_{n\to\infty} \sigma_n=0 \quad\mbox{and}\quad\lim
_{n\to
\infty}n \sigma_n^2=+\infty.
\]
Let $m$ be a probability measure on $\R$ such that $\langle
m,x^2\rangle>1$ and $\langle m,x^8\rangle<+\infty$. If the initial
random variables $(X^{1,n}_0,\ldots,X^{n,n}_0)$ are i.i.d. according
to $m$, then the processes\vspace*{-2pt} $((X^{1,n}_{\lfloor
t/\sigma_n^2\rfloor})_{t\geq0},\ldots,\allowbreak  (X^{n,n}_{\lfloor
t/\sigma_n^2\rfloor})_{t\geq0})$ are $Q$-chaotic where $Q$ denotes
the law of the Ornstein--Uhlenbeck process
%
%
\begin{equation}
\label{eq:OU} \mathrm{d}Y_t=\mathrm{d}B_t-
\frac{Y_t}{2} \,\mathrm{d}t
\end{equation}
with the initial position $Y_0$ distributed according to $m$ and
independent from the Brownian motion $(B_t)_{t\geq0}$. Moreover, the
limiting mean acceptance rate is $1$.
\end{proposition}
%
%
\begin{remark}
For a more general potential $V$, if the initial random variables
$(X^{1,n}_0,\ldots,X^{n,n}_0)$ are exchangeable and $m$-chaotic with
$\langle m,(V')^2V''+V^{(4)}-2V^{(3)}V'-(V'')^2\rangle>0$, one expects
the limit in law to be the one of the solution of
$Y_t=Y_0+B_t-\int_0^t\frac{V'(Y_s)}{2}\,\mathrm{d}s$. But, unlike in
the Gaussian
case, it is not clear that
$\E[\{(V')^2V''+V^{(4)}-2V^{(3)}V'-(V'')^2\}(Y_t)]>0$ for all $t\geq
0$. Therefore, setting
$T=\inf\{t>0\dvtx\E[\{(V')^2V''+V^{(4)}-2V^{(3)}V'-(V'')^2\}
(Y_t)]=0\}$
with the convention\vspace*{-1pt} $\inf\emptyset=+\infty$ and denoting by $Q^T$ the
law of $(Y_t)_{t\in[0,T)}$, one actually expects the processes
$((X^{1,n}_{\lfloor t/\sigma_n^2\rfloor})_{t\in[0,T)},\ldots
,(X^{n,n}_{\lfloor t/\sigma_n^2\rfloor})_{t\in[0,T)})$ to be $Q^T$-chaotic.
\end{remark}
\begin{pf*}{Proof of Proposition~\ref{propmalagausfav}}
As\vspace*{-2pt} in the proof of Proposition~\ref{propvarconst},
let $(Y^{1,n},\ldots,Y^{n,n})$ with $Y^{i,n}_0=X^{i,n}_0$ and
$Y^{i,n}_{k+1}= (1-\frac{\sigma_n^2}{2} )Y^{i,n}_k+\sigma
_n G^i_{k+1}$
denote the processes obtained when all moves are accepted in the MALA
algorithm \eqref{eq:MALA}. The processes $ (Y^{1,n}_{\lfloor
t/\sigma_n^2\rfloor},\ldots,Y^{n,n}_{\lfloor
t/\sigma_n^2\rfloor} )$ are
independent and identically distributed and their common distribution
converges weakly to $Q$ by the strong convergence analysis of the
Euler scheme applied to \eqref{eq:OU}. Hence, to conclude the proof,
it is enough to check that for fixed $T>0$,
$\lim_{n\to\infty}\P (\bigcap_{k=1}^{\lfloor
T/\sigma_n^2\rfloor}\tilde{\mathcal A}^n_k )=1$, where, as in
the proof of Proposition~\ref{propvarconst},
\begin{eqnarray*}
&&\tilde{\mathcal A}^n_{k+1}= \Biggl\{U_{k+1}\leq
\exp\Biggl\{\sum_{i=1}^n \biggl(V
\bigl(Y^{i,n}_k\bigr)-V\bigl(Y^{i,n}_{k+1}
\bigr)
\\
&&\hphantom{\tilde{\mathcal A}^n_{k+1}= \Biggl\{U_{k+1}\leq
\exp\Biggl\{\sum_{i=1}^n \biggl(}{}+\frac{1}{2} \biggl[\bigl(G^i_{k+1}
\bigr)^2-\biggl(G^i_{k+1}-\frac{\sigma
_n}{2}
\bigl(V'\bigl(Y^{i,n}_k\bigr)+V'
\bigl(Y^{i,n}_{k+1}\bigr)\bigr)\biggr)^2 \biggr]\biggr)
\Biggr\} \Biggr\}.
\end{eqnarray*}
To do so, we use an upper-bound sharper than \eqref{contprobrej}. Let
us introduce (using \eqref{eq:accept_gauss}):
\begin{eqnarray*}
S^n_k&=&\frac{8}{n
\sigma_n^4}\sum
_{i=1}^n \biggl\{V\bigl(Y^{i,n}_k
\bigr)-V\bigl(Y^{i,n}_{k+1}\bigr)+\frac
{1}{2} \biggl[
\bigl(G^i_{k+1}\bigr)^2-\biggl(G^i_{k+1}-
\frac{\sigma
_n}{2}\bigl(V'\bigl(Y^{i,n}_k
\bigr)+V'\bigl(Y^{i,n}_{k+1}\bigr)\bigr)
\biggr)^2 \biggr] \biggr\}
\\
&=&\frac{2}{n\sigma_n} \sum_{i=1}^n
R^{i,n}_k,
\end{eqnarray*}
where the random variables
\[
\biggl(R^{i,n}_k= \biggl(\frac{\sigma_n}{2}-
\frac{\sigma
_n^3}{8} \biggr) \bigl(Y^{i,n}_k
\bigr)^2+ \biggl(\frac{\sigma_n^2}{2}-1 \biggr)Y^{i,n}_kG^i_{k+1}-
\frac{\sigma_n}{2}\bigl(G^i_{k+1}\bigr)^2
\biggr)_{1\leq
i\leq n}
\]
are independent and identically distributed.
Then, we have
%
%
\begin{eqnarray}
\label{precontprobrej} \P\bigl(\bigl\{\tilde{\mathcal A}^n_{k+1}
\bigr\}^c\bigr)&=&\E \bigl( \bigl(1-\mathrm{e}^{
({n\sigma_n^4}/{8})S^n_k}
\bigr)^+ \bigr)\leq-\E \biggl(\frac
{n\sigma
_n^4}{8}S^n_k1_{\{S^n_k\leq0\}}
\biggr)
\nonumber
\\[-8pt]
\\[-8pt]
&=&-\frac{\sigma_n^3}{4}\E \Biggl(\sum_{i=1}^nR^{i,n}_k1_{\{\sum
_{i=1}^nR^{i,n}_k\leq0\}}
\Biggr).
\nonumber
\end{eqnarray}
We\vspace*{-1pt} need to estimate the moments of the random variables $R^{i,n}_k$. To
do so, we assume from now on that $n$ is large enough so that $\sigma
_n<\sqrt{2}$ and we first estimate the moments of
\[
Y^{i,n}_k= \biggl(1-\frac{\sigma_n^2}{2}
\biggr)^kX^{i,n}_0+\sigma _n\sum
_{j=1}^k \biggl(1-\frac{\sigma_n^2}{2}
\biggr)^{k-j}G^{i}_j.
\]
One has, using the fact that
$\sigma_n\sum_{j=1}^k (1-\frac{\sigma_n^2}{2} )^{k-j}G^{i}_j
\sim{\mathcal N}_1 (0,\frac{1-(1-\sigma_n^2/2)^{2k}}{1-\sigma
_n^2/4} )$,
\begin{eqnarray*}
\E\bigl(
\bigl(Y^{i,n}_k\bigr)^8\bigr)&=& \biggl(1-
\frac{\sigma_n^2}{2} \biggr)^{8k}\bigl\langle m, x^8\bigr
\rangle+28 \biggl(1-\frac{\sigma_n^2}{2} \biggr)^{6k}\frac
{1-(1-\sigma_n^2/2)^{2k}}{1-\sigma_n^2/4}
\bigl\langle m, x^6\bigr\rangle
\\
&&{}+210 \biggl(1-\frac{\sigma_n^2}{2} \biggr)^{4k} \biggl(
\frac
{1-(1-\sigma_n^2/2)^{2k}}{1-\sigma_n^2/4} \biggr)^2\bigl\langle m, x^4\bigr
\rangle
\\
&&{} +420 \biggl(1-\frac{\sigma_n^2}{2} \biggr)^{2k} \biggl(
\frac
{1-(1-\sigma_n^2/2)^{2k}}{1-\sigma_n^2/4} \biggr)^3\bigl\langle m, x^2\bigr
\rangle
\\
&&{}+105 \biggl(\frac{1-(1-\sigma_n^2/2)^{2k}}{1-\sigma
_n^2/4} \biggr)^4
\\
&\leq&\bigl\langle m, x^8\bigr\rangle+56\bigl\langle m,
x^6\bigr\rangle+840\bigl\langle m, x^4\bigr\rangle+3360
\bigl\langle m, x^2\bigr\rangle+1680.
\end{eqnarray*}
Therefore, $\sup_{n:\sigma_n<\sqrt{2}}\sup_{k\geq
0}\E((R^{i,n}_k)^4)<+\infty$. Moreover, for $n$ large enough so that
$\sigma_n^2\leq\frac{2(\langle m, x^2\rangle-1)}{\langle m,
x^2\rangle}$ (so that we also have $\sigma_n^2 < 2$),
\begin{eqnarray*}
\biggl(1-\frac{\sigma_n^2}{4} \biggr)\E\bigl(\bigl(Y^{i,n}_k
\bigr)^2\bigr)&=&1+ \biggl(1-\frac{\sigma_n^2}{2} \biggr)^{2k}
\biggl( \biggl(1-\frac{\sigma
_n^2}{4} \biggr)\bigl\langle m, x^2\bigr
\rangle-1 \biggr)
\\
&\geq&1+\mathrm{e}^{-
({2k\sigma_n^2}/({2-\sigma_n^2}))}\frac{\langle m, x^2\rangle-1}{2}
 \geq 1+\mathrm{e}^{-\langle m,x^2\rangle T}\frac{\langle m,
x^2\rangle-1}{2},
\end{eqnarray*}
where the latter inequality holds for $k\leq\lfloor T/\sigma
_n^2\rfloor$.
From now on, we suppose that $n$ is large enough so that $\sigma_n^2
\le\frac{2(\langle m, x^2\rangle-1)}{\langle m, x^2\rangle}$ and we
fix $k\leq\lfloor T/\sigma_n^2\rfloor$. Setting $c_T=(\mathrm
{e}^{-\langle
m,x^2\rangle T})\frac{\langle m, x^2\rangle-1}{4}$, one has
\[
\E\bigl(R^{i,n}_k\bigr)=\frac{\sigma_n}{2} \biggl(
\biggl(1-\frac{\sigma
_n^2}{4} \biggr)\E\bigl(\bigl(Y^{i,n}_k
\bigr)^2\bigr)-1 \biggr)\geq c_T\sigma_n.
\]
Therefore (using in particular the fact that $\E(R^{i,n}_k) \ge0$),
%
%
\begin{eqnarray}
\label{espsum} -\E \Biggl(\sum_{i=1}^nR^{i,n}_k1_{\{\sum_{i=1}^nR^{i,n}_k\leq
0\}}
\Biggr)&=&-n\E\bigl(R^{1,n}_k\bigr)\P \Biggl(\sum
_{i=1}^nR^{i,n}_k\leq 0
\Biggr)
\nonumber
\\
&&{} -\E \Biggl(\sum_{i=1}^n
\bigl(R^{i,n}_k-\E\bigl(R^{i,n}_k
\bigr)\bigr)1_{\{\sum
_{i=1}^n(R^{i,n}_k-\E(R^{i,n}_k))\leq
-n\E(R^{1,n}_k)\}} \Biggr)
\nonumber
\\
&\leq& 0+\frac{\E ( (\sum_{i=1}^n(R^{i,n}_k-\E(R^{i,n}_k))
)^4 )}{(n\E(R^{1,n}_k))^3}
\\
&=&\frac{3n(n-1)\operatorname{Var}^2((R^{1,n}_k))+n\E
((R^{1,n}_k-\E(R^{1,n}_k))^4
)}{(n\E(R^{1,n}_k))^3}
\nonumber
\\
&\leq&\frac{(3n^2+13n)\E((R^{i,n}_k)^4)}{c_T^3n^3\sigma_n^3}\leq \frac{
C_T}{n\sigma_n^3},
\nonumber
\end{eqnarray}
where $C_T$ is some constant not depending on $n$ and $k$. With \eqref
{precontprobrej}, we deduce that
%
%
\begin{equation}
\label{majopun} \P \Biggl(\bigcup_{k=0}^{\lfloor
T/\sigma_n^2\rfloor-1}
\bigl\{\tilde{\mathcal A}^n_{k+1}\bigr\}^c
\Biggr)\leq\sum_{k=0}^{\lfloor
T/\sigma_n^2\rfloor-1}\P \bigl(\bigl\{
\tilde{\mathcal A}^n_{k+1}\bigr\} ^c \bigr)\leq
\frac{TC_T}{4n\sigma_n^2}.
\end{equation}
Since $\lim_{n\to\infty}n\sigma_n^2=+\infty$, we conclude that
$\lim_{n\to\infty}\P (\bigcap_{k=1}^{\lfloor
T/\sigma_n^2\rfloor}\tilde{\mathcal A}^n_k )=1$.
\end{pf*}
%
%
%
\begin{remark}
In the case when $\lim_{n\to\infty} n\sigma_n^2=0$ and the initial
conditions $(X^{1,n}_0,\ldots,X^{n,n}_0)$ are i.i.d. according to $m$
such that $\langle m, x^4\rangle<+\infty$, then, whatever the sign of
$\langle m,x^2\rangle-1$, the processes $((X^{1,n}_{\lfloor t/\sigma
_n^2\rfloor})_{t\geq0},\ldots,(X^{n,n}_{\lfloor t/\sigma_n^2\rfloor
})_{t\geq0})$ are $Q$-chaotic where $Q$ denotes the law\vspace*{-2pt} of the
Ornstein--Uhlenbeck process $Y_t=Y_0+B_t-\int_0^t\frac{Y_s}{2}\,
\mathrm{d}s$ with
the initial position $Y_0$ distributed according to $m$ and independent
from the Brownian motion $(B_t)_{t\geq0}$.

Indeed, for $n$ large enough so that $\sigma_n<\sqrt{2}$, one may
check that $\sup_{k\geq0}\E((Y^{i,n}_k)^4)<C$ and replace \eqref
{espsum} by the estimation
\[
-\E \Biggl(\sum_{i=1}^nR^{i,n}_k1_{\{\sum_{i=1}^nR^{i,n}_k\leq0\}
}
\Biggr)\leq n\bigl |\E\bigl(R^{1,n}_k\bigr)\bigr |+\E^{1/2}
\Biggl( \Biggl(\sum_{i=1}^n
\bigl(R^{i,n}_k-\E\bigl(R^{i,n}_k
\bigr)\bigr) \Biggr)^2 \Biggr)\leq Cn\sigma _n+C\sqrt{n},
\]
so that
\[
\P \Biggl(\bigcup_{k=0}^{\lfloor
T/\sigma_n^2\rfloor-1}\bigl\{
\tilde{A}^n_{k+1}\bigr\}^c \Biggr)\leq
C_T \bigl( n\sigma_n^2 + \sqrt{n}
\sigma_n \bigr)
\]
which converges to zero when $n$ goes to infinity.
\end{remark}

\begin{appendix}
\section{Proof of Lemma \texorpdfstring{\protect\ref{lem:fF}}{1}}\label{sec:fF}

Let us define for $x\in\R$,
%
%
\begin{eqnarray}
\label{eq:def_f}f(x)&=&\exp\bigl(x^2/2\bigr)\Phi(x),
\\
\label{eq:def_h}h(x)&=&x f(x)=x\exp{ \biggl(\frac{x^2}{2} \biggr)}\Phi(x).
\end{eqnarray}
The derivative of $f$ is
\[
f'(x)=\frac{1}{\sqrt{2\uppi}}+x\exp{ \biggl(\frac{x^2}{2} \biggr)}
\Phi(x).
\]
For $x \ge0$, $f'(x)>0$. For $x<0$, using the upper-bound in \eqref
{contPhi}, we also obtain
$f'(x)> 0$. Therefore, the function $f$ is increasing.

Since $h'(x)={ (1+x^2 )}\exp{ (\frac{x^2}{2}
)}\Phi(x)+\frac
{x}{\sqrt{2\uppi}}$,
it is obvious that $h'(x)>0$ for $x\geq0$. For $x<0$ this comes for
the lower-bound in \eqref{contPhi}.

By definitions of $\Gamma$ and $\cG$, we get
%
%
\begin{equation}
\label{eq:lemf} \frac{\Gamma(a,b,\ell)-2\cG(a,b,\ell)}{\ell^2}=\Phi{ \biggl(\frac {-\ell b}{2\sqrt{a}} \biggr)}-
\exp{ \biggl(\frac{\ell
^2}{2}(a-b) \biggr)}\Phi{ \biggl(\frac{\ell b}{2\sqrt{a}}-\ell
\sqrt {a} \biggr)}. 
\end{equation}
Using the identity
\[
\exp{ \biggl(\frac{\ell^2}{2}(a-b) \biggr)}=\exp{ \biggl[\frac{\ell
^2}{2}{
\biggl(\frac{b}{2\sqrt{a}}-\sqrt{a} \biggr)}^2 \biggr]}\exp { \biggl(
\frac{-\ell ^2b^2}{8a} \biggr)}\eqsp,
\]
the right-hand side of \eqref{eq:lemf} can be rewritten in terms of
$f$ (defined by \eqref{eq:def_f})
\[
\frac{\Gamma(a,b,\ell)-2\cG(a,b,\ell)}{\ell^2}=\exp{ \biggl(\frac {-\ell^2b^2}{8a} \biggr)} { \biggl[f{
\biggl(\frac{-\ell
b}{2\sqrt{a}} \biggr)}-f{ \biggl(\frac{\ell(b-2a)}{2\sqrt{a}} \biggr)}
\biggr]}\eqsp.
\]
Now it is clear that
\[
\sign\bigl(\Gamma(a,b,\ell)-2\cG(a,b,\ell)\bigr)=\sign{ \biggl[f{ \biggl(
\frac {-\ell b}{2\sqrt{a}} \biggr)}-f{ \biggl(\frac{\ell
(b-2a)}{2\sqrt{a}} \biggr)} \biggr]}\eqsp.
\]
Recall that the function $f$ is increasing and thus
$\sign(\Gamma(a,b,\ell)-2\cG(a,b,\ell))=\sign(a-b)$.

Similarly,
\[
F(a,b,\ell) = %
\cases{ 2\ell\sqrt{a}\exp{ \biggl(\displaystyle
\frac{-\ell^2b^2}{8a} \biggr)}\displaystyle\frac{h{ ({\ell(b-2a)}/({2\sqrt{a}})
)}-h{ ({-\ell b}/({2\sqrt {a}}) )}}{b-a},&\quad\mbox{if }$a
\neq b$,\vspace*{2pt}
\cr
2\ell^2\exp{ \biggl(\displaystyle\frac{-\ell^2a}{8}
\biggr)}h' \biggl(-\displaystyle\frac{\ell\sqrt {a}}{2} \biggr),&\quad
\mbox{if }$a=b$. } %
\]
This\vspace*{1pt} shows the continuity of $F$, and the positivity of $F$ is a
consequence of the positivity of $h'$.

Setting for $(a,b)\in\R_+^*\times\R$, $\chi(a,b,\ell)=\ell\frac
{b-2a}{2\sqrt{a}}$ and $\zeta(a,b,\ell)=-\frac{\ell b}{2\sqrt{a}}$
and for $(x,y)\in\R^2$,
\[
\psi(x,y)= %
\cases{ \displaystyle\frac{-(x+y)(f(x)-f(y))}{(f(x)+f(y))(x-y)},&\quad\mbox {if
}$y\neq x$,\vspace*{2pt}
\cr
-\displaystyle\frac{xf'(x)}{f(x)},&\quad\mbox{otherwise}, }
\]
one has
\[
F(a,b,\ell)= %
\cases{ \Gamma(0,b,\ell),&\quad\mbox{if }$a=0$,
\cr
\Gamma(a,b,\ell) \bigl(1-\psi\bigl(\chi(a,b,\ell),\zeta(a,b,\ell)\bigr)\bigr),&
\quad \mbox{otherwise}. } %
\]
%
By \cite{JLM1}, Lemma~2, Equation (3.2), the function $(a,b)\mapsto
\Gamma(a,b,\ell)$ is bounded from below by a positive
constant on $[0,+\infty]\times[-M,M]$. To show \eqref{lem:F}, it is then
sufficient to show that
$\sup_{(a,b)\in(0,M]\times\R} \psi(\chi(a,b,\ell),\zeta
(a,b,\ell
))<1$.

When $x>y$, since $h$ is increasing, $yf(y)<xf(x)$ which implies
$-(x+y)(f(x)-f(y))<(f(x)+f(y))(x-y)$ and therefore $\psi(x,y)<1$. This
inequality remains valid for $y>x$ by symmetry of $\psi$ and for $y=x$
since $xf'(x)+f(x)=h'(x)>0$.

For $(x,y)\in\R^2$, with $x>0$ and $-\ell\sqrt{M} \leq x+y\leq0$, (so
that $x-y\ge2x >0$)
one has $0\leq-\frac{x+y}{x-y}\leq\frac{\ell\sqrt{M}}{2x}$ and
$0<\frac{f(x)-f(y)}{f(x)+f(y)}<1$ so that
$\psi(x,y)\leq\frac{\ell\sqrt{M}}{2x}$. With the symmetry of\vspace*{-1pt} $\psi
$, one
deduces that $\sup_{(x,y):-\ell\sqrt{M}\leq x+y\leq0,x\vee y\geq
\ell\sqrt{M}}\psi(x,y)\leq\frac{1}{2}$. Since $f$ is ${\mathcal C}^1$
and positive, one easily checks that $\psi$ is continuous on
$\R^2$. As $\psi<1$ and $\{(x,y)\dvtx-\ell\sqrt{M} \leq x+y\leq0,
x\vee
y\leq\ell\sqrt{M}\}$ is compact, one obtains that
$\sup_{(x,y):-\ell\sqrt{M} \leq x+y\leq0}\psi(x,y)<1$.

As for $(a,b)
\in\R_+^*\times\R$, $\chi(a,b,\ell)+\zeta(a,b,\ell)=-\ell\sqrt{a}$,
one concludes that
\[
\sup_{(a,b)\in(0,M]\times\R} \psi\bigl(\chi
(a,b,\ell),\zeta(a,b,\ell
)\bigr)<1.
\]


\section{Proof of Lemma \texorpdfstring{\protect\ref{lem:lstar}}{2}}\label{seclstar}

Recall first that the function $(s,\ell) \mapsto F_1(s,\ell)$ is
${\mathcal
C}^\infty$ on $\R_+ \times\R_+$. It is easily checked that for any
$s\geq0$, $F_1(s,0)=0$ and $\lim_{\ell\to
\infty}F_1(s,\ell)=0$. With \eqref{lem:F} and the continuity of
$\ell\mapsto F_1(s,\ell)$, one deduces the existence of a point $\ell
^\star(s)>0$ such
that $F_1(s,\ell^\star(s))=\max_{\ell\ge0} F_1(s,\ell)$.

When $s=0$, $F_1(0,\ell)=\ell^2 \exp (- \frac{\ell
^2}{2} )$. This
function admits a unique maximum at point $\ell^\star(0)=\sqrt{2}$. For
further use, we observe that
%
%
\begin{equation}
\label{eq:d2F1_s0} \frac{\partial^2 F_1}{\partial\ell^2}\bigl(0,\ell^\star(0)\bigr) \neq0.
\end{equation}

In the case $s=1$, we compute the derivatives
\begin{eqnarray*}
\frac{\partial F_1}{\partial
\ell}(1,\ell)&=&\bigl(4\ell+2\ell^3\bigr)\Phi \biggl(-
\frac{\ell}{2} \biggr)-\frac{4\ell^2}{\sqrt{2\uppi}}\exp \biggl(-\frac{\ell
^2}{8}
\biggr),
\\
\frac{\partial^2 F_1}{\partial
\ell^2}(1,\ell)&=&\bigl(4+6\ell^2\bigr)\Phi \biggl(-
\frac{\ell}{2} \biggr)-\frac{10\ell}{\sqrt{2\uppi}}\exp \biggl(-\frac{\ell^2}{8}
\biggr).
\end{eqnarray*}
As a consequence, at a critical point of $\ell\mapsto F_1(1,\ell)$,
%
%
\begin{equation}
\label{eq:d2F_s=1} \frac{\partial^2 F_1}{\partial
\ell^2}(1,\ell)=\bigl(\ell^2-6\bigr)\Phi
\biggl(-\frac{\ell}{2} \biggr).
\end{equation}
We deduce that any local maximum belongs to $(0,\sqrt{6}]$ and any local
minimum to $[\sqrt{6},+\infty)$. Since there is a local minimum
(resp. maximum) between two distinct local maxima (resp. minima), we
conclude that $\ell\mapsto F_1(1,\ell)$ admits a unique local maximum which
is also a global maximum and belongs to $(0,\sqrt{6}]$ and no local
minimum on $(0,+\infty)$. For further use, we observe that
$\frac{\partial F_1}{\partial\ell}(1,\sqrt{6}) \neq0$ and thus
(from \eqref{eq:d2F_s=1})
%
%
\begin{equation}
\label{eq:d2F1_s1} \frac{\partial^2 F_1}{\partial\ell^2}\bigl(1,\ell^\star(1)\bigr) \neq0.
\end{equation}

Let us now consider the case $s \in(0,1) \cup(1,\infty)$.
The partial derivative of $F_1$ with respect to $\ell$ is:
%
%
\begin{equation}
\label{dlf1} \frac{\partial F_1}{\partial\ell}(s,\ell)= \biggl(\frac{2}{\ell} -
\ell(1-s) \biggr) F_1(s,\ell) + \ell^2 \biggl( - \sqrt{
\frac
{2s}{\uppi}} \exp \biggl(-\frac{\ell^2}{8s} \biggr) + \ell\Phi \biggl(-
\frac{\ell}{2\sqrt{s}} \biggr) \biggr).
\end{equation}
Of course, $\frac{\partial F_1}{\partial\ell}(s,\ell^\star(s))=0$.
Then, \emph{at any critical point} of $\ell\mapsto F(s,\ell)$, we
have (using the fact that $\frac{\partial F_1}{\partial\ell}(s,\ell)=0$)
$\frac{\partial^2 F_1}{\partial\ell^2}(s,\ell)=\tilde{\rho
}(s,\ell)$ where
\[
\tilde{\rho}(s,\ell)= \biggl(-\frac{2}{\ell^2} -1+s \biggr)
F_1(s,\ell) - 2\ell\sqrt{\frac{2s}{\uppi}} \exp \biggl(-
\frac
{\ell
^2}{8s} \biggr) + 3 \ell^2 \Phi \biggl(-
\frac{\ell}{2\sqrt {s}} \biggr)
\]
so that $\frac{\partial^2 F_1}{\partial\ell^2}(s,\ell)=\rho
(s,\ell)$ with
(using again $\frac{\partial F_1}{\partial\ell}(s,\ell)=0$ to
eliminate $F_1(s,\ell)$)
\begin{eqnarray*}
\rho(s,\ell) &=& \ell\frac{-\ell^2+s\ell^2 +6}{\ell^2 - s \ell^2 -2}\sqrt {\frac{2s}{\uppi}} \exp
\biggl(-\frac{\ell^2}{8s} \biggr) + 2 \ell^2 \frac{\ell^2- s\ell^2 -4}{\ell^2 - s \ell^2 -2}
\Phi \biggl(-\frac
{\ell}{2\sqrt{s}} \biggr)
\\
&=& 2 \ell^2 \frac{\ell^2- s\ell^2 -4}{\ell^2 - s \ell^2 -2} \chi (s,\ell),
\end{eqnarray*}
where
\[
\chi(s,\ell)=\Phi \biggl(-\frac{\ell}{2\sqrt{s}} \biggr)- \frac
{1}{\ell}
\frac{\ell^2-s\ell^2 -6}{\ell^2- s\ell^2 -4}\sqrt{\frac
{s}{2\uppi}} \exp \biggl(-\frac{\ell^2}{8s}
\biggr).
\]

\subsection{The case $s>1$}

Let us assume $s>1$. In this section, we will prove that the function
$\ell\mapsto\rho(s,\ell)$ is negative on some interval
$(0,\tilde{\ell})$ and positive on $(\tilde{\ell},\infty)$, which is
equivalent to show that $\ell
\mapsto\chi(s,\ell)$ is negative on $(0,\tilde{\ell})$ and
positive on $(\tilde{\ell},\infty)$, since the ratio $ \frac{\ell
^2- s\ell^2 -4}{\ell^2 - s
\ell^2 -2}$ is positive. This implies that $\ell\mapsto
F(s,\ell)$ has a unique global maximum at point
$\ell^\star(s)$. Indeed, if $\ell^\star_1(s)<\ell^\star_2(s)$ are
two points in
$\arg\max_{\ell\ge0} F_1(s,\ell)$, then, $\ell^\star_2(s) <
\tilde{\ell}$ and we reach a contradiction by noticing that there is
necessarily a local minimum of $\ell\mapsto
F(s,\ell)$ in the interval $(\ell^\star_1(s),\ell^\star_2(s))$.

We note that
%
%
\begin{equation}
\label{eq:dchidl} \frac{\partial\chi}{\partial\ell}(s,\ell) = \frac{1}{\sqrt {2\uppi
s}}
\frac{1}{\ell^2}\frac{1}{(\ell^2(1-s)-4)^2} \exp \biggl(-\frac
{\ell^2}{8s} \biggr) P
\bigl(s, \ell^2 \bigr),
\end{equation}
where
\begin{eqnarray*}
P(s,y)&=&-\frac{(1-s)^2}{4} y^3 + (1-s) \biggl(s(1-s)+
\frac{3} 2 \biggr)y^2-\bigl(2+14s(1-s)\bigr)y + 24 s
\\
&=&-\frac{(1-s)^2}{4} \biggl( y- \frac{2}{1-s} \biggr) \biggl(
y^2 - 4 \biggl( \frac{1}{1-s}+s \biggr)y + \frac{48 s}{1-s}
\biggr).
\end{eqnarray*}
We will show that $y \mapsto P(s,y)$ is positive on some interval
$(0,\bar{\ell})$ and negative on $(\bar{\ell},\infty)$. This means
that $\ell
\mapsto\chi(s,\ell)$ is increasing on $(0,\bar{\ell})$ and
decreasing on
$(\bar{\ell},\infty)$. Since $\lim_{\ell\to0} \chi(s,\ell)=-
\infty$,
$\lim_{\ell\to\infty} \chi(s,\ell)=0$ and $\ell\mapsto\chi
(s,\ell)$ is a
${\mathcal C}^\infty$ function, this implies that $\ell\mapsto\chi
(s,\ell)$ is negative on some interval $(0,\tilde{\ell})$ and
positive on $(\tilde{\ell},\infty)$, which concludes the proof.

Let us now study the polynomial $y \mapsto P(s,y)$. Let us introduce
\[
Q(s,y)= y^2 - 4 \biggl( \frac{1}{1-s}+s \biggr)y +
\frac{48 s}{1-s}.
\]
The discriminant of $y \mapsto Q(s,y)$ is
\[
\Delta(s)=\frac{16}{(1-s)^2} \bigl( s^2(1-s)^2 -10
s(1-s) + 1 \bigr).
\]
Since $s>1$, and thus $s(1-s)<0$, then $\Delta(s)>0$. The polynomial
$y\mapsto Q(s,y)$ has two roots:
\[
y_+=2 \biggl( \frac{1}{1-s}+s \biggr) +\frac{2}{|1-s|} \bigl(
s^2(1-s)^2 -10 s(1-s) + 1 \bigr)^{1/2}
\]
and
\[
y_-=2 \biggl( \frac{1}{1-s}+s \biggr) -\frac{2}{|1-s|} \bigl(
s^2(1-s)^2 -10 s(1-s) + 1 \bigr)^{1/2}.
\]
Then, $Q(s,y)<0$ if and only if $y \in(y_-,y_+)$. The roots of $y
\mapsto P(s,y)$ are $\{y_-,y_+,y_0\}$ where
\[
y_0=\frac{2}{1-s}.
\]
We notice that $y_- < y_+$ and $y_+>y_0$. We observe that
\begin{eqnarray*}
y_- < y_0 \quad& \iff&\quad2 \biggl( \frac{1}{1-s}+s
\biggr) -\frac{2}{|1-s|} \bigl( s^2(1-s)^2 -10 s(1-s)
+ 1 \bigr)^{1/2} < \frac{2}{1-s}
\\[-2pt]
& \iff&\quad s< \frac{1}{|1-s|} \bigl( s^2(1-s)^2
-10 s(1-s) + 1 \bigr)^{1/2}
\\[-2pt]
& \iff&\quad s(1-s)< \frac{1}{10}.
\end{eqnarray*}

Thus, since $s>1$, we have
\[
y_-<y_0<0<y_+,
\]
and $y \mapsto P(s,y)$ changes its sign at each of its roots
$\{y_-,y_+,y_0\}$. Since $\lim_{y \to\infty} P(s,y) = -\infty$, we
deduce that $P(s,y)>0$ for $y \in(0,y_+)$ and $P(s,y)<0$ for $y \in
(y_+,\infty)$. This concludes the proof in the case $s>1$.

\subsection{The case $s<1$}

First, we observe that the maximum of $\ell\mapsto F_1(s,\ell)$ is
necessarily in
$ (0,\ell_0 )$ where
\[
\ell_0=\sqrt{y_0}=\sqrt{\frac{2}{1-s}}.
\]
Indeed, if $\ell^2\ge\frac{2}{1-s}$,
we have (using the fact that $F_1 \ge0$ and the upper bound in the
classical inequality \eqref{contPhi}):
\begin{eqnarray*}
\frac{\partial F_1}{\partial\ell}(s,\ell) &=& \biggl(\frac{2}{\ell} - \ell(1-s) \biggr)
F_1(s,\ell) + \ell ^2 \biggl( - \sqrt{
\frac{2s}{\uppi}} \exp \biggl(-\frac{\ell^2}{8s} \biggr) + \ell\Phi \biggl(-
\frac{\ell}{2\sqrt{s}} \biggr) \biggr)
\\[-2pt]
&<& \ell^2 \biggl( - \sqrt{\frac{2s}{\uppi}} \exp \biggl(-
\frac{\ell^2}{8s} \biggr) + \sqrt{\frac{2s}{\uppi}} \exp \biggl(-
\frac{\ell^2}{8s} \biggr) \biggr)
\\[-2pt]
&<&0.\vadjust{\goodbreak}
\end{eqnarray*}
This shows in particular that $\ell^\star(s) \in(0,\ell_0)$. In all what
follows, we only study the function $\ell\mapsto F_1(s,\ell)$ for
\[
\ell\in(0,\ell_0).
\]
We need to prove that $\ell\mapsto F_1(s,\ell)$ admits a unique global
maximum on $ (0,\ell_0 )$. A sufficient condition is that
$\ell
\mapsto\rho(s,\ell)$ is negative for $\ell< \ell_0$.

Notice that the function $\ell\mapsto
\chi(s,\ell)$ is ${\mathcal C}^\infty([0,\ell_0))$, has the same
sign as
$\rho(s,\ell)$ and that $\lim_{\ell\to0}
\chi(s,\ell)=-\infty$ while
%
%
\begin{eqnarray}
\label{eq:majochi} \chi(s,\ell_0)&=&\Phi \biggl(-\frac{\ell_0}{2\sqrt{s}}
\biggr)- \frac{1}{\ell_0}\frac{\ell_0^2(1-s) -6}{\ell_0^2( 1- s)
-4}\sqrt{\frac{s}{2\uppi}} \exp
\biggl(-\frac{\ell_0^2}{8s} \biggr)
\nonumber
\\[-8pt]
\\[-8pt]
&=&\Phi \biggl(-\frac{\ell_0}{2\sqrt{s}} \biggr)- \frac{1}{\ell_0}\sqrt{
\frac{2s}{\uppi}} \exp \biggl(-\frac{\ell
_0^2}{8s} \biggr)
\nonumber
\end{eqnarray}
which is negative, using the upper bound in the classical inequality
\eqref{contPhi}.

Let us now study the sign of $\chi(s,\ell)$. As in the previous case, we
first study the sign of $\frac{\partial\chi}{\partial\ell}$, namely
the sign of $P$. We distinguish between two cases.

If $s(1-s) \ge5-\sqrt{24}$, then $\Delta(s) \le0$, so that $Q(s,y)
>0$ for $y<y_0$. This implies that $P(s,y)>0$ for $y<y_0$. Therefore,
in view of \eqref{eq:dchidl},
$\frac{\partial\chi}{\partial\ell}(s,\ell)>0$ for
$\ell<\ell_0$. Thus, in this case, $\ell\mapsto\chi(s,\ell)$ is
increasing
from $0$ to $\ell_0$, going from $-\infty$ to $\chi(s,\ell_0)$
which is
negative. In conclusion, $\ell\mapsto\chi(s,\ell)$ is negative on
$(0,\ell_0)$, and $\ell\mapsto F_1(s,\ell)$ admits a unique
global maximum.

Now, if $s(1-s)<5-\sqrt{24}$, $\Delta(s)>0$, so that $y \mapsto
Q(s,y)$ has two roots $y_+ > y_-$. We recall that $y_-<y_0 \iff
s(1-s)< \frac{1}{10}$ and notice that $\frac{1}{10}<5-\sqrt{24}$. Let
us thus distinguish between two subcases.

If $s(1-s)\in [ \frac{1}{10}, 5-\sqrt{24}  )$, then $0 <
y_0 \le y_- < y_+$. The polynomial $y \mapsto P(s,y)$
changes its sign at each of its roots $\{y_0,y_-,y_+\}$, and $\lim_{y
\to\infty} P(s,y) = -\infty$. Thus,
in this case, $\ell\mapsto\chi(s,\ell)$ is increasing from $0$ to
$\ell_0$, going from $-\infty$ to $\chi(s,\ell_0)$ which is
negative. In
conclusion, $\chi(s,\ell)$ is negative, and $\ell\mapsto F_1(s,\ell
)$ admits a
unique global maximum.

The last subcase to consider is $s(1-s)<\frac{1}{10}$, which is
equivalent to
\[
s \in(0,s_0) \cup(s_1,1)
\]
with
\[
s_0=\tfrac{1}{2} \Bigl( 1 - \sqrt{\tfrac{3}{5}} \Bigr)
\quad \mbox{and}\quad s_1=\tfrac{1}{2} \Bigl( 1 + \sqrt{
\tfrac{3}{5}} \Bigr).
\]
In this case, $0 < y_- < y_0 < y_+$. Indeed (using the fact that $s<1$),
\begin{eqnarray*}
y_->0 \quad& \iff&\quad \bigl( 1+s(1-s) \bigr) > \bigl( s^2(1-s)^2
-10 s(1-s) + 1 \bigr)^{1/2}
\\
& \iff&\quad s(1-s)>0,
\end{eqnarray*}
which is true. The polynomial $y \mapsto P(s,y)$ changes its sign at
each of its roots
$\{y_-,y_0,y_+\}$, and $\lim_{y \to\infty} P(s,y) = -\infty$. Let us
denote $\ell_-=\sqrt{y_-}$. Thus,
in this case, $\ell\mapsto\chi(s,\ell)$ is increasing from $0$ to
$\ell_-$ (going from $-\infty$ to $\chi(s,\ell_-)$) and then
decreasing from
$\ell_-$ to $\ell_0$ (going from $\chi(s,\ell_-)$ to $\chi(s,\ell
_0)$, which is negative). Thus, if
$\chi(s,\ell_-)<0$, then $\chi(s,\ell)$ is negative, and $\ell
\mapsto
F_1(s,\ell)$ admits a unique global maximum.

In conclusion, $\ell\mapsto
F_1(s,\ell)$ admits at least one local maximum and at most two local
maxima. The function $\ell\mapsto
F_1(s,\ell)$ admits two local maxima $\ell^\star_1<\ell^\star_2$
if and only if
$\chi(s,\ell_-)\geq0$, in which case $\ell^\star_1<\ell_-<\ell
^\star_2$, and $\frac{\partial F_1}{\partial\ell}(s,\ell^\star
_1)=\frac{\partial F_1}{\partial\ell}(s,\ell^\star_2)=0$.

\subsubsection{The case $s\in(0,s_0)$}

Let us assume
the existence of $\underline{s}\in(0,s_0)$ such that
$\ell\mapsto F_1(\underline{s},\ell)$ admits two local maxima
$\ell^\star_1(\underline{s})<\ell^\star_2(\underline{s})$ and let
us show
that
%
%
\begin{equation}
\label{eq:aa} \exists(s,\ell)\in[\underline{s},s_0]\times\R_+^*,
\qquad\frac
{\partial F_1}{\partial\ell}(s,\ell)=\frac{\partial^2
F_1}{\partial
\ell^2}(s,\ell)=0.
\end{equation}
If $\frac{\partial^2 F_1}{\partial
\ell^2}(\underline{s},\ell^\star_1(\underline{s}))=0$ or $\frac
{\partial^2
F_1}{\partial\ell^2}(\underline{s},\ell^\star_2(\underline
{s}))=0$, we are done.
Otherwise, we may apply the implicit function theorem to construct for
$i\in\{1,2\}$ a continuous curve $\ell^\star_i(s)$ on a maximal\vspace*{-2pt}
interval $[\underline{s},\bar{s}_i)$ with $\bar{s}_i>\underline{s}$
such that for
$s\in[\underline{s},\bar{s}_i)$, $\frac{\partial F_1}{\partial
\ell}(s,\ell^\star_i(s))=0$ and $\frac{\partial^2 F_1}{\partial
\ell^2}(s,\ell^\star_i(s))<0$. In case $\min(\bar{s}_1,\bar
{s}_2)>s_0$, then, since by the uniqueness part of the implicit function
theorem, $\forall s\in[\underline{s},\min(\bar{s}_1,\bar{s}_2))$,
$\ell^\star_1(s)<\ell^\star_2(s)$, we contradict the fact that
$\ell\mapsto F_1(s_0,\ell)$ admits a unique
local maximum. Thus, choosing $i\in\{1,2\}$ such that $\bar{s}_i=\min
(\bar{s}_1,\bar{s}_2)$, one has $\bar{s}_i\le s_0$.
Since $\ell^\star_i(s)<\ell_0(s)=\sqrt{\frac{2}{1-s}}$, we may
find an increasing
sequence $(s_n)_{n\in{\mathbb N}}$ of elements of
$[\underline{s},\overline{s}_i)$ converging to $\overline{s}_i$ and
such that $\ell^\star_i(s_n)$ converges to some limit denoted by
$\ell^\star_i(\overline{s}_i)$ as $n\to\infty$. By continuity of
$\frac{\partial F_1}{\partial\ell}(s,\ell)$ and $\frac{\partial^2
F_1}{\partial\ell^2}(s,\ell)$, one has $\frac{\partial
F_1}{\partial
\ell}(\overline{s}_i,\ell^\star_i(\overline{s}_i))=0$ and
$\frac{\partial^2 F_1}{\partial
\ell^2}(\overline{s}_i,\ell^\star_i(\overline{s}_i))\leq0$.\vspace*{1pt} Let
us now
consider $\ell^\star_{3-i}(\overline{s}_i)$, defined as the limit of a
converging subsequence of $(\ell^\star_{3-i}(s_n))_n$ in case
$\overline{s}_{3-i}=\overline{s}_{i}$. If
$\ell^\star_{1}(\overline{s}_i)=\ell^\star_{2}(\overline{s}_i)$,
then from
the existence of a local minimum
$\ell\in(\ell^\star_{1}(s_n),\ell^\star_{2}(s_n))$ such that
$\frac{\partial^2
F_1}{\partial\ell^2}(s_n,\ell)\geq0$, we conclude that $\frac
{\partial^2
F_1}{\partial
\ell^2}(\overline{s}_i,\ell^\star_{i}(\overline{s}_i))=0$. If
$\ell^\star_{1}(\overline{s}_i)<\ell^\star_{2}(\overline{s}_i)$
and\vspace*{-2pt} both
$\frac{\partial^2 F_1}{\partial
\ell^2}(\overline{s}_i,\ell^\star_{1}(\overline{s}_i))$ and
$\frac{\partial^2 F_1}{\partial
\ell^2}(\overline{s}_i,\ell^\star_{2}(\overline{s}_i))$ are negative,
then, using the implicit function\vspace*{1pt} theorem, we contradict the
maximality of $\overline{s}_i$. This concludes the proof
of \eqref{eq:aa}.


Let us consider\vspace*{-2pt} a point $(s,\ell)$ such that $\frac{\partial
F_1}{\partial
\ell}(s,\ell)=\frac{\partial^2 F_1}{\partial
\ell^2}(s,\ell)=0$, where $s\in[0,s_0] \cup[s_1,1]$ and $\ell^2 <
\frac
{2}{1-s}$. From $\frac{\partial F_1}{\partial
\ell}(s,\ell)=0$, we get:
\[
F_1(s,\ell) = - \frac{\ell^3}{2 - \ell^2 (1-s)} \biggl( - \sqrt {
\frac{2s}{\uppi}} \exp \biggl(-\frac{\ell^2}{8s} \biggr) + \ell \Phi \biggl(-
\frac{\ell}{2\sqrt{s}} \biggr) \biggr).
\]
From $\frac{\partial^2 F_1}{\partial
\ell^2}(s,\ell)=0$, which implies $\chi(s,\ell)=0$ (since $\frac
{\partial F_1}{\partial
\ell}(s,\ell)=0$), we get:
%
%
\begin{equation}
\label{eq:phiexp} \Phi \biggl(-\frac{\ell}{2\sqrt{s}} \biggr)= \frac{1}{\ell}
\frac
{\ell^2(1-s) -6}{\ell^2(1- s) -4}\sqrt{\frac{s}{2\uppi}} \exp \biggl(-\frac{\ell^2}{8s}
\biggr).
\end{equation}
By combining these two relations, we have
\begin{eqnarray*}
F_1(s,\ell) &=& - \frac{\ell^3}{2 - \ell^2 (1-s)} \biggl( - \sqrt{
\frac{2s}{\uppi}} \exp \biggl(-\frac{\ell^2}{8s} \biggr) +
\frac{\ell^2(1-s) -6}{\ell^2(1- s) -4}\sqrt{\frac{s}{2\uppi}} \exp \biggl(-\frac{\ell^2}{8s}
\biggr) \biggr)
\\
&=& - \frac{\ell^3}{\ell^2(1- s) -4} \sqrt{\frac{s}{2\uppi}} \exp \biggl(-
\frac{\ell^2}{8s} \biggr).
\end{eqnarray*}
Finally, using the expression for $F_1(s,\ell)$, we get:
\[
\Phi \biggl(-\frac{\ell}{2\sqrt{s}} \biggr)+(1-2s)\exp \biggl(\frac
{\ell^2(s-1)}{2}
\biggr)\Phi \biggl(\frac{\ell}{2\sqrt{s}}-\ell \sqrt{s} \biggr)= \frac{\ell(1-s)}{4-\ell^2(1- s)}
\sqrt{\frac
{s}{2\uppi}} \exp \biggl(-\frac{\ell^2}{8s} \biggr).
\]
Using again \eqref{eq:phiexp}, this yields
\begin{eqnarray*}
&&\frac{1}{\ell}\frac{\ell^2(1-s) -6}{\ell^2(1- s)
-4}\sqrt{\frac{s}{2\uppi}} \exp \biggl(-
\frac{\ell^2}{8s} \biggr) +(1-2s)\exp \biggl(\frac{\ell^2(s-1)}{2} \biggr)\Phi
\biggl(\frac
{\ell}{2\sqrt{s}}-\ell\sqrt{s} \biggr)
\\
&&\quad= \frac{\ell(1-s)}{4-\ell^2(1- s)} \sqrt{\frac{s}{2\uppi
}} \exp \biggl(-
\frac{\ell^2}{8s} \biggr),
\end{eqnarray*}
which implies
%
%
\begin{equation}
\label{eq:2} (1-2s)\Phi \biggl(\frac{\ell}{2\sqrt{s}}-\ell\sqrt{s} \biggr)=
\frac{2}{\ell} \frac{\ell^2(1-s)-3}{4-\ell^2(1- s)} \sqrt{\frac
{s}{2\uppi}} \exp \biggl(-
\frac{\ell^2}{8s} (1-2s)^2 \biggr).
\end{equation}

We notice that the right-hand side is negative, so that this equation
has no solution if $1-2s >0$, which leads to a contradiction with
\eqref{eq:aa} in the case
$s \in[0,s_0]$. In conclusion, in the case $s \in(0,s_0)$, $\ell
\mapsto
F_1(s,\ell)$ admits only one local maximum at point $\ell^\star(s)$,
which is
also a global maximum.

\subsubsection{The case $s\in(s_1,1)$}

In the case $s \in(s_1,1)$, we need another argument.
%
%
\begin{Lemma}\label{lem:ll}
Let us consider $s \in(s_1,1)$ and $\ell\in[0,\ell_0(s)]$ such
that $\frac{\partial F_1}{\partial
\ell}(s,\ell)=\frac{\partial^2 F_1}{\partial
\ell^2}(s,\ell)=0$. Then, $\ell< \ell_-(s)$.
\end{Lemma}
\begin{pf}
We know from the previous computations that $(s,\ell)$
satisfies \eqref{eq:2}. Using
the lower bound in the classical inequality \eqref{contPhi}, we get
\[
\Phi \biggl(\frac{\ell}{2\sqrt{s}}-\ell\sqrt{s} \biggr) = \Phi \biggl(-
\frac{\ell(2s-1)}{2\sqrt{s}} \biggr) > \frac{
{\ell(2s-1)}/({2\sqrt{s}})}{1+{\ell^2(2s-1)^2}/({4s})} \exp \biggl( -
\frac{\ell^2(2s-1)^2}{8s} \biggr)\frac{1}{\sqrt{2
\uppi}}.
\]
From \eqref{eq:2}, we thus obtain (since $1-2s < 0$)
\begin{eqnarray*}
&&\frac{2}{\ell} \frac{\ell^2(1-s)-3}{4-\ell^2(1- s)} \sqrt{\frac
{s}{2\uppi}} \exp \biggl(-
\frac{\ell^2}{8s} (1-2s)^2 \biggr)
\\
&&\quad<(1-2s)\frac{{\ell
(2s-1)}/({2\sqrt{s}})}{1+{\ell^2(2s-1)^2}/({4s})} \exp
\biggl( -\frac{\ell^2(2s-1)^2}{8s} \biggr)\frac{1}{\sqrt{2
\uppi}}
\end{eqnarray*}
which implies
\[
\frac{\ell^2(1-s)-3}{4-\ell^2(1- s)} <- \frac{\ell^2
(2s-1)^2}{4s+\ell^2(2s-1)^2 }
\]
and then (since $\ell^2 (1-s) < 2$)
\[
\bigl(\ell^2(1-s)-3\bigr) \bigl(4s+\ell^2(2s-1)^2
\bigr) < -\ell^2 (2s-1)^2\bigl(4-\ell^2(1- s)
\bigr).
\]
This implies that
\[
\ell^2< 12s.
\]
On the other hand, it is easy to check that
\[
(\ell_-)^2 > 12 s.
\]
Indeed
\begin{eqnarray*}
(\ell_-)^2 > 12 s \quad& \iff&\quad2 \biggl( \frac{1}{1-s}+s
\biggr) -\frac{2}{|1-s|} \bigl( s^2(1-s)^2 -10 s(1-s)
+ 1 \bigr)^{1/2} > 12 s
\\
& \iff&\quad 1 -5 s (1-s) > \bigl( s^2(1-s)^2 -10
s(1-s) + 1 \bigr)^{1/2}
\\
& \iff&\quad 1 -10 s (1-s) + 25 s^2 (1-s)^2>
s^2(1-s)^2 -10 s(1-s) + 1
\end{eqnarray*}
which is obviously true. Thus, \eqref{eq:2} implies $\ell<\ell_-$.
\end{pf}

Let us now assume the existence of $\underline{s}\in(s_1,1)$ such that
$\ell\mapsto F_1(\underline{s},\ell)$ admits two local maxima
$\ell^\star_1(\underline{s})<\ell^\star_2(\underline{s})$.
We recall that necessarily, $\chi(\underline{s},\ell_-(\underline
{s}))\geq
0$ and $\ell^*_1(\underline{s})<\ell_-(\underline{s})<\ell^\star
_2(\underline{s})$. Lemma~\ref{lem:dchids} below shows that $\frac{d}{ds}
\chi(s,\ell_-(s))>0$ for $s \in(s_1,1)$. This implies that $\forall
s\in(\underline{s},1]$, $
\chi(s,\ell_-(s))> 0$. Using the implicit function theorem, we can
construct, a continuous curve $\ell^\star_2(s)$ on a maximum interval
of the form $s\in[\underline{s},\overline{s})$ with $\overline
{s}>\underline{s}$ such that\vspace*{-1pt} for $s \in[\underline{s},\overline
{s})$, $\frac{\partial F_1}{\partial\ell} (s,\ell^\star_2(s))=0$,
$\frac{\partial^2 F_1}{\partial\ell^2} (s,\ell^\star_2(s))
<0$ and thus $\chi(s,\ell^\star_2(s))
<0$. Due to the respective signs of the continuous function $\chi
(s,\ell)$ on the two continuous curves $s\mapsto\ell^\star_2(s)$ and
$s\mapsto\ell_-(s)$, these curves cannot intersect on
$[\underline{s},\min(\overline{s},1))$. Therefore, $\forall
s\in[\underline{s},\min(\overline{s},1))$, $\ell_2^\star(s)>\ell
_-(s)$. We
now distinguish between three cases.

If $\overline{s}>1$, then $\ell^\star_2(1)\geq\ell_-(1)=\sqrt {12}$ whereas $\frac{\partial F_1}{\partial\ell} (1,\ell^\star
_2(1))=0$ and $\frac{\partial^2 F_1}{\partial\ell^2} (1,\ell^\star
_2(1))<0$ so that we contradict \eqref{eq:d2F_s=1}.\vspace*{-1pt}

If $\overline{s}<1$, then since
$\ell^\star_2(s)<\ell_0(s)=\sqrt{\frac{2}{1-s}}$, we may find an increasing
sequence $(s_n)_{n\in{\mathbb N}}$ of elements of
$[\underline{s},\overline{s})$ converging to $\overline{s}$ and such
that $\ell^\star_2(s_n)$ converges to some limit denoted\vspace*{-1pt} by
$\ell^\star_2(\overline{s})$ and which belongs to
$[\ell_-(\overline{s}),\ell_0(\overline{s})]$. By continuity of
$\frac{\partial F_1}{\partial\ell}(s,\ell)$ and $\frac{\partial^2
F_1}{\partial\ell^2}(s,\ell)$, one\vspace*{-1pt} has $\frac{\partial
F_1}{\partial
\ell}(\overline{s},\ell^\star_2(\overline{s}))=0$, $\frac
{\partial^2
F_1}{\partial\ell^2}(\overline{s},\ell^\star_2(\overline{s}))\leq
0$ and
thus $\chi(\overline{s},\ell^\star_2(\overline{s}))\leq0$. This\vspace*{-1pt} implies
that $\ell^\star_2(\overline{s})>\ell_-(\overline{s})$ since
$\chi(\overline{s}, \ell_-(\overline{s})) > 0$. In turn, this
implies, by
Lemma~\ref{lem:ll}, that $\frac{\partial^2 F_1}{\partial
\ell^2}(\overline{s},\ell^\star_2(\overline{s}))<0$. Combining the
implicit function theorem with the uniqueness of local maxima of
$\ell\mapsto F_1(s,\ell)$ for $\ell\ge\ell_-(s)$, we contradict
the maximality of $\overline{s}$.

Let us finally consider the case $\overline{s}=1$. We are
going to check that $\frac{\partial F_1}{\partial\ell}(s,\ell)$ is negative
for $\ell$ large uniformly in $s\in(\frac{1}{2},1)$ (see
Lemma~\ref{lem:df1neg}) so that
$\ell^\star_2(s)$ remains bounded in the limit $s \to1$. This implies
that we may find an increasing sequence $(s_n)_{n\in{\mathbb N}}$ of
elements of
$[\underline{s},1)$ converging to $1$ and such
that $\ell^\star_2(s_n)$ converges to some limit denoted by
$\ell^\star_2(1) \geq\ell_-(1)=\sqrt{12}$. By continuity of
$\frac{\partial F_1}{\partial\ell}(s,\ell)$ and\vspace*{1pt} $\frac{\partial^2
F_1}{\partial\ell^2}(s,\ell)$, one has $\frac{\partial
F_1}{\partial
\ell}(1,\ell^\star_2(1))=0$ and $\frac{\partial^2
F_1}{\partial\ell^2}(1,\ell^\star_2(1))\leq0$ but this contradicts
\eqref{eq:d2F_s=1}, and concludes the proof of Lemma~\ref{lem:lstar}.

%
\begin{Lemma}\label{lem:df1neg}
There exists $L>0$ and $\alpha< 0$ such that, for all $\ell> L$ and for
all $s \in(s_1, 1)$, $\frac{\partial F_1}{\partial\ell}(s,\ell)<
\alpha$.
\end{Lemma}
\begin{pf} Let
$s\in(s_1,1)$. By \eqref{dlf1} and nonnegativity of $F_1$, one has
\begin{eqnarray*}
\frac{\partial F_1}{\partial\ell}(s,\ell)&\leq&\frac{2}{\ell} F_1(s,\ell) +
\ell^2 \biggl( - \sqrt{\frac{2s}{\uppi}} \exp \biggl(-
\frac{\ell^2}{8s} \biggr) + \ell\Phi \biggl(-\frac{\ell}{2\sqrt {s}} \biggr) \biggr)
\\
&=&\frac{2\ell}{1-s} \biggl(1+\frac{\ell^2}{2}(1-s)+(1-2s)\exp { \biggl(
\frac{\ell^2(s-1)}{2} \biggr)} \biggr)\Phi \biggl(-\frac
{\ell}{2\sqrt {s}} \biggr)
\\
&&{} -\frac{2\ell(2s-1)}{1-s}\exp{ \biggl(\frac{\ell
^2(s-1)}{2} \biggr)}\int
_{{\ell(2s-1)}/({2\sqrt{s}})}^{{\ell}/({2\sqrt{s}})}\exp \biggl(-\frac{x^2}{2} \biggr)
\frac{\mathrm{d}x}{\sqrt{2\uppi
}}
\\
&&{}-\frac{2\ell
^2\sqrt{s}}{\sqrt{2\uppi}} \exp \biggl(-\frac{\ell^2}{8s} \biggr).
\end{eqnarray*}
Using two integrations by parts, one obtains
\begin{eqnarray*}
\int_{{\ell(2s-1)}/({2\sqrt{s}})}^{{\ell}/({2\sqrt
{s}})}\mathrm{e}^{-{x^2}/{2}}\,
\mathrm{d}x&\geq& \biggl(\frac{2\sqrt {s}}{\ell
(2s-1)}-\frac{8s^{3/2}}{\ell^3(2s-1)^3} \biggr)\exp
\biggl(-\frac
{\ell^2(2s-1)^2}{8s} \biggr)
\\
&&{}- \biggl(\frac{2\sqrt{s}}{\ell}-
\frac
{8s^{3/2}}{\ell^3} \biggr)\exp \biggl(-\frac{\ell^2}{8s} \biggr)
\end{eqnarray*}
and
\[
\Phi \biggl(-\frac{\ell}{2\sqrt{s}} \biggr)=\frac{1}{\sqrt{2\uppi
}} \biggl(
\frac{2\sqrt{s}}{\ell}-\frac{8s^{3/2}}{\ell^3}+{\mathcal O} \biggl(\frac{1}{\ell^5}
\biggr) \biggr)\exp \biggl(-\frac{\ell
^2}{8s} \biggr),
\]
with the term ${\mathcal O} (\frac{1}{\ell^5} )$ uniform
in $s\in(s_1,1)$.
Using the fact that
\begin{eqnarray*}
&&\frac{2\ell}{1-s} \biggl(1+\frac{\ell^2}{2}(1-s)+(1-2s)\exp { \biggl(
\frac{\ell^2(s-1)}{2} \biggr)} \biggr)
\\
&&\quad=\ell^3+4\ell\exp{ \biggl(\frac{\ell^2(s-1)}{2} \biggr)}+
\frac{2\ell
}{1-s} \biggl(1-\exp{ \biggl(\frac{\ell^2(s-1)}{2} \biggr)} \biggr)
\end{eqnarray*}
we get, since $s < 1$,
\[
0\leq\frac{2\ell}{1-s} \biggl(1+\frac{\ell^2}{2}(1-s)+(1-2s)\exp { \biggl(
\frac{\ell^2(s-1)}{2} \biggr)} \biggr) \leq2\ell^3+4\ell.
\]
Thus, we get
\begin{eqnarray*}
&& \sqrt{2\uppi}\exp \biggl(\frac{\ell^2}{8s} \biggr) \frac
{\partial
F_1}{\partial\ell}(s,
\ell)
\\
&&\quad\le \biggl( \ell^3+4\ell\exp{ \biggl(\frac{\ell
^2(s-1)}{2}
\biggr)}
 +\frac
{2\ell}{1-s} \biggl(1-\exp{ \biggl(\frac{\ell^2(s-1)}{2} \biggr)}
\biggr) \biggr)
\\
&&\qquad{}\times \biggl(\frac{2\sqrt{s}}{\ell}-\frac{8s^{3/2}}{\ell^3}+{\mathcal O}
\biggl(\frac{1}{\ell^5} \biggr) \biggr)
\\
&&\qquad{} -\frac{2\ell(2s-1)}{1-s} \biggl(\frac{2\sqrt{s}}{\ell(2s-1)}-\frac{8s^{3/2}}{\ell
^3(2s-1)^3}
\biggr)
\\
&&\qquad{}+\frac{2\ell(2s-1)}{1-s}\exp{ \biggl(\frac{\ell^2(s-1)}{2} \biggr)} \biggl(
\frac{2\sqrt{s}}{\ell}-\frac{8s^{3/2}}{\ell^3} \biggr)-2\ell ^2\sqrt{s}
\\
&&\quad= - 8 s^{3/2} + 8 \sqrt{s} \exp{ \biggl(\frac{\ell
^2(s-1)}{2}
\biggr)} + \frac{4
\sqrt{s}}{1-s} \biggl(1-\exp{ \biggl(\frac{\ell^2(s-1)}{2} \biggr)}
\biggr)
\\
&&\qquad{}- \frac{16
s^{3/2}}{\ell^2(1-s)} \biggl(1-\exp{ \biggl(\frac{\ell
^2(s-1)}{2} \biggr)}
\biggr)
\\
&&\qquad{} - \frac{4 \sqrt{s}}{1-s} + \frac{16 s^{3/2}}{\ell
^2(2s-1)^2(1-s)} + \exp{ \biggl(
\frac{\ell^2(s-1)}{2} \biggr)} \frac{4\sqrt{s}(2s-1)}{1-s}
\\
&&\qquad{}-\exp{ \biggl(\frac{\ell^2(s-1)}{2}
\biggr)} \frac
{16(2s-1)s^{3/2}}{\ell
^2(1-s)} +{\mathcal O} \biggl(\frac{1}{\ell^2} \biggr)
\\
&&\quad= - 8 s^{3/2} + \frac{16
s^{3/2} 2 (1-s)}{\ell^2(1-s)} \exp{ \biggl(
\frac{\ell
^2(s-1)}{2} \biggr)} + \frac{16 s^{3/2} 4 s (1-s)}{\ell^2(2s-1)^2(1-s)}+{\mathcal O} \biggl(
\frac{1}{\ell^2} \biggr).
\end{eqnarray*}
Therefore, one concludes that
\[
\sqrt{2\uppi}\exp \biggl(\frac{\ell^2}{8s} \biggr)\frac{\partial
F_1}{\partial\ell}(s,\ell)
\leq-8s^{3/2}+{\mathcal O} \biggl(\frac
{1}{\ell^2} \biggr),
\]
which indeed shows that $\frac{\partial F_1}{\partial\ell}(s,\ell)$ is
negative for $\ell$ large uniformly in $s\in(s_1,1)$.
\end{pf}

To conclude the proof, we need to prove the following lemma which has
been used above.
%
%
\begin{Lemma}\label{lem:dchids}
The function $s \mapsto\frac{\mathrm{d}}{\mathrm{d}s} \chi(s,\ell
_-(s))$ is positive for
$s \in(s_1,1)$.
\end{Lemma}
\begin{pf} Let us consider the derivative $\frac{\mathrm{d}}{\mathrm
{d}s} \chi
(s,\ell_-(s))$. Using the fact that $\frac{\partial
\chi}{\partial\ell}(s,\ell_-(s))=0$, we obtain that
\[
\frac{\mathrm{d}}{\mathrm{d}s} \chi\bigl(s,\ell_-(s)\bigr) = \frac{\partial\chi}{\partial s}\bigl(s,
\ell_-(s)\bigr) 
=\frac{1}{\sqrt{2 \uppi}} \exp \biggl( - \frac{\ell_-^2}{8s} \biggr)
\frac{1}{(\ell_-^2- s\ell_-^2
-4)^2 s^{3/2} \ell_-} \xi(s),
\]
where
\[
\xi(s)= \frac{\ell_-^2 (\ell_-^2(1- s)
-4)^2}{4} + 2 \ell_-^2 s^2 - \biggl(
\frac{s}{2} + \frac{\ell
_-^2}{8} \biggr) \bigl(\ell_-^2(1-s)
-6\bigr) \bigl(\ell_-^2(1- s) -4\bigr),
\]
where, here and in the following, $\ell_-$ should be understood as
$\ell_-(s)$. Notice that $\frac{\mathrm{d}}{\mathrm{d}s} \chi
(s,\ell_-(s))$ has the
same sign
as $\xi(s)$. By simple computations, we get:
\begin{eqnarray*}
\xi(s)&=& \frac{\ell_-^6(1- s)^2 - 8 \ell_-^4(1- s) + 16\ell
_-^2}{4} + 2 \ell_-^2 s^2
\\
&&{}- \biggl(
\frac{s}{2} + \frac{\ell_-^2}{8} \biggr) \bigl(\ell_-^4(1-s)^2
- 10 \ell_-^2(1- s) + 24\bigr)
\\
&=& \frac{\ell_-^6(1- s)^2}{8} - \frac{3}{4} \ell_-^4(1- s) + \ell
_-^2 + 2 \ell_-^2 s^2 - \frac{s}{2}
\ell_-^4(1-s)^2 + 5 s \ell_-^2(1- s) - 12 s.
\end{eqnarray*}
By using the fact that $Q(s,\ell_-^2)=0$, namely $\ell_-^4= 4
( \frac{1}{1-s}+s ) \ell_-^2 - \frac{48 s}{1-s}$ to rewrite
the term proportional to $\ell_-^6$, we obtain
\[
\xi(s) = - s \ell_-^2 (1-s) - \tfrac{1}{4}
\ell_-^4(1- s) + \ell_-^2 + 2 \ell_-^2
s^2 - 12 s
\]
so that, using again $\ell_-^4= 4
( \frac{1}{1-s}+s ) \ell_-^2 - \frac{48 s}{1-s}$ to rewrite
the term proportional to $\ell_-^4$,
\[
\xi(s) = 2 s \ell_-^2 ( 2s - 1 )
\]
which is positive for $s \in(s_1,1)$. This concludes the proof.
\end{pf}

\end{appendix}

\section*{Acknowledgements}

This work is supported by the French National
Research Agency under the grants ANR-08-BLAN-0218 (BigMC),
ANR-09-BLAN-0216-01 (MEGAS) and ANR-12-BLAN-Stab.


%

\printhistory

\end{document}